\documentclass[aps,prx,twocolumn,floatfix,10pt,superscriptaddress]{revtex4-2} 

\usepackage{url}
\usepackage{amsmath,amsfonts,amssymb,graphicx,bbm,times}
\usepackage{dsfont}
\usepackage{listings}
\usepackage[usenames,dvipsnames,table,xcdraw]{xc olor}   
\usepackage{hyperref}
\usepackage{comment}
\usepackage{cleveref} 

\usepackage[utf8]{inputenc} 

\usepackage[normalem]{ulem} 

\hypersetup{colorlinks=true,
						linkcolor=blue,      
						citecolor=blue,            
						filecolor=blue,             
						urlcolor=cyan              
} 

\preprint{\bf PREPRINT} 	

\begin{document}

\def\>{\rangle}
\def\<{\langle}
\def\({\left(}
\def\){\right)}
\newcommand{\ket}[1]{\left|#1\right>}
\newcommand{\bra}[1]{\left<#1\right|}
\newcommand{\braket}[2]{\<#1|#2\>}
\newcommand{\ketbra}[2]{\left|#1\right>\!\left<#2\right|}
\newcommand{\proj}[1]{|#1\>\!\<#1|}
\newcommand{\avg}[1]{\< #1 \>}
\renewcommand{\tensor}{\otimes}
\newcommand{\einfuegen}[1]{\textcolor{PineGreen}{#1}}
\newcommand{\streichen}[1]{\textcolor{red}{\sout{#1}}}
\newcommand{\todo}[1]{\textcolor{blue}{(ToDo: #1)}}
\newcommand{\transpose}[1]{{#1}^t}
\newcommand{\om}[1]{\textcolor{red}{#1}}
\newcommand{\jl}[1]{\textcolor{red}{#1}}
\newcommand{\mbp}[1]{\textcolor{green}{#1}}

\title{Driving Force and Nonequilibrium Vibronic Dynamics\\ in Charge Separation of Strongly Bound Electron-Hole Pairs}
\author{Alejandro D. Somoza} 
\email{alejandro.somoza@dlr.de}
\affiliation{Insitut f\"ur Theoretische Physik and IQST, Albert-Einstein-Allee 11, Universit\"at Ulm, D-89069 Ulm, Germany}
\affiliation{German Aerospace Center (DLR), Institute of Engineering Thermodynamics, Wilhelm-Runge Straße 10, 89081 Ulm}
\author{Nicola Lorenzoni}
\affiliation{Insitut f\"ur Theoretische Physik and IQST, Albert-Einstein-Allee 11, Universit\"at Ulm, D-89069 Ulm, Germany}
\author{James Lim}
\affiliation{Insitut f\"ur Theoretische Physik and IQST, Albert-Einstein-Allee 11, Universit\"at Ulm, D-89069 Ulm, Germany}
\author{Susana F. Huelga}
\affiliation{Insitut f\"ur Theoretische Physik and IQST, Albert-Einstein-Allee 11, Universit\"at Ulm, D-89069 Ulm, Germany}
\author{Martin B. Plenio}
\email{martin.plenio@uni-ulm.de}
\affiliation{Insitut f\"ur Theoretische Physik and IQST, Albert-Einstein-Allee 11, Universit\"at Ulm, D-89069 Ulm, Germany}

\begin{abstract}
Electron-hole pairs in organic photovoltaics dissociate efficiently despite their Coulomb-binding energy exceeding thermal energy at room temperature. The electronic states involved in charge separation couple to structured vibrational environments containing multiple underdamped modes. The non-perturbative simulations of such large, spatially extended electronic-vibrational (vibronic) systems remains an outstanding challenge. Current methods bypass this difficulty by considering effective one-dimensional Coulomb potentials or unstructured environments. Here we extend and apply a recently developed method for the non-perturbative simulation of open quantum systems to the dynamics of charge separation in one, two and three- dimensional donor-acceptor networks. This allows us to identify the precise conditions in which underdamped vibrational motion induces efficient long-range charge separation. Our analysis provides a comprehensive picture of ultrafast charge separation by showing how different mechanisms driven either by electronic or vibronic couplings are well differentiated for a wide range of driving forces and how entropic effects become apparent in large vibronic systems. These results allow us to quantify the relative importance of electronic and vibronic contributions in organic photovoltaics and provide a toolbox for the design of efficient charge separation pathways in artificial nanostructures.
\end{abstract}

\date{\today}
\maketitle


\section{Introduction}

\begin{figure*}
	\begin{center}
	\includegraphics[width=0.95\textwidth]{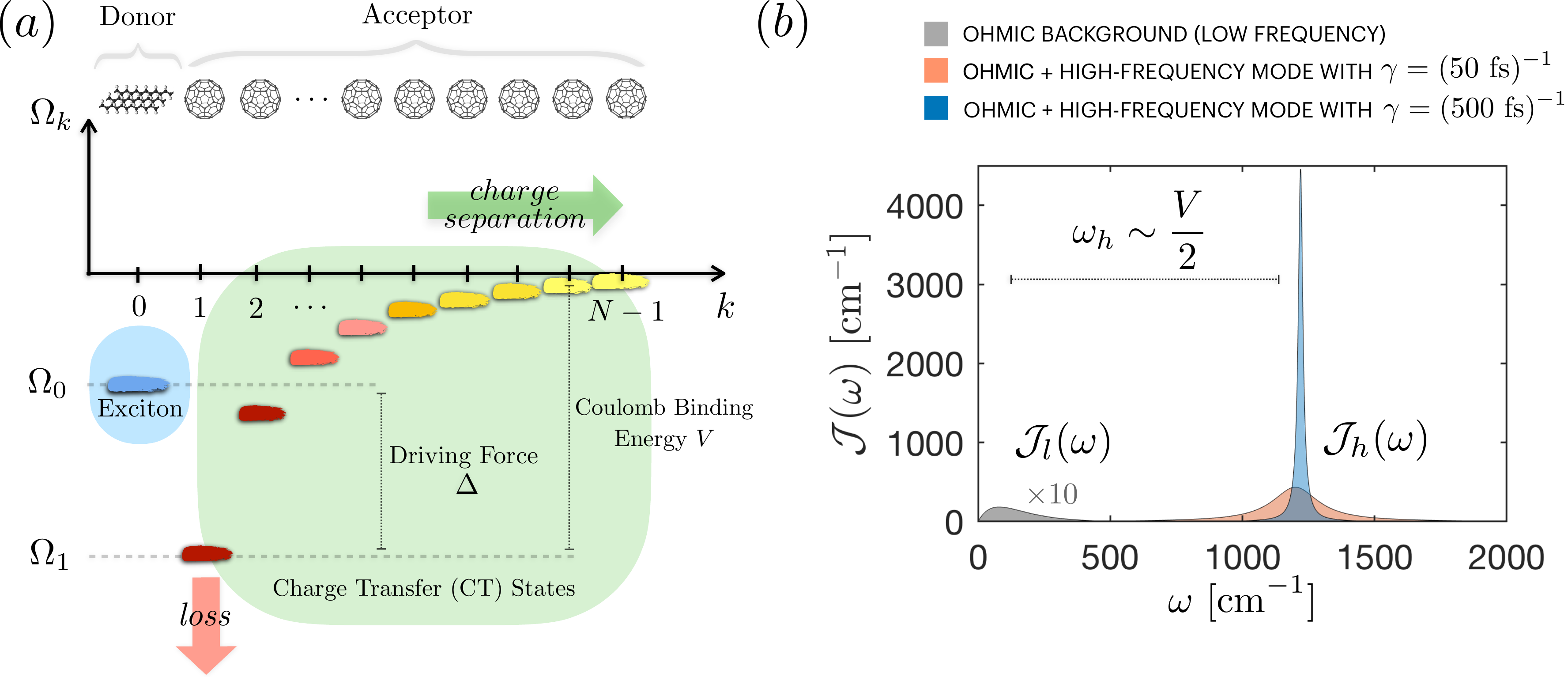}
	\caption{Coulomb potential and vibrational environments. (a) Schematic representation of a one-dimensional chain consisting of a donor and $(N-1)$ acceptors. The Coulomb-binding energy of electron and hole is modelled by $\Omega_k=-V/k$ with $V=0.3\,$eV for $k\ge 1$. The energy-gap between exciton and interfacial CT states is defined as driving force $\Delta=\Omega_0-\Omega_1$. (b) Vibrational environments consist of low-frequency phonon baths with room temperature energy scales ($k_B T\approx 200\,{\rm cm}^{-1}$) and high-frequency vibrational modes. In this work, the low-frequency phonons are modelled by an Ohmic spectral density ${\cal J}_{l}(\omega)$ with an exponential cutoff, while the high-frequency modes are described by Lorentzian spectral densities ${\cal J}_{h}(\omega)$ centered at vibrational frequency $\omega_h=1200\,{\rm cm}^{-1}$ (or $1500\,{\rm cm}^{-1}$). The vibrational damping rate of the high-frequency modes is taken to be $\gamma=(50\,{\rm fs})^{-1}$ or $(500\,{\rm fs})^{-1}$, as shown in red and blue, respectively, to investigate the role of non-equilbrium vibrational motion in long-range charge separation.}
		\label{fig:intro1} 
		\end{center}
	\end{figure*}

When a solar cell made out of an inorganic semiconductor like silicon is exposed to light, electrons can be readily extracted from the valence band to the conduction band and then captured at the electrodes.  If, however, light is absorbed by carbon-based materials, photons produce strongly bound electron-hole pairs called excitons, which are collective optical excitations that may be delocalized across several molecular units \cite{Scholes2006}. Excitons are charge neutral, namely the electron and the hole occupy, respectively, the lowest unoccupied molecular orbital (LUMO) and the highest occupied molecular orbital (HOMO) of the same molecular unit, and require dissociation in order to produce a current \cite{Hedley2017}. In contrast, the charge-transfer (CT) states describe partially separated electron-hole pairs where an electron and a hole occupy, respectively, the LUMO and HOMO levels that belong to different molecular sites. The transfer from exciton to CT states is thus suitable to describe the dynamics of electron transfer \cite{Vandewal2016,Coropceanu2019}. In photosynthetic organisms excitons are split in pigment-protein complexes called reaction centers \cite{Valkunas_book,Huelga2013,Romero2017}. In organic photovoltaics (OPV), blends of materials with different electron affinities are used to provide an energetic landscape that is favourable to charge separation at the interface \cite{Bredas2009}. These devices exhibit ultrafast, long-range charge separation with high quantum efficiencies  \cite{Park2009,Clarke2010,Gao2014}. This means that a large proportion of absorbed photons produces excitons or strongly bound CT states that are successfully dissociated. Some of these electron-hole pairs however thermalize towards the lowest-energy CT state localized at the interface, which is for this reason considered an energetic trap that leads to non-radiative electron-hole recombination \cite{Faist2011,Nan2016,Azzouzi2018,Eisner2019}, as schematically shown in Fig.\ref{fig:intro1}(a). This localization process is predominantly mediated by high-frequency vibrational modes that can bridge the energy gap between high-lying exciton/CT states and the lowest-energy interfacial CT state. The energy loss associated to this process is typically larger than $0.6\,$eV per photon \cite{Burke2015,Menke2018Review,Liu2019PRApp}, leading to a low power conversion efficiency in OPV with respect to their inorganic counterparts that results in a small open circuit voltage \cite{Azzouzi2019}.

Although energetically costly, dissociation of strongly bound electron-hole pairs  \cite{Clarke2010,Troisi2011Chem,Gao2014} takes place despite the much lower thermal energy at room temperature. The energy of bound CT states is largely dependent on the offset between the LUMO of the acceptor and the HOMO of the donor \cite{Scharber2006,Veldman2009,Hallerman2009}. Fixing the acceptor and employing different donor materials (or vice versa) is a popular strategy to investigate the energetics at the interface and achieve a high voltage, small energy losses and sufficient photocurrent density \cite{Deibel2010,Coffey2012,Wang2018,Qian2018,Yang2018,Azzouzi2019}. Surprisingly, some of these blends show ultrafast and efficient exciton dissociation despite having small or no apparent driving force \cite{Zhang2014,Liu2016NatEner,Meng2018,Cui2019,Eisner2019,Liu2019PRApp,Liu2020,Zhang2020,Zhong2020}. 
The driving force is a crucial parameter in charge separation and refers to the energy difference between exciton and interfacial CT state (see $\Delta$ in Fig.\ref{fig:intro1}(a)). Hybridization between exciton and CT states has been thought to be behind the successful ultrafast charge separation of these promising materials, which are often based on small molecules (oligomers) with acceptor-donor-acceptor structures that have reached power conversion efficiencies of up to $17\,\%$ \cite{Chen2018NatComms,Meng2018,Eisner2019,Firdaus2019}. This represents an astonishing 50$\,\%$ increase in the state-of-the-art performance of organic photovoltaics in less than a decade. From finite molecular clusters to periodic molecular solids, ultrafast long-range charge separation has appeared across a wide variety of photovoltaic platforms, but the underlying mechanism has not been understood fully, leading some to advocate for a deeper analysis of charge separation processes \cite{Karki2020,Alvertis2020}.

Some experimental studies rule out thermal activation as an important mechanism for charge separation in a large number of photovoltaic devices \cite{Pensack2009,Vandewal2014,Bassler2015}. In contrast, the vibronic coupling to underdamped vibrational modes is presumed to enable coherent charge separation \cite{Falke2014,Song2014,Rafiq2015,DeSio2016,Bredas2017,DeSio2018,DeSio2021}, which requires non-perturbative simulation tools for a reliable description of the vibronic interaction between exciton/CT states and molecular vibrations. However, in many theoretical studies on the charge separation in extended systems, a broad and unstructured environmental spectral density has been considered \cite{Smith2015,Lee2015,Kato2018} to reduce simulation costs, neglecting the ubiquitous presence of underdamped vibrational modes in organic molecules and their role in charge separation. In addition, the non-Markovian vibronic effects proposed to suppress the localization of electron-hole pairs at the interfaces, e.g. suggested in Ref.\cite{Kato2018}, are found to be well described by a Markovian quantum master equation, as shown in Appendix \ref{appendix:PRL2018Ishizaki}, due to weak vibronic coupling strength and no underdamped modes considered in simulations. This indicates that a vibronic mechanism inferred solely on the basis of non-perturbative numerical results without the subsequent formulation of an accurate physical mechanism may lead to ambiguities in the interpretation of the underlying mechanism. Some first-principles numerical methods have been employed to simulate vibronic charge separation \cite{Tamura2013JACS,Tamura2013JPCC,Hughes2014,Chenel2014,Rotllant2015,Chaudhuri2017,Polkehn2018,Bian2020}, where underdamped vibrational modes are considered. However, the interpretation of simulated results is a non-trivial issue here. For instance, in Ref.\cite{Tamura2013JACS}, an effective one-dimensional Coulomb potential is considered where electron-hole binding energy is assumed to be reduced by instantaneous electron delocalization in three-dimensional acceptor aggregates and as a result the electronic coupling being responsible for a hole transfer becomes larger in magnitude than the detunings in energy levels of the effective potential. In Appendix \ref{appendix:JACS2013Burghardt}, we show how, in this case, deactivating completely the vibrational environment has little impact in charge separation dynamics. This leads us to conclude that the ultrafast long-range charge separation observed in Ref.\cite{Tamura2013JACS} is not necessarily enhanced by vibronic couplings, but merely induced by the weak Coulomb-binding energy.  Other theoretical studies have focused on intermolecular modes as the relevant vibrations behind charge separation \cite{Yao2016a}, while  sometimes, intramolecular modes are attributed a hampering role \cite{Duan2020}. This is, as we will demonstrate, in sharp contrast to our findings, as intramolecular modes can induce both effects.

Given the heterogeneity of donor-acceptor materials and model parameters employed across the literature, we aim to discern the underlying mechanisms of charge separation dynamics as a function of the driving force and the structure of vibrational environments based on non-perturbative simulations and detailed reduced model analysis. We determine under what conditions underdamped vibrational motion induces efficient long-range charge separation in the presence of strong Coulomb-binding energy $V\sim 0.3\,$eV. To this end we consider one, two and three-dimensional donor-acceptor networks, instead of effective one-dimensional Coulomb potentials, by using our non-perturbative simulation method called dissipation assisted matrix product factorization (DAMPF) \cite{Tamascelli2018,Somoza2019,Tamascelli2019,Mascherpa2020}, to investigate how coherent vibronic couplings promote long-range charge separation in high-dimensional multi-site systems. We show that there are two available mechanisms for ultrafast long-range charge separation in donor-acceptor interfaces. For low driving forces $\Delta\sim 0.15\,$eV, the transitions between near-resonant exciton and delocalised CT states occur on a sub-ps time scale even if vibronic couplings are not considered. For high driving forces $\Delta\sim 0.3\,$eV, the vibronic coupling of underdamped high-frequency vibrational modes with frequencies $\omega_h\sim 0.15\,$eV induces the transitions between exciton and CT states delocalised over multiple acceptors. Here a vibrationally cold exciton can interact resonantly with vibrationally hot lower-energy CT states and, subsequently, also with vibrationally cold high-energy CT states. The charge separation process becomes significantly inefficient in this case when vibronic couplings are ignored in simulations, hinting the genuine vibronic effects induced by underdamped vibrational modes. For both low and high driving forces, we demonstrate that the time scale of the charge localization towards the donor-acceptor interfaces is determined by the lifetime of the high-frequency vibrational modes, as strongly damped modes promote the transitions to the lowest-energy interfacial CT state. These results demonstrate that experimentally measured long-lived vibrational and vibronic coherences in OPV \cite{Falke2014,Song2014,Rafiq2015,DeSio2016,Bredas2017,DeSio2018,DeSio2021} may have a functional relevance in charge separation processes.

\begin{figure*}
	\begin{center}
		\includegraphics[width=0.9\textwidth]{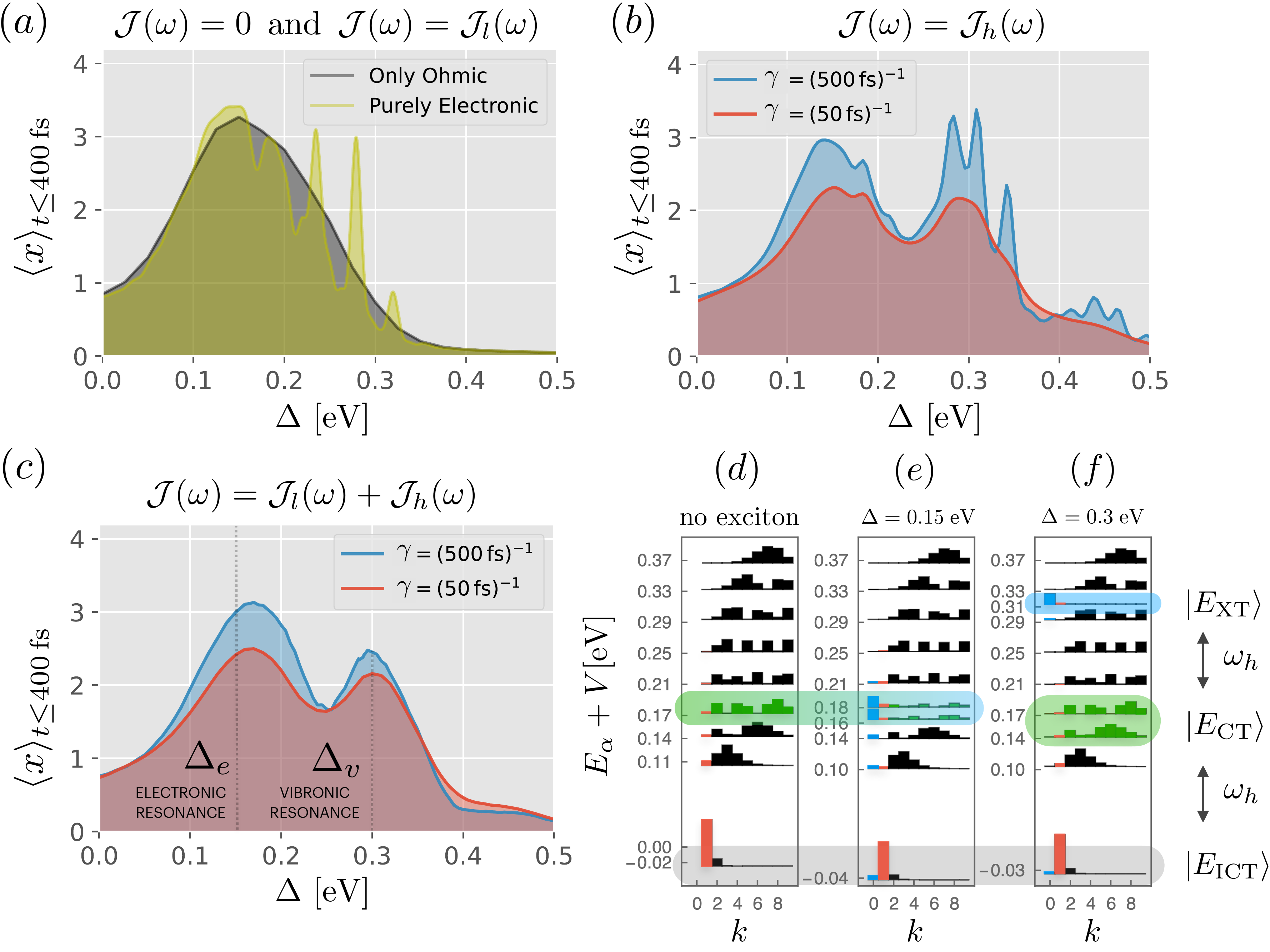}
	    \caption{Charge separation under different vibrational environments and electronic eigenstate structures. (a) Time-averaged mean electron-hole distance $\langle x\rangle_{t\le 400\,{\rm fs}}$ is displayed as a function of driving force $\Delta$ when vibrational environments are absent (${\cal J}(\omega)=0)$ or only low-frequency phonon baths are present (${\cal J}(\omega)={\cal J}_{l}(\omega)$), shown in yellow and gray, respectively. Here we consider a linear chain consisting of a donor and nine acceptors ($N=10$ sites). (b-c) Time-averaged mean electron-hole distance when electronic states are coupled to (b) high-frequency vibrational modes only (${\cal J}(\omega)={\cal J}_{h}(\omega)$) or to (c) the total vibrational environments (${\cal J}(\omega)={\cal J}_{l}(\omega)+{\cal J}_{h}(\omega)$). Here the vibrational damping rate of the high-frequency modes with frequency $\omega_h=1200\,{\rm cm}^{-1}$ is taken to be $\gamma=(50\,{\rm fs})^{-1}$ or $(500\,{\rm fs})^{-1}$, shown in red and blue, respectively. (d) Electronic eigenstates in the absence of exciton-CT couplings where probability distributions for finding an electron at the $k$-th acceptor are vertically shifted depending on electronic energy-levels $E_{\alpha}$. (e-f) Electronic eigenstates in the presence of exciton-CT couplings where driving force is taken to be (e) $\Delta_e=0.15\,$eV or (f) $\Delta_v=0.3\,$eV. In (e), hybrid exciton-CT states contributing to initial charge separation dynamics are colored in blue/green. In (f), the exciton and delocalised CT states, governing initial charge separation via a vibronic mixing, are highlighted in blue and green, respectively. In (d-f), the probabilities for finding an electron at donor/acceptor interface are shown in red.}
		\label{fig:intro2} 
	\end{center}
\end{figure*}

\section{Model}
To investigate the influence of underdamped vibrational motion on the charge separation of strongly bound electron-hole pairs, we consider a one-dimensional chain consisting of $N$ sites, composed of an electron donor in contact with a chain of $(N-1)$ electron acceptors, as schematically shown in Fig.\ref{fig:intro1}(a). Two and three-dimensional donor-acceptor networks will be considered later. The electronic Hamiltonian is modeled by 
\begin{equation}
H_{e}=\sum_{k=0}^{N-1}\Omega_{k}\ket{k}\bra{k}+\sum_{k=0}^{N-2}J_{k,k+1}(\ket{k}\bra{k+1}+h.c.), 
\label{eq:He}
\end{equation}
where $h.c.$ denotes the Hermitian conjugate. Here $\ket{0}$ denotes an exciton state localized at the donor, while $\ket{k}$ with $k\ge 1$ is a CT state with an electron localized at the $k$-th acceptor. For simplicity, we assume that the hole is fixed at the donor within the time scale of our simulations due to its lower mobility with respect to the electron \cite{Gelinas2014Science,Kato2018,Smith2015}. The energy-levels of the CT states take into account the Coulomb attraction between electron and hole, modelled by $\Omega_k=-V/k$ with $V=0.3\,{\rm eV}$ for $k\ge 1$. We take  $J_{k,k+1}=500\,{\rm cm}^{-1}\approx 0.06\,{\rm eV}$ for the electronic coupling being responsible for an electron transfer, a common value found in {acceptor} aggregates {such as} fullerene derivatives \cite{Tamura2013JACS,Tamura2013JPCC,Nan2015}. The exciton energy $\Omega_0$ depends on the molecular properties of the donor \cite{Scharber2006,Dennler2009,Veldman2009,Clarke2010}, which will be considered a free variable parametrized by the driving force $\Delta=\Omega_0-\Omega_1$, as shown in Fig.\ref{fig:intro1}(a). 

For simplicity, we assume that each electronic state $\ket{k}$ is coupled to an independent vibrational environment that is initially in a thermal state at room temperature. The vibrational Hamiltonian is written as
\begin{equation}
H_{v}=\sum_{k=0}^{N-1}\sum_{q}\omega_q b_{k,q}^{\dagger}b_{k,q},
\end{equation}
with $b_{k,q}$ ($b_{k,q}^{\dagger}$) describing the annihilation (creation) operator of a vibrational mode with frequency $\omega_q$ that is locally coupled to the electronic state $\ket{k}$. The vibronic interaction is modeled by
\begin{equation}
H_{e-v}=\sum_{k=0}^{N-1}\ket{k}\bra{k}\sum_{q}\omega_q \sqrt{s_q}(b_{k,q}+b_{k,q}^{\dagger}),
 \end{equation}
where the vibronic coupling strength is quantified by the Huang-Rhys (HR) factors $s_q$. The vibrational environments are fully characterized by a phonon spectral density ${\cal J}(\omega)=\sum_{q}\omega_{q}^{2}s_{q}\delta(\omega-\omega_{q})$ with $\delta(\omega)$ denoting the Dirac delta function. According to first-principles calculations of functionalized fullerene  electron acceptors, the vibrational environment consists of multiple low-frequency modes, with vibrational frequencies smaller than the thermal energy at room temperature ($k_B T\approx 200\,{\rm cm}^{-1}\approx 0.025\,{\rm eV}$), and a few discrete modes with high vibrational frequencies of the order of $\sim 1000\,{\rm cm}^{-1}$ and HR factors $\lesssim 0.1$ \cite{Cheung2010,Ide2014,Chen2018,Chen2018AdvMater}. Motivated by these observations, we consider a phonon spectral density ${\cal J}(\omega)={\cal J}_{l}(\omega)+{\cal J}_{h}(\omega)$ where ${\cal J}_{l}(\omega)=\frac{\lambda_l}{\omega_l}\omega e^{-\omega/\omega_l}$, with $\omega_l= 80\,{\rm cm}^{-1}$ and $\lambda_l=50\,{\rm cm}^{-1}$, describing a low-frequency phonon spectrum (see gray curve in Fig.\ref{fig:intro1}(b)). The high-frequency vibrational modes are modeled by a Lorentzian function ${\cal J}_{h}(\omega)=\frac{4 \omega_h s_h \gamma (\omega_{h}^{2}+\gamma^{2})}{\pi} \omega ((\omega+\omega_h)^{2}+\gamma^{2})^{-1} ((\omega-\omega_h)^{2}+\gamma^{2})^{-1}$ with vibrational frequency $\omega_h=1200\,{\rm cm}^{-1}\approx V/2=0.15\,$eV and HR factor $s_h=0.1$. Here the reorganization energy of the high-frequency mode, defined by $\int_{0}^{\infty}d\omega {\cal J}_{h}(\omega)\omega^{-1}=w_h s_h$, is independent of its vibrational damping rate $\gamma$ (see red and blue curves in Fig.\ref{fig:intro1}(b)).

In order to tackle the problem of simulating large vibronic systems, we have extended DAMPF \cite{Somoza2019}, where a continuous vibrational environment is described by a finite number of oscillators undergoing Markovian dissipation (pseudomodes) and a tensor network formalism is used. With DAMPF the reduced electronic system dynamics can be simulated in a numerically accurate manner for highly structured phonon spectral densities by fitting the corresponding bath correlation functions via an optimal set of parameters of either coupled or uncoupled pseudomodes \cite{Tamascelli2018,Somoza2019,Tamascelli2019,Mascherpa2020}. The extended DAMPF-method opens the door to non-perturbative simulations of many body systems consisting of several tens of sites coupled to structured environments in one, two- and three spatial dimensions, as will be demonstrated in this work. More details about the method and the explicit equation of motion in terms of pseudomodes can be found in Appendix \ref{appendix:method}.

\section{Results}

\subsection{Driving Force and Vibrational Environments}

Here we investigate the charge separation dynamics on a sub-ps time scale simulated by DAMPF. For simplicity, we consider a linear chain consisting of a donor and nine acceptors ($N=10$). Longer one-dimensional chains and higher-dimensional donor/acceptor networks will be considered later. We assume that an exciton state $\ket{0}$ localised at the donor site is created at the initial time $t=0$ and then an electron transfer through the acceptors induces the transitions from the exciton to the CT states $\ket{k}$ with $k\ge 1$. The mean distance between electron and hole is considered a figure of merit for charge separation, defined by $\langle x(t)\rangle=\sum_{k=0}^{N-1}kP_{k}(t)$ with $P_k(t)$ representing the populations of the exciton and CT states $\ket{k}$ at time $t$, with the assumption that the distance between nearby sites is uniform. To investigate how the initial charge separation dynamics depends on the exciton energy $\Omega_0$ and the structure of vibrational environments, we analyse the time-averaged electron-hole distance, defined by $\langle x\rangle_{t\le T}=\frac{1}{T}\int_{0}^{T}dt\langle x(t)\rangle$ with $T=400\,{\rm fs}$, as a function of the driving force $\Delta=\Omega_0+V$ for various environmental structures. The role of high-frequency vibrational modes and their non-equilibrium dynamics in charge separation processes is identified by considering (i) no environments (${\cal J}(\omega)=0$), (ii) low-frequency phonon baths (${\cal J}(\omega)={\cal J}_{l}(\omega)$, see gray curve in Fig.\ref{fig:intro1}(b)), (iii) high-frequency vibrational modes with controlled damping rates $\gamma\in\{(50\,{\rm fs})^{-1},(500\,{\rm fs})^{-1}\}$ (${\cal J}(\omega)={\cal J}_{h}(\omega)$, see red and blue curves in Fig.\ref{fig:intro1}(b)), and (iv) the total vibrational environments including both low-frequency phonon baths and high-frequency vibrational modes (${\cal J}(\omega)={\cal J}_{l}(\omega)+{\cal J}_{h}(\omega)$).

\begin{figure}
\includegraphics[width=0.48\textwidth]{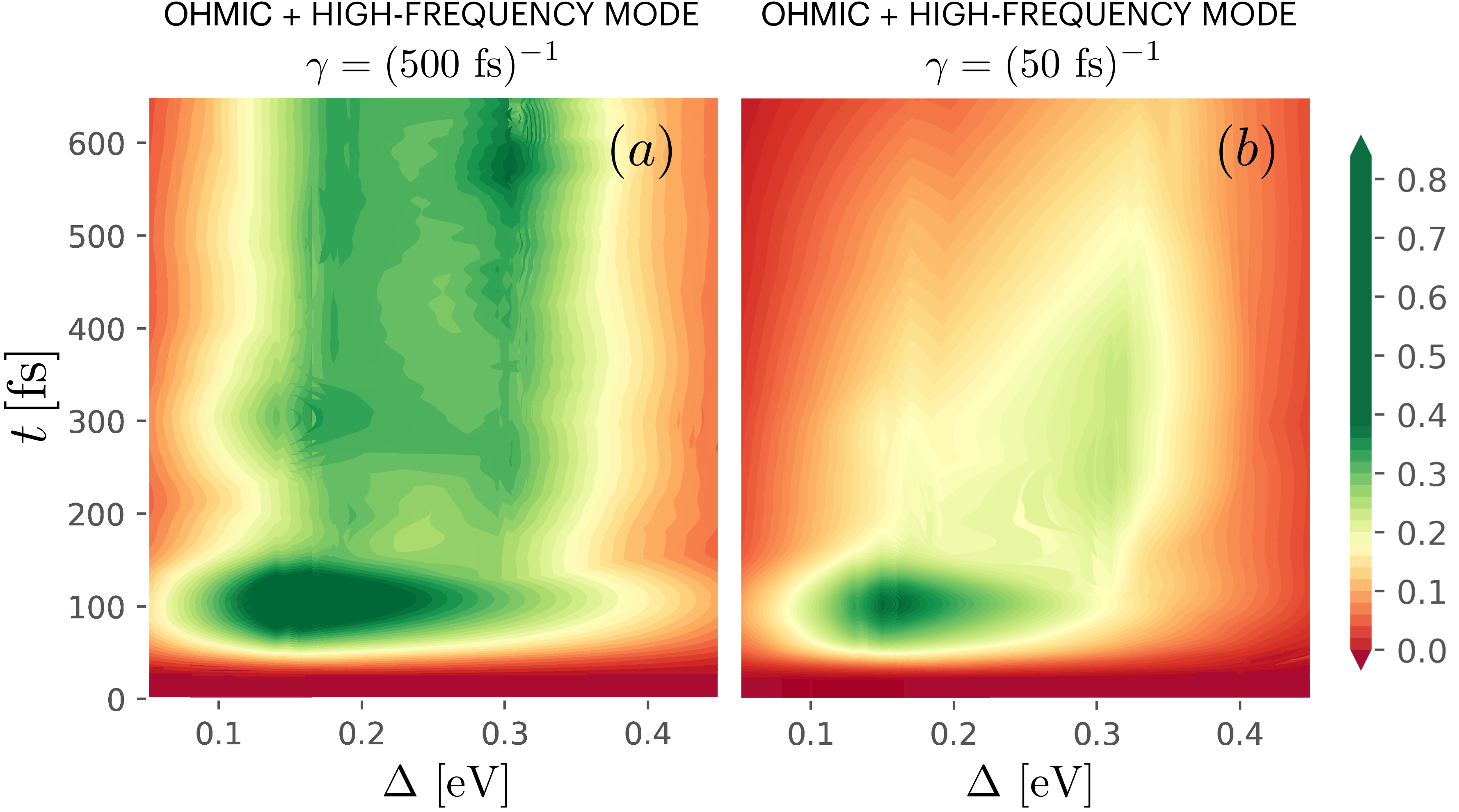}
\caption{Long-range charge separation dynamics. (a-b) Total probability for separating an electron-hole pair more than four molecular units, $\sum_{k=5}^{9}P_{k}(t)$, is shown as a function of time $t$ and driving force $\Delta$. Here a linear chain consisting of $N=10$ sites is considered where electronic states are coupled to total vibrational environments (${\cal J}(\omega)={\cal J}_{l}(\omega)+{\cal J}_{h}(\omega)$). The damping rate of the high-frequency vibrational modes with $\omega_h=1200\,{\rm cm}^{-1}$ is taken to be (a) $\gamma=(500\,{\rm fs})^{-1}$ or (b) $\gamma=(50\,{\rm fs})^{-1}$.}
\label{fig:dynamics_delta}
\end{figure}

In Fig.\ref{fig:intro2}(a), the time-averaged electron-hole distance is shown as a function of the driving force $\Delta$ when vibrational environments are not considered ($\mathcal{J}(\omega)=0$). In this case, the charge separation dynamics is purely electronic and the mean electron-hole distance shows multiple peaks for $\Delta\lesssim 0.3\,{\rm eV}$. When electronic states are only coupled to low-frequency phonon baths (${\cal J}(\omega)={\cal J}_{l}(\omega)$), these peaks are smeared out, resulting in a smooth, broad single peak centered at $\Delta_e \approx 0.15 \;\text{eV}$. In Fig.\ref{fig:intro2}(b) where the electronic states are coupled to high-frequency vibrational modes (${\cal J}(\omega)={\cal J}_{h}(\omega)$), the time-averaged electron-hole distance is displayed for different vibrational damping rates $\gamma=(50\,{\rm fs})^{-1}$ and $\gamma=(500\,{\rm fs})^{-1}$, shown in red and blue, respectively. With $\omega_h$ denoting the vibrational frequency of the high-frequency modes, the electron-hole distance is maximized at $\Delta_e\approx 0.15\,{\rm eV}$, $\Delta_e+\omega_h\approx 0.3\,{\rm eV}$, $\Delta_e+2\omega_h\approx 0.45\,{\rm eV}$, making the charge separation process efficient for a broader range of the driving force $\Delta$ when compared to the cases that the high-frequency modes are ignored (see Fig.\ref{fig:intro2}(a)). It is notable that the electron-hole distance is larger for the lower damping rate $\gamma=(500\,{\rm fs})^{-1}$ of the high-frequency vibrational modes than for the higher damping rate $\gamma=(50\,{\rm fs})^{-1}$. These results imply that non-equilibrium vibrational dynamics can promote long-range charge separation. This observation still holds even if the low-frequency phonon baths are considered in addition to the high-frequency vibrational modes (${\cal J}(\omega)={\cal J}_{l}(\omega)+{\cal J}_{h}(\omega)$), as shown in Fig.\ref{fig:intro2}(c), where the electron-hole distance is maximized at $\Delta_e \approx 0.15 \;\text{eV}$ and $\Delta_v = \Delta_e + \omega_h \approx 0.3 \;\text{eV}$. We note that the electron-hole distance at low driving forces $\Delta\sim\Delta_e$ is insensitive to the presence of vibrational environments, while at high driving forces  $\Delta\sim\Delta_v$, the charge separation process becomes significantly inefficient when the high-frequency vibrational modes are ignored. These results suggest that vibrational environments may play an essential role in the long-range charge separation at high driving forces, while the exciton dissociation at low driving forces may be governed by electronic interactions.

So far the time-averaged mean electron-hole distance has been considered to identify under what conditions the charge separation on a sub-ps time scale becomes efficient. However, it does not show how much populations of the CT states with well-separated electron-hole pairs are generated and how quickly the long-range electron-hole separation takes place. In Fig.\ref{fig:dynamics_delta}, we show the population dynamics of the CT states where electron and hole are separated more than four molecular units, defined by $\sum_{k=5}^{9}P_k(t)$, for the case that electronic states are coupled to the total vibrational environments (${\cal J}(\omega)={\cal J}_{l}(\omega)+{\cal J}_{h}(\omega)$). When the high-frequency vibrational modes are weakly damped with $\gamma=(500\,{\rm fs})^{-1}$, the electron is transferred to the second half of the acceptor chain within 100\,fs and then the long-range electron-hole separation is sustained on a sub-ps time scale for a wide range of the driving forces $\Delta$, as shown in Fig.\ref{fig:dynamics_delta}(a). When the high-frequency modes are strongly damped with $\gamma=(50\,{\rm fs})^{-1}$, for low driving forces around $\Delta_e\approx 0.15\,$eV the long-range charge separation occurs within 100\,fs, but the electron is quickly transferred back to the donor-acceptor interface, as shown in Fig.\ref{fig:dynamics_delta}(b). For high driving forces around $\Delta_v\approx 0.3\,$eV, the long-range charge separation and subsequent localization towards the interface take place on a slower time scale when compared to the case of the low driving forces. These results demonstrate that underdamped vibrational motion can promote long-range charge separation when the excess energy $\Delta-V$, defined by the energy difference between exciton state and fully separated free charge carriers, is negative or close to zero \cite{Menke2018,Zhang2020,Liu2020,Hinrichsen2020,Zhong2020}.

\begin{figure*}
\includegraphics[width=0.8\textwidth]{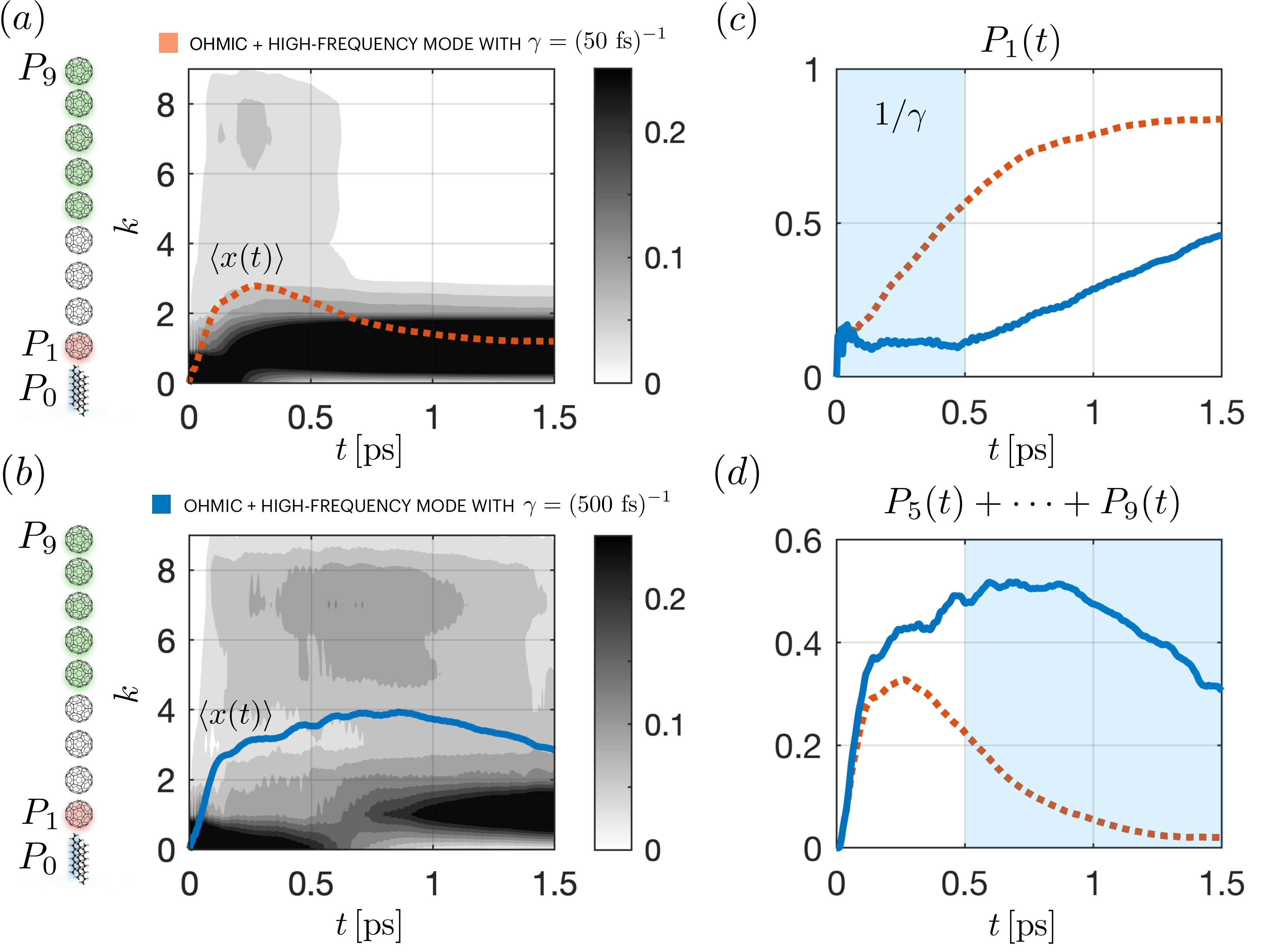}
\caption{Vibration-assisted exciton dissociation and charge localization towards donor/acceptor interfaces. (a) With a hole fixed at donor, the probability distribution for finding an electron at the donor ($k=0$, corresponding to exciton) or at the $k$-th acceptor ($k\ge 1$) is displayed as a function of time $t$, with mean electron-hole distance $\langle x(t)\rangle$ shown in red. With high driving force $\Delta_v=0.3\,$eV, here we consider a linear chain consisting of $N=10$ sites and total vibrational environments (${\cal J}(\omega)={\cal J}_{l}(\omega)+{\cal J}_{h}(\omega)$) including strongly damped high-frequency modes with $\omega_h=1200\,{\rm cm}^{-1}$ and $\gamma=(50\,{\rm fs})^{-1}$.  (b) Charge separation dynamics when the high-frequency vibrational modes are weakly damped with $\gamma=(500\,{\rm fs})^{-1}$. (c) Population dynamics of interfacial CT state $\ket{1}$. (d) Probability for separating an electron-hole pair more than four molecular units, $\sum_{k=5}^{9}P_{k}(t)$, shown as a function of time $t$. In both (c) and (d), the strongly (weakly) damped case with $\gamma=(50\,{\rm fs})^{-1}$ ($\gamma=(500\,{\rm fs})^{-1}$ is shown in red (blue).}
\label{fig:population_dynamics}
\end{figure*}

\subsection{Electronic mixing at low driving forces}\label{section:results_electronic_mixing}

The long-range charge separation observed in DAMPF simulations can be rationalized by analysing the energy-levels and delocalization lengths of the exciton and CT states. In Fig.\ref{fig:intro2}(d), we consider the eigenstates $|E_{\alpha}^{\rm (CT)}\rangle$ of the electronic Hamiltonian where the exciton state $\ket{0}$ and its coupling $J_{0,1}$ to the CT states are ignored, namely $H_{\rm CT}=\sum_{k=1}^{N-1}\Omega_k\ket{k}\bra{k}+\sum_{k=1}^{N-2}J_{k,k+1}(\ket{k}\bra{k+1}+h.c.)$. With a hole fixed at the donor site, the probability distributions $|\langle k|E_{\alpha}^{\rm (CT)}\rangle|^{2}$ for finding an electron at the $k$-th acceptor site is displayed, which are vertically shifted by $E_{\alpha}^{\rm (CT)}+V$ with $E_{\alpha}^{\rm (CT)}$ representing the eigenvalues of $H_{\rm CT}$. The lowest-energy CT eigenstate is mainly localised at the interface due to the strong Coulomb-binding energy considered in simulations ($\Omega_2-\Omega_1=V/2=0.15\,{\rm eV}>J_{1,2}\approx 0.06\,{\rm eV}$). The other higher-energy CT eigenstates are significantly delocalised in the acceptor domain with smaller populations $|\langle 1|E_{\alpha}^{\rm (CT)}\rangle|^{2}$ at the interface for higher energies $E_{\alpha}^{\rm (CT)}$, as highlighted in red.

We now consider the full electronic Hamiltonian $H_{e}$ including the exciton state at the low driving force $\Delta_e \approx 0.15\,{\rm eV}$ where efficient long-range charge separation occurs even in the absence of vibrational environments (see Fig.\ref{fig:intro2}(a)). The exciton state $\ket{0}$ is coupled to the eigenstates $|E_{\alpha}^{\rm (CT)}\rangle$ of $H_{\rm CT}$ via the electronic coupling $H_i = J_{0,1}(\ket{0}\bra{1}+h.c.)$ at the interface, leading to the exciton-CT couplings in the form $\bra{0}H_i|E_{\alpha}^{\rm (CT)} \rangle = J_{0,1}\langle 1|E_{\alpha}^{\rm (CT)} \rangle$. This implies that the transition between exciton and CT state $|E_{\alpha}^{\rm (CT)}\rangle$ is enhanced when the exciton energy $\Omega_0$ is near-resonant with the CT energy $E_{\alpha}^{\rm (CT)}$ and the CT state has sufficiently high population $|\langle 1|E_{\alpha}^{\rm (CT)}|^{2}$ at the interface (see red bars in Fig.\ref{fig:intro2}(d)). For $\Delta_e=\Omega_0+V= 0.15\,{\rm eV}$, the exciton state can be strongly mixed with a near-resonant CT state delocalised over multiple acceptor sites (see Fig.\ref{fig:intro2}(d)), leading to two hybrid exciton-CT eigenstates of the total electronic Hamiltonian $H_e$, described by the superpositions of $\ket{0}$ and multiple $\ket{k}$ with $k\ge 1$ (see Fig.\ref{fig:intro2}(e)). This indicates that the multiple peaks in the time-averaged electron-hole distance $\langle x\rangle_{t\leq 400\,{\rm fs}}$ shown in Fig.\ref{fig:intro2}(a) originate from the resonances between exciton and CT states $|E_{\alpha}^{\rm (CT)}\rangle$. Here the high-lying CT states with energies $E_{\alpha}^{\rm (CT)}+V\gtrsim 0.3\,{\rm eV}$ do not show long-range electron-hole separation, as the interfacial electronic couplings $J_{0,1}\langle 1|E_{\alpha}^{\rm (CT)} \rangle$ are not strong enough to induce notable transitions between exciton and CT states within the time scale $T=400\,{\rm fs}$ considered in Fig.\ref{fig:intro2}(a). These high-energy CT states can be populated via a near-resonant exciton state, but the corresponding purely electronic charge separation occurs on a slower ps time scale, as shown in Appendix \ref{appendix:onlyel}, and therefore this process can be significantly affected by low-frequency phonon baths. This is contrary to the charge separation at the low driving force $\Delta_e\approx 0.15\,{\rm eV}$, which takes place within 100\,fs and therefore the early electronic dynamics is weakly affected by vibrational environments. We note that when this analysis is applied to the charge separation model in Ref.\cite{Tamura2013JACS}, it can be shown that an exciton state is strongly mixed with near-resonant CT states delocalised in an effective one-dimensional Coulomb potential and as a result the ultrafast long-range charge separation reported in Ref.\cite{Tamura2013JACS} can be well described by a purely electronic model where vibrational environments are ignored (see Appendix \ref{appendix:JACS2013Burghardt}).

\subsection{Vibronic mixing at high driving forces}

Contrary to the case of $\Delta_e=0.15\,{\rm eV}$, the eigenstates of the full electronic Hamiltonian $H_e$ with $\Delta_v=0.3\,{\rm eV}$ show a weak mixing between exciton and CT states, as displayed in Fig.\ref{fig:intro2}(f), where the eigenstate $\ket{E_{\rm XT}}$ with the most strong excitonic character $|\langle 0|E_{\rm XT}\rangle|\approx 1$, marked in blue, has small amplitudes around the interface, $|\langle k|E_{\rm XT}\rangle|\ll 1$ for $k\ge 1$. Here the energy-gaps between the exciton state $\ket{E_{\rm XT}}$, shown in blue, and lower-energy eigenstates $\ket{E_{\rm CT}}$ with strong CT characters, shown in green, are near-resonant with the vibrational frequency of the high-frequency modes, $E_{\rm XT}-E_{\rm CT}\approx \omega_h$. Therefore, the vibrationally cold exciton state $\ket{E_{\rm XT},0_v}$ can resonantly interact with vibrationally hot CT states $\ket{E_{\rm CT},1_v}$ where one of the high-frequency modes is singly excited. Here the CT states are delocalised in the acceptor domain, but have non-negligible amplitudes around the interface, leading to a moderate vibronic coupling to the exciton state, $\bra{E_{\rm XT}}H_{e-v}\ket{E_{\rm CT}}=\sum_{k=0}^{N-1}\langle E_{\rm XT}|k\rangle \langle k| E_{\rm CT}\rangle \omega_h \sqrt{s_h}(b_{k,h}+b_{k,h}^{\dagger})$ with $b_{k,h}$ ($b_{k,h}^{\dagger}$) denoting the annihilation (creation) operator of the high-frequency vibrational mode locally coupled to electronic state $\ket{k}$. The other high-lying CT states $\ket{E_{\rm CT}'}$ near-resonant with the exciton state, $E_{\rm CT}'\approx E_{\rm XT}$, may have relatively small amplitudes around the interface, so the direct vibronic coupling to the exciton state could be small. However, the transitions from the exciton $\ket{E_{\rm XT},0_v}$ to the vibrationally hot low-lying CT states $\ket{E_{\rm CT},1_v}$ can allow subsequent transitions to vibrationally cold high-lying CT states $\ket{E_{\rm CT}',0_v}$, as the delocalised CT states $\ket{E_{\rm CT}}$ and $\ket{E_{\rm CT}'}$ are spatially overlapped. Such consecutive transitions are mediated by vibrational excitations and can delay the process of charge localization at donor-acceptor interfaces if the damping rate of the high-frequency vibrational modes is sufficiently lower than the transition rates amongst exciton and CT states. This picture is in line with the vibronic eigenstate analysis where the high-frequency modes are included as a part of system Hamiltonian in addition to the electronic states, as summarised in Appendix \ref{appendix:reduced_model}.

\begin{figure*}
\includegraphics[width=0.9\textwidth]{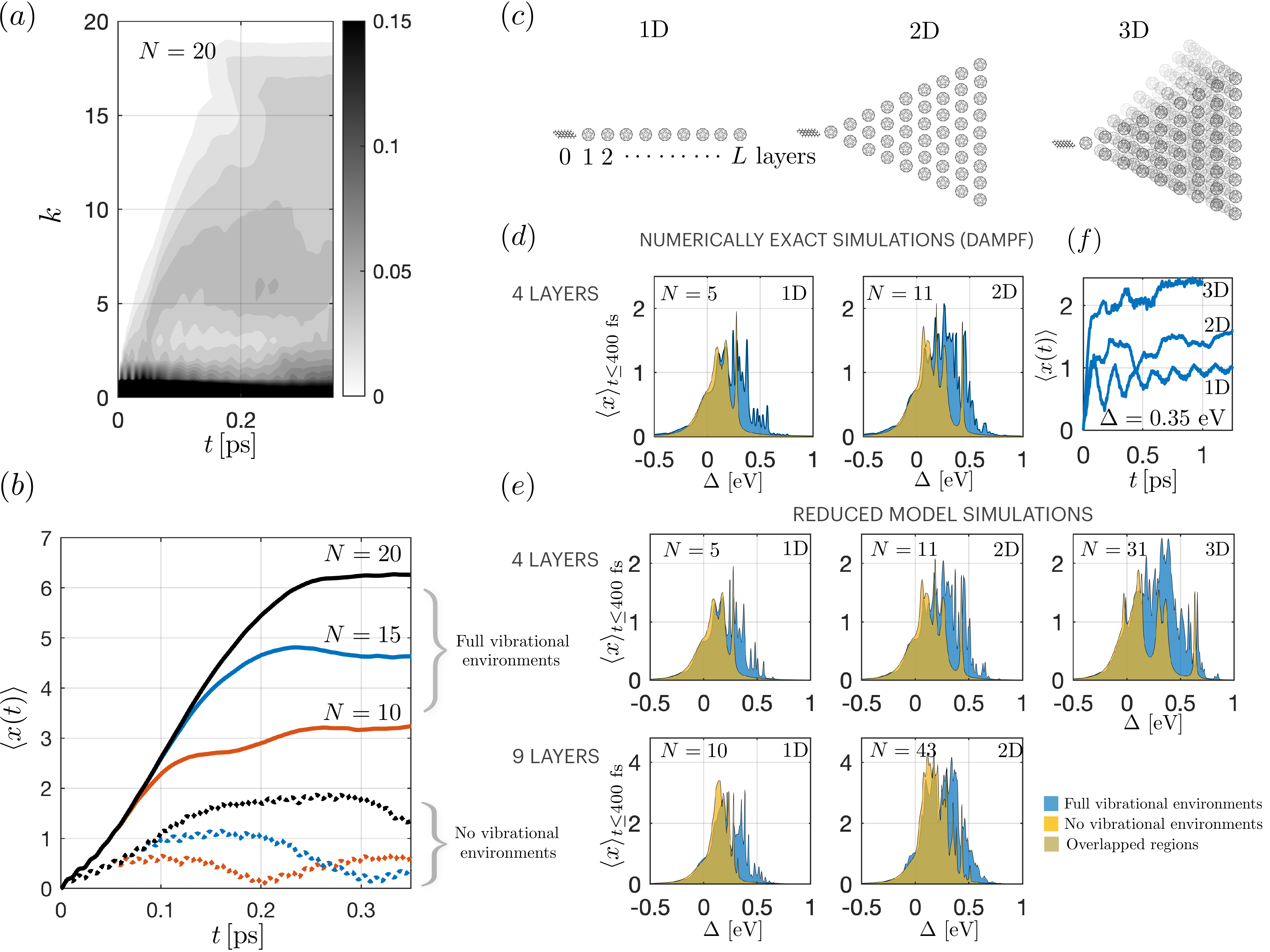}
\caption{Charge separation in large vibronic systems in one, two and three spatial dimensions. (a) With a hole fixed at donor, the probability distribution for finding an electron at the donor ($k=0$) or at the $k$-th acceptor ($k\ge 1$) is shown as a function of time $t$ for a linear chain consisting of $N=20$ sites. Here electronic states are coupled to total vibrational environments (${\cal J}(\omega)={\cal J}_{l}(\omega)+{\cal J}_{h}(\omega)$) including weakly damped high-frequqency modes with $\omega_h=1500\,{\rm cm}^{-1}$ and $\gamma=(500\,{\rm fs})^{-1}$. (b) Mean electron-hole distances $\langle x(t)\rangle$ of linear chains composed of $N\in\{10,20,30\}$ sites, shown in $\{$red,blue,black$\}$, respectively. Here we compare the case of the full environments, shown in solid lines, with that of no vibrational environments (${\cal J}(\omega)=0$), shown in dashed lines. (c) Schematic representation of one-, two- and three-dimensional donor-acceptor networks where a donor is coupled to acceptor aggregates. (d) Time-averaged mean electron-hole distances $\langle x\rangle_{t\le 400\,{\rm fs}}$ of one- and two-dimensional networks consisting of $L=4$ acceptor layers, displayed as a function of driving force $\Delta$ and computed by DAMPF \cite{Somoza2019}. Here the full and no environment cases are shown in blue and yellow, respectively. Note that the overlapped regions are shown in darker yellow, indicating that the no environment case is well covered by the full environment case. (e) Time-averaged mean electron-hole distances of one-, two- and three-dimensional networks consisting of $L=4$ or $L=9$ acceptor layers, simulated by a reduced vibronic model (see Appendix \ref{appendix:reduced_model} for more details). For the three-dimensional network with $L=4$, $\langle x\rangle_{t\le 400\,{\rm fs}}$ is maximized at high driving forces $\Delta\approx 0.35\,$eV. (f) For $\Delta=0.35\,$eV, time evolution of mean electron-hole distance $\langle x(t)\rangle$ is shown for one-, two- and three-dimensional networks with $L=4$, computed by DAMPF.}
\label{fig:size_entropy}
\end{figure*}

\subsection{Functional Relevance of Long-lived Vibrational Motion}

So far we have discussed the underlying mechanisms behind long-range charge separation on a sub-ps time scale. We now investigate how subsequent charge localization towards the donor-acceptor interface depends on the lifetimes of high-frequency vibrational modes to demonstrate that non-equilibrium vibrational dynamics can maintain long-range electron-hole separation.

In Fig.\ref{fig:population_dynamics}(a) and (b), where the high-frequency modes are strongly and weakly damped, respectively, with $\gamma=(50\,{\rm fs})^{-1}$ and $\gamma=(500\,{\rm fs})^{-1}$, the population dynamics $P_{k}(t)$ of the exciton $\ket{0}$ and CT states $\ket{k}$ with $k\ge 1$ is shown as a function of time $t$ up to 1.5\,ps in addition to the mean electron-hole distance $\langle x(t)\rangle$. Here we consider the high driving force $\Delta_v=0.3\,{\rm eV}$ where the vibronic transition from exciton $\ket{E_{\rm XT},0_v}$ to delocalised CT states $\ket{E_{\rm CT},1_v}$ takes place. When the high-frequency modes are strongly damped, the vibrationally hot CT states $\ket{E_{\rm CT},1_v}$ quickly dissipate to $\ket{E_{\rm CT},0_v}$, leading to subsequent vibronic transitions to vibrationally hot interfacial CT states $\ket{E_{\rm ICT},1_v}$ (see Fig.\ref{fig:intro2}(f)). After that, the vibrational damping of the high-frequency modes generates the population of the lowest-energy interfacial CT state $\ket{E_{\rm ICT},0_v}$ and makes the electron-hole pair trapped at the interface, as shown in Fig.\ref{fig:population_dynamics}(a). When the high-frequency vibrational modes are weakly damped, the mean electron-hole distance is maximized at $\sim 700\,{\rm fs}$, as shown in Fig.\ref{fig:population_dynamics}(b), and then the population $P_{1}(t)$ of the CT state $\ket{1}$ localised around the interface starts to be increased. This localized interfacial state $\ket{1}$ has been considered an energetic trap that leads to non-radiative losses \cite{Azzouzi2019}. In Fig.\ref{fig:population_dynamics}(c), the population dynamics of $P_{1}(t)$ is shown in red and blue, respectively, for $\gamma=(50\,{\rm fs})^{-1}$ and $\gamma=(500\,{\rm fs})^{-1}$. In the strongly damped case, $P_1(t)$ rapidly increases in time and then saturates at $\sim 0.9$ on a picosecond time scale. This is contrary to the weakly damped case where $P_1(t)$ is quickly saturated at $\sim 0.1$ within 100\,fs and then does not increase until $\sim 500\,{\rm fs}$, demonstrating that the charge localization towards the interface can be delayed by the underdamped nature of the high-frequency vibrational modes. The delayed charge localization makes long-range electron-hole separation maintained on a picosecond time scale, as shown in Fig.\ref{fig:population_dynamics}(d) where $\sum_{k=5}^{9}P_{k}(t)$ is plotted. These results suggest that long-lived vibrational and vibronic coherences observed in nonlinear optical spectra of organic solar cells \cite{Song2014,DeSio2016,DeSio2018} may have a functional relevance in long-range charge separation.

\subsection{Large Vibronic Systems}

So far we have considered a one-dimensional chain consisting of $N=10$ sites. Here we investigate the charge separation dynamics in larger multi-site systems, including longer linear chains, and donor-acceptor networks in two and three spatial dimensions.

For the linear chains consisting of a donor and $(N-1)$ acceptors, we consider the total vibrational environments including low-frequency phonon baths and high-frequency vibrational modes with $\gamma=(500\,{\rm fs})^{-1}$ (${\cal J}(\omega)={\cal J}_{l}(\omega)+{\cal J}_{h}(\omega)$). The driving force is taken to be $\Delta_v=0.3\,{\rm eV}$, for which long-range charge separation occurs mediated by vibronic couplings in the case of $N=10$ sites. In Fig.\ref{fig:size_entropy}(a), a longer linear chain is considered with $N=20$ and the population dynamics $P_{k}(t)$ of the exciton and CT states $\ket{k}$ is shown. It is notable that an electron-hole pair is separated more than ten molecular units within $\sim 200$\,fs. Interestingly, with a hole fixed at the donor site, the probability distributions $P_{k}(t)$ for finding an electron at the $k$-th acceptor are strongly delocalised over the entire acceptor chain, which are maximized at $k\approx 6$ and locally minimized at $k\approx 3$. This implies that an exciton state is vibronically mixed with strongly delocalised CT states, as the detunings $\Omega_{k+1}-\Omega_k=V (k(k+1))^{-1}$ in the energy-levels of the Coulomb potential become smaller in magnitude than the electronic coupling $J_{k,k+1}=500\,{\rm cm}^{-1}$ being responsible for an electron transfer when $V=0.3\,{\rm eV}$ and $k>1$. This is in line with the dynamics of the mean electron-hole distance $\langle x(t)\rangle$ of the linear chains consisting of $N\in\{10,15,20\}$ sites, shown in solid lines in Fig.\ref{fig:size_entropy}(b), where the exciton dissociation becomes more efficient for longer acceptor chains. The mean electron-hole distance is decreased when the energy-levels $\Omega_k$ of the Coulomb potential are randomly generated based on independent Gaussian distributions, as the delocalization lengths of the CT states are reduced on average (see Appendix \ref{appendix:static_disorder}). Importantly, for the high driving force $\Delta_v=0.3\,{\rm eV}$, the charge separation process becomes significantly less efficient when vibrational environments are not considered (${\cal J}(\omega)=0$), as shown in dashed lines in Fig.\ref{fig:size_entropy}(b). These results suggest that non-equilibrium vibrational dynamics in ordered donor/acceptor aggregates can promote long-range charge separation.

From the perspective of the microcanonical ensemble, the number of charge-separated states becomes much larger than that of interfacial CT states as the dimension of donor-acceptor aggregates is increased \cite{Gregg2011}. The statistical advantage results in an entropic drive that further promotes charge separation \cite{Ono2016} and is relevant in two- and three-dimensional donor-acceptor networks in the thermodynamic limit. To corroborate these ideas, we consider a variety of donor-acceptor networks with different sizes and dimensions. In Fig.\ref{fig:size_entropy}(c), the schematic representations of one-, two- and three-dimensional donor-acceptor networks considered in our simulations are displayed where the size of each network is quantified by the number $L$ of acceptor layers. In the one-dimensional chains, the number of acceptors in each layer is unity, while in the two-dimensional triangular (three-dimensional pyramidal) structures, the number of acceptors in each layer increases linearly (quadratically) as a function of the minimum distance to the donor site. We assume that the distances between nearby sites are uniform and the corresponding nearest-neighbour electron-transfer couplings are taken to be $500\,{\rm cm}^{-1}$. The electronic Hamiltonian is described by the exciton and CT states $\ket{k}$ where a hole is fixed at the donor while an electron is localized at the $k$-th acceptor. The corresponding CT energy is modelled by $\Omega_k = -V/|{\bf r}_0-{\bf r}_k|$ with $V=0.3\,{\rm eV}$ where ${\bf r}_0$ and ${\bf r}_k$ denote, respectively, the positions of the donor and $k$-th acceptor with the distance between nearby sites taken to be unity and dimensionless. To increase the size of the donor-acceptor networks that can be considered in simulations, we only consider the high-frequency vibrational modes (${\cal J}(\omega)={\cal J}_{h}(\omega)$) with $\omega_h=1500\,{\rm cm}^{-1}$, $s_h=0.1$ and $\gamma=(500\,{\rm fs})^{-1}$.

In Fig.\ref{fig:size_entropy}(d), the time-averaged electron-hole distance $\langle x\rangle_{t\le 400\,{\rm fs}}$ simulated by DAMPF is shown as a function of the driving force $\Delta$ for one- and two-dimensional networks with $L=4$. Here we consider the minimum distance between donor and each acceptor layer in the computation of the mean electron-hole distance, instead of the distances between donor and individual acceptors. We compare the case that the high-frequency vibrational modes are coupled to electronic states (${\cal J}(\omega)={\cal J}_h(\omega)$), shown in blue, with that of no vibrational environments (${\cal J}(\omega)=0$), shown in yellow (slightly darker in the overlapped regions). Note that vibronic couplings make charge separation efficient for a broader range of the driving force $\Delta$ in both one- and two-dimensional networks, and that long-range charge separation is further enhanced in the higher-dimensional network. To simulate larger vibronic systems, in Fig.\ref{fig:size_entropy}(e), we consider a reduced vibronic model constructed within vibrational subspaces describing up to four vibrational excitations distributed amongst the high-frequency vibrational modes in the polaron basis (see Appendix \ref{appendix:reduced_model} for more details). For $L=4$, the simulated results obtained by the reduced models of one- and two-dimensional networks are qualitatively similar to the numerically exact DAMPF results shown in Fig.\ref{fig:size_entropy}(d). The reduced model results demonstrate that long-range charge separation can be enhanced by considering a three-dimensional donor-acceptor network with $L=4$, or by increasing the number of layers to $L=9$ in the one- and two-dimensional cases. In Fig.\ref{fig:size_entropy}(f), the dynamics of the mean electron-hole distance $\langle x(t)\rangle$ of the one-, two- and three-dimensional systems with $L=4$, computed by DAMPF, is shown for a high driving force $\Delta=0.35\,{\rm eV}$ where the time-averaged electron-hole distance of the three-dimensional system shown in Fig.\ref{fig:size_entropy}(e) is maximized. These results demonstrate that long-range charge separation can be enhanced by considering higher-dimensional multi-site systems with vibronic couplings.

\section{Conclusions}
We have extended the non-perturbative simulation method DAMPF to provide access to charge separation dynamics of a strongly bound electron-hole pair in one-, two- and three-dimensional donor-acceptor networks where a donor is coupled to acceptor aggregates. By controlling the driving force and the structure of vibrational environments, we identified two distinct mechanisms for long-range charge separation. The first mechanism, activated at low driving forces, is characterized by hybrid exciton-CT states where long-range exciton dissociation takes place on a sub-100\,fs time scale, which is not assisted by underdamped high-frequency vibrational modes. In the second mechanism, dominating charge separation at high driving forces, the exciton-CT hybridization occurs and it is mediated by vibronic interaction with underdamped high-frequency vibrational modes, leading to efficient charge separation for a broad range of driving forces. For both mechanisms, we have demonstrated that long-range charge separation is significantly suppressed when the high-frequency vibrational modes are strongly damped or delocalization lengths of the CT states are reduced by static disorder in the energy-levels of Coulomb potentials. These results suggest that non-equilibrium vibrational motion can promote long-range charge separation in ordered donor-acceptor aggregates.

The formulation and analysis of a reduced model whose validity became accessible to numerical corroboration thanks to the extension of the numerically exact simulation tool DAMPF allows us to identify unambiguously the mechanisms that underlie charge separation dynamics. The methods employed here can be applied to more realistic models where multiple donors are coupled to acceptor aggregates, without introducing effective one-dimensional Coulomb potentials, and vibrational environments are highly structured, which deserves a separate investigation. We expect our findings to help open up the engineering of vibrational environments for efficient long-range charge separation in organic solar cells and the identification of charge separation processes in other systems such as photosynthetic reaction centers and other biological processes driven by electron transfer.

\section*{Acknowledgments}
This work was supported by the ERC Synergy grant HyperQ (grant no. 856432), the BMBF project PhoQuant (grant no. 13N16110) under funding program quantum technologies - from basic research to market, and an IQST PhD fellowship. The authors acknowledge support by the state of Baden-Württemberg through bwHPC and the German Research Foundation (DFG) through grant no. INST 40/575-1 FUGG (JUSTUS 2 cluster)

\appendix

\section{Markovian quantum master equation}\label{appendix:PRL2018Ishizaki}

\begin{figure}
\includegraphics[width=0.45\textwidth]{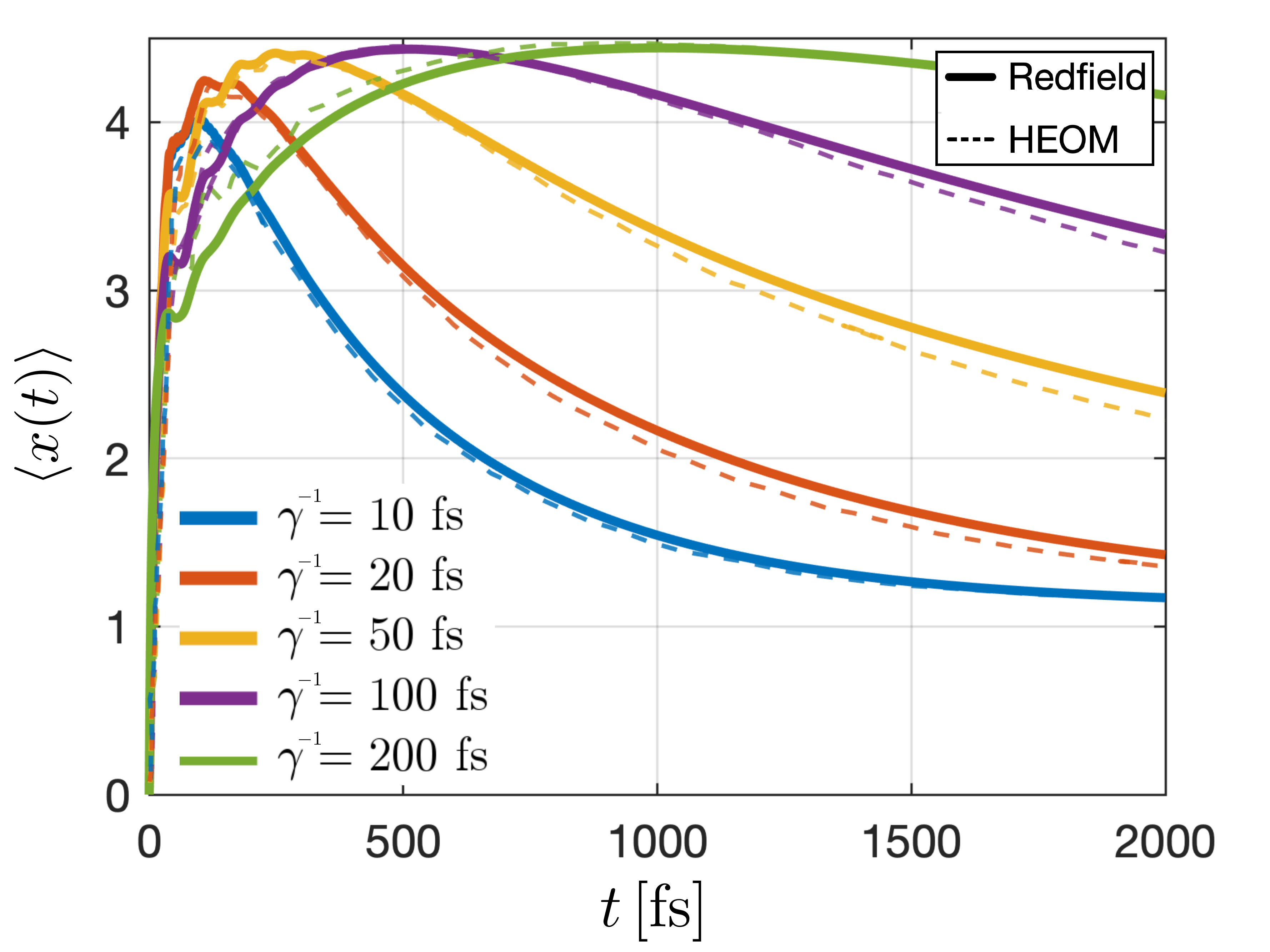}
\caption{Charge separation dynamics in the presence of weakly coupled unstructured vibrational environments. The mean electron-hole distances $\langle x(t)\rangle$, computed by a non-perturbative method (HEOM), are shown in dashed lines, while the predictions of a perturbative method (Redfield equation) are shown in solid lines. Here we consider several values of bath relaxation times $\gamma^{-1}\in\{10,20,50,100,200\}\,$fs (see blue, red, yellow, purple, green lines, respectively), considered in Ref.\cite{Kato2018}.}
\label{figSA}
\end{figure}

In Ref.\cite{Kato2018}, a linear chain consisting of a donor and ten acceptors ($N=11$ sites) was considered where each electronic state $\ket{k}$ is coupled to a broad phonon spectrum, modelled by an Ohmic spectral density with a Lorentz-Drude cutoff function, ${\cal J}(\omega)=\frac{2}{\pi}\lambda \gamma \omega (\omega^2+\gamma^2)^{-1}$, where $\lambda=\int_{0}^{\infty}d\omega {\cal J}(\omega)\omega^{-1}$ is the reoganization energy quantifying overall environmental coupling strength and $\gamma$ determines the width of the phonon spectral density. Here underdamped high-frequency vibrational modes are not considered, and as a result, quantum correlations between electronic states and vibrational environments decay within a 100\,fs time scale for the values of $\gamma$ considered in Ref.\cite{Kato2018}. The energy-levels of the CT states $\ket{k}$ with $k\ge 1$ were modelled by $\Omega_k=-V/k$ with $V=0.3$\,eV, as is the case of our work. To take into account non-Markovian environmental effects, hierarchical equations of motion (HEOM) \cite{Tanimura1989,Tanimura2006} were used in Ref.\cite{Kato2018}. 
However, electronic couplings ($J_{0,1}=0.15\,$eV and $J_{k,k+1}=0.10\,$eV for $k\ge 1$) were assumed to be several times stronger than the overall environmental coupling strength ($\lambda=0.02$\,eV), already suggesting that the influence of the broad phonon spectral density considered in Ref.\cite{Kato2018} on reduced electronic dynamics could be well described by a Markovian quantum master equation where non-Markovian environmental effects are ignored \cite{RivasSusana2012Book}.
Fig.\ref{figSA} shows that the mean electron-hole distances $\langle x(t)\rangle$ computed by HEOM, shown in dashed lines, are well matched to the perturbative electronic dynamics computed by the Redfield equation, shown in solid lines, for several values of $\gamma$ considered in Ref.\cite{Kato2018}. These results demonstrate that the electronic dynamics reported in Ref.\cite{Kato2018} can be well reproduced by a Markovian noise model and therefore non-Markovian effects may not play an important role in charge separation processes for the model parameters considered in Ref.\cite{Kato2018}.

\section{Weak Coulomb-binding energy}\label{appendix:JACS2013Burghardt}

\begin{figure}
\includegraphics[width=0.49\textwidth]{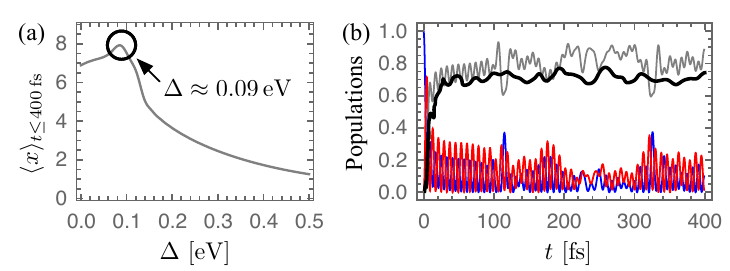}
\caption{Electronic charge separation of weakly bound electron-hole pair. (a) Time-averaged mean electron-hole distance $\langle x\rangle_{t\le 400\,{\rm fs}}$ is shown as a function of driving force $\Delta$, where a linear chain consisting of $N=20$ sites is considered with electronic parameters from Ref.\cite{Tamura2013JACS}. Here ${\cal J}(\omega)=0$ (no vibrational modes), and $\Delta\approx 0.09\,{\rm eV}$ is the value considered in Ref.\cite{Tamura2013JACS}. (b) The population dynamics of $P_{0}(t)$, $P_{1}(t)$ and $\sum_{k=2}^{19}P_{k}(t)$, in the absence of vibrational modes, are shown in blue, red and grey, respectively. The transient of $\sum_{k=2}^{19}P_{k}(t)$ in the presence of vibrational environments estimated by first-principles methods is shown in black with data taken from Ref.\cite{Tamura2013JACS}.}
\label{figSB}
\end{figure}

Here we show that long-range charge separation can take place even in the absence of vibrational environments when electronic couplings are sufficiently stronger than detunings in the energy-levels of Coulomb potentials. In this case, vibronic enhancement of charge separation can be negligible even if vibrational environments are modelled based on first-principles methods.

In Ref.\cite{Tamura2013JACS}, where vibronic effects in charge separation process were investigated by using first-principles methods, a hole transfer through a linear chain of donors was simulated with the assumption that an electron is delocalized in acceptor aggregates. It was suggested that the electron delocalisation can reduce the Coulomb attraction between electron and hole. This leads to an effective one-dimensional model where the detunings $|\Omega_k-\Omega_{k+1}|$ in the energy-levels of the CT states $\ket{k}$ are smaller in magnitude than electronic couplings, $J_{0,1}=0.2\,{\rm eV}$ and $J_{k,k+1}=0.12\,{\rm eV}$ for $1\le k\le 18$. For these electronic parameters, the time-averaged mean electron-hole distance $\langle x\rangle_{t\le 400\,{\rm fs}}$ in the absence of vibrational environments (${\cal J}(\omega)=0$) is shown as a function of driving force $\Delta$ in Fig.\ref{figSB}(a), which is maximised at $\Delta\approx 0.09\,{\rm eV}$, being close to the driving force considered in numerical simulations of Ref.\cite{Tamura2013JACS}. For $\Delta\approx 0.09\,{\rm eV}$, the population dynamics of $P_{0}(t)$, $P_{1}(t)$ and $\sum_{k=2}^{19}P_{k}(t)$ in the absence of vibrational environments are shown in blue, red and grey lines, respectively, in Fig.\ref{figSB}(b), which are qualitatively similar to the results of Ref.\cite{Tamura2013JACS} where electronic-vibrational couplings were considered in simulations (see a black line in Fig.\ref{figSB}(b) displaying the transient of $\sum_{k=2}^{19}P_{k}(t)$ shown in Ref.\cite{Tamura2013JACS}). These results demonstrate that ultrafast long-range charge separation can take place even if it is not assisted by vibronic couplings, contrary to the vibronic mechanism proposed in Ref.\cite{Tamura2013JACS}, when electronic couplings are sufficiently strong.

\section{DAMPF}\label{appendix:method}

In this work, non-perturbative simulations of charge separation dynamics have been carried out by using dissipation-assisted matrix product factorization (DAMPF) method \cite{Somoza2019} to take into account non-Markovian effects induced by a continuous spectrum of vibrational environments. The DAMPF method relies on a tensor network formalism and pseudomode theory \cite{Tamascelli2018,Mascherpa2020}, which enables one to consider a finite number of oscillators under Markovian noise (pseudomodes) to mimic in a numerically accurate way the action of the original continuous vibrational environments on reduced electronic system dynamics. In this work, vibrational environments consist of low-frequency phonon baths and high-frequency vibrational modes. The low-frequency phonon bath, modelled by an Ohmic spectral density with an exponential cutoff function, has been described by five pseudomodes coupled to each other. On the contrary, every high-frequency vibrational mode has been described by an independent pseudomode, which can well describe a narrow Lorentzian spectral density. In this case, the vibrational Hamiltonians of the coupled and uncoupled pseudomodes, respectively, are described by
\begin{align}
    H_v^{(l)} = &\sum_{k=0}^{N-1}\sum_{q=1}^{5}(\omega_q a^\dagger_{k,q}a_{k,q}+g_q a_{k,q}a^\dagger_{k,q+1} + h.c.),\\
    H_v^{(h)} = &\sum_{k=0}^{N-1}\omega_{h} b^\dagger_{k,h}b_{k,h},
\end{align}
where $a_{k,q}$ and $b_{k,h}$ represent pseudomodes coupled to an electronic state $\ket{k}$, and $g_q$ is the coupling between pseudomodes $a_{k,q}$ and $a_{k,q+1}$ with $q+1$ reduced to 1 when $q=5$. The vibronic couplings to the pseudomodes are described by
\begin{align}
    H_{e-v}&=\sum_{k=0}^{N-1}\ket{k}\bra{k}\sum_{q=1}^{5}(c_{q}a_{k,q}+c_{q}^{*}a_{k,q}^{\dagger})\\
    &\quad+\sum_{k=0}^{N-1}\ket{k}\bra{k}\omega_{h}\sqrt{s_h}(b_{k,h}+b_{k,h}^{\dagger}).
\end{align}
The time evolution of a vibronic density matrix $\rho$ is governed by a Lindblad equation in the form
\begin{align}
    \dot{\rho} &= -i[H_e + H_v^{(l)} + H_v^{(h)} + H_{e-v}, \rho]\\
    &\quad+\sum_{k=0}^{N-1}\sum_{q=1}^{5}\Big[\gamma_q(n(\omega_q) + 1)(a_{k,q}\rho a_{k,q}^\dagger - \frac{1}{2}\{a_{k,q}^\dagger a_{k,q},\rho\})\nonumber\\
    &\quad\qquad\qquad\qquad\,+\gamma_q n(\omega_q)(a_{k,q}^\dagger\rho a_{k,q} - \frac{1}{2}\{a_{k,q} a_{k,q}^\dagger,\rho\})\Big]\nonumber\\
    &\quad+\sum_{k=0}^{N-1}\Big[\gamma_h(n(\omega_h) + 1)(b_{k,h}\rho b_{k,h}^\dagger - \frac{1}{2}\{b_{k,h}^\dagger b_{k,h},\rho\})\nonumber\\
    &\quad\qquad\qquad\,\,\,+\gamma_h n(\omega_h)(b_{k,h}^\dagger\rho b_{k,h} - \frac{1}{2}\{b_{k,h} b_{k,h}^\dagger,\rho\})\Big],\nonumber
\end{align}
with $n(\omega)=(\exp(\hbar\omega/k_B T)-1)^{-1}$. The parameters $\omega_q$, $g_q$, $c_q$ and $\gamma_q$ of the coupled pseudomodes can be found in Table IV of Ref.\cite{Mascherpa2020} with a center frequency $\omega_l=80$ cm$^{-1}$. The vibrational frequency $\omega_h$ of the high-frequency modes is approximately an order of magnitude larger than the thermal energy $k_B T$ at room temperature, for which $n(\omega_h)\ll 1$ and $\gamma_h$ is reduced to the vibrational damping rate $\gamma\in\{(50\,{\rm fs})^{-1},(500\,{\rm fs})^{-1}\}$ considered in the main text.

In DAMPF, where the number of uncoupled (coupled) pseudomodes is denoted by $Q_u=N$ ($Q_c=5N$), we consider a vibronic density matrix $\rho$ represented in the form
\begin{align}
    \rho = &\sum_{k,k'=0}^{N-1} \ket{k}\bra{k'}\otimes \sum_{i_1}\sum_{i_2}\cdots\sum_{i_{Q_c}}\sum_{i_{Q_u}}\\
    &A^{(k,k')}_{(1;i_1)}A^{(k,k')}_{(2;i_2)}\dots A^{(k,k')}_{(Q_c;i_{Q_c})}A^{(k,k')}_{(Q_u;i_{Q_u})} x_{i_1}\otimes\dots\otimes x_{i_{Q_u}},\nonumber
\end{align}
describing a collection of $N^2$ matrix product states (MPSs), via a vectorization of the density matrix, conditional to the populations $\ket{k}\bra{k}$ or coherences $\ket{k}\bra{k'}$ with $k\neq k'$ of electronic states $\ket{k}$. This representation retains thus all quantum correlations that have a clear electronic origin, restricting the factorization at the level of different vibrational environments. We found that the simulation costs of DAMPF required to achieve convergence in reduced electronic dynamics are significantly higher for genuine vibronic dynamics occurring at high driving forces $\Delta_v\approx 0.3$\,eV than for electronic dynamics taking place at low driving forces $\Delta_e\approx 0.15$\,eV. This implies that strong correlations between electronic states and vibrational modes, quantified by bond dimensions, occur when a vibronic mixing induces long-range charge separation.

\section{Purely coherent electronic dynamics}\label{appendix:onlyel}

\begin{figure}
\includegraphics[width=0.48\textwidth]{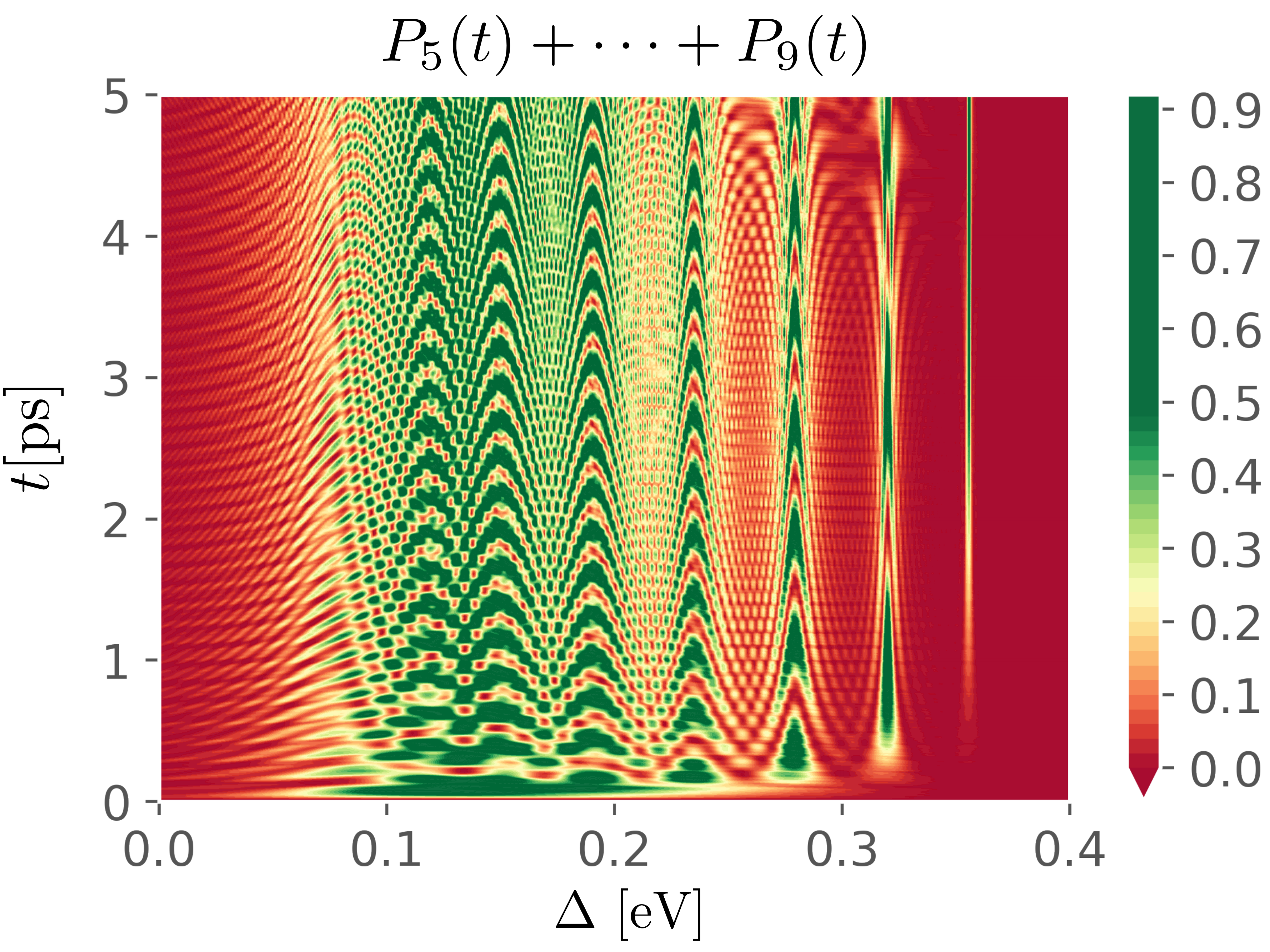}
\caption{Charge separation in the absence of vibrational environments. For a linear chain consisting of $N=10$ sites, the probability for separating an electron-hole pair more than four molecular units, $\sum_{k=5}^{9}P_{k}(t)$, is shown as a function of time $t$ and driving force $\Delta$.}
\label{figSD}
\end{figure}

In Fig.\ref{figSD}, the probability for separating an electron-hole pair more than four molecular units, defined by $\sum_{k=5}^{9}P_{k}(t)$, is shown as a function of time $t$ and driving force $\Delta$ for a linear chain composed of a donor and nine acceptors ($N=10$) where vibrational environments are not considered (${\cal J}(\omega)=0$). Long-range charge separation takes place at discrete values of the driving force $\Delta$ when exciton $\ket{0}$ is near-resonant with a delocalised CT state $|E_{\alpha}^{({\rm CT})}\rangle$. Here oscillatory features of $\sum_{k=5}^{9}P_{k}(t)$ on a picosecond time scale are due to the absence of noise induced by vibrational environments (${\cal J}(\omega)=0$). Note that the period of oscillations is longer at higher driving force $\Delta$, demonstrating that the coupling $\bra{0}H_i|E_{\alpha}^{\rm (CT)} \rangle = J_{0,1}\langle 1|E_{\alpha}^{\rm (CT)} \rangle$ between exciton and CT states is weaker for higher-energy CT state. This implies that long-range charge separation at higher driving force $\Delta$ requires a smaller energy-gap between exciton and CT states since the degree of exciton-CT mixing is determined by the ratio of the coupling and detuning between $\ket{0}$ and $|E_{\alpha}^{({\rm CT})}\rangle$, leading to narrower vertical line shapes at higher driving forces in Fig.\ref{figSD}.

\section{Vibronic eigenstate analysis}\label{appendix:reduced_model}

To analyse vibronic mixing of exciton and delocalized CT states at high driving force $\Delta_v=0.3\,$eV, we consider a reduced vibronic model where electronic states are coupled to high-frequency vibrational modes. For simplicity, the damping of the high-frequency modes and the vibronic coupling to low-frequency phonon baths are not considered. The reduced model Hamiltonian consists of three parts, $H_r=H_e+H_{v}'+H_{e-v}'$, defined by
\begin{align}
	&H_{e}=\sum_{k}\Omega_k \ket{k}\bra{k}+\sum_{k,k'}J_{k,k'}\ket{k}\bra{k'},\\
	&H_{v}'=\sum_{k}\omega_{h}b_{k,h}^{\dagger}b_{k,h},\\
	&H_{e-v}'=\sum_{k}\ket{k}\bra{k}\omega_h\sqrt{s_h}(b_{k,h}+b_{k,h}^{\dagger}).
\end{align}
To minimize the number of vibrational states required to achieve the numerical convergence in reduced electronic dynamics, we consider a displaced vibrational basis defined by unitary displacement operator conditional to electronic states (polaron transformation), $U=\sum_{k}\ket{k}\bra{k}D_k$ with $D_k=\exp(\sqrt{s_h}(b_{k,h}^{\dagger}-b_{k,h}))$. In the polaron basis, the reduced vibronic Hamiltonian is expressed as
\begin{align}
	UH_{r}U^{\dagger}&=\sum_{k}(\Omega_k-\omega_h s_h)\ket{k}\bra{k}+\sum_{k}\omega_{h}b_{k,h}^{\dagger}b_{k,h}\\
	&\quad+\sum_{k,k'}J_{k,k'}\ket{k}\bra{k'}D_k D_{k'}^\dagger.\nonumber
\end{align}
In simulations, we consider a vibrational subspace spanned by up to $N_v$ vibrational excitations distributed amongst multiple high-frequency modes $b_{k,h}$. When $N_v=0$, only the global vibrational ground state $\ket{0_v}$ is considered where all the high-frequency modes are in their vibrational ground states. When $N_v=1$, singly excited vibrational states $|1_{v}^{(k)}\rangle$ are considered, in addition to $\ket{0_v}$, where only the $k$-th high-frequency mode $b_{k,h}$ is singly excited, while all the other modes are in their vibrational ground states. When $N_v=2$, doubly excited vibrational states are considered, in addition to $\ket{0_v}$ and $|1_{v}^{(k)}\rangle$, where only a single vibrational mode is doubly excited ($|2_{v}^{(k)}\rangle$) or two different modes are singly excited at the same time ($|1_{v}^{(k)},1_{v}^{(k')}\rangle$).

\begin{figure}
\includegraphics[width=0.48\textwidth]{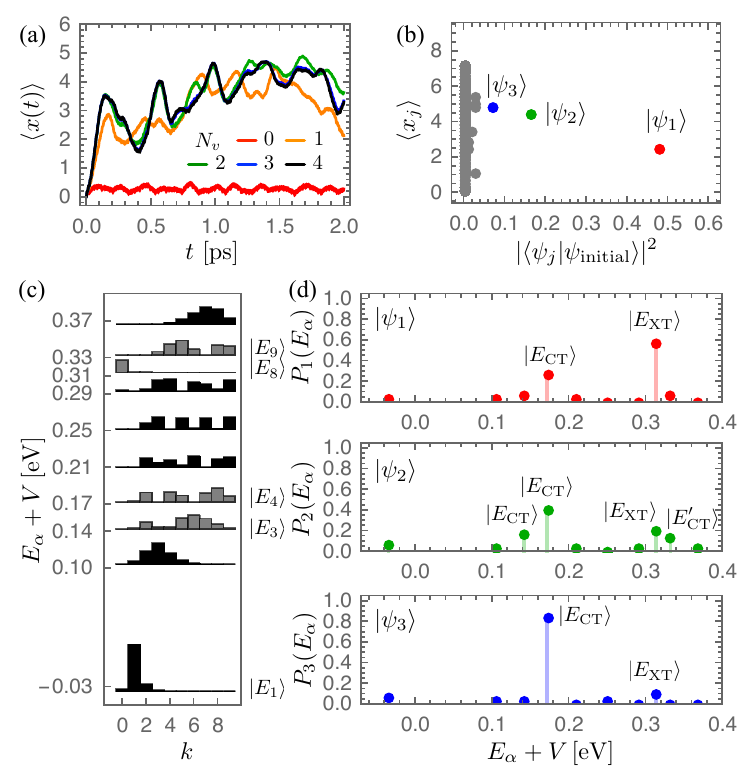}
\caption{Vibronic eigenstate analysis. (a) Transient of mean electron-hole distance $\langle x(t)\rangle$ of a linear chain consisting of a donor and nine acceptors where $\Delta_v=0.3\,{\rm eV}$. Here we consider a reduced vibronic model where up to $N_v$ vibrational excitations are considered in total. (b) Electron-hole distance $\langle x_j\rangle$ of vibronic eigenstate $\ket{\psi_j}$ is displayed as a function of the overlap $|\langle \psi_j|\psi_{\rm initial}\rangle|^{2}$ with vibrationally cold initial exciton state $\ket{\psi_{\rm initial}}$. (c) Electronic eigenstate structure. (d) For vibronic eigenstates that have large overlaps with the initial state, denoted by $\ket{\psi_{j=1,2,3}}$ in (b), the populations $P_{j}(E_{\alpha})$ in electronic eigenbasis are shown, which are dominated by exciton state $\ket{E_{\rm XT}}=\ket{E_8}$, low-lying CT states $\ket{E_{\rm CT}}\in\{\ket{E_3},\ket{E_4}\}$ and high-lying CT state $\ket{E_{\rm CT}'}=\ket{E_9}$.}
\label{figSE}
\end{figure}

In Fig.\ref{figSE}, we consider a linear chain consisting of $N=10$ sites and $\Delta_v=0.3\,{\rm eV}$ where the vibronic mixing of exciton and delocalized CT states is induced by high-frequency vibrational modes with $\omega_h=1200\,{\rm cm}^{-1}$ and $s_h=0.1$. In Fig.\ref{figSE}(a), the mean electron-hole distance $\langle x(t)\rangle$ computed by the reduced Hamiltonian is displayed where the total number $N_v$ of vibrational excitations is increased from 0 to 4. When vibrational excitations are not considered in simulations ($N_v=0$), the mean electron-hole distance remains below 1 up to $2\,{\rm ps}$, as shown in red. When vibrational excitations are considered in simulations ($N_v\ge 1$), charge separation process is significantly enhanced by vibronic couplings. The electronic dynamics shows convergence when $N_v\ge 3$ where a blue line for $N_v=3$ is well overlapped with a black line for $N_v=4$. The qualitative features of the numerically converged electronic dynamics can be well reproduced by approximate results for $N_v=1$ and $N_v=2$, as shown in orange and green, respectively.

\begin{table}
\caption{Vibronic eigenstate analysis. For vibronic eigenstates $\ket{\psi_{j=1,2,3}}$ shown in Fig.\ref{figSE}(b), populations $P_{j}(E_{\alpha},N_{v})$ in electronic eigenbasis are summarized where $N_v$ denotes the total number of vibrational excitations distributed amongst high-frequency modes. $P_{j}(E_{\alpha},N_{v})< 0.01$ are not shown here.}
\label{tableS1}
\begin{ruledtabular}
\begin{tabular}{llll}
$|\psi_1\rangle$ & & $|\psi_2\rangle$ & \\
\hline
$|E_{\alpha},N_{v}\rangle$ & $P_{1}(E_{\alpha},N_{v})$ & $|E_{\alpha},N_{v}\rangle$ & $P_{2}(E_{\alpha},N_{v})$ \\
\hline
$|E_8,0_v\rangle$ & 0.55 & $|E_4,1_v\rangle$ & 0.41 \\
$|E_4,1_v\rangle$ & 0.26 & $|E_8,0_v\rangle$ & 0.20 \\
$|E_3,1_v\rangle$ & 0.06 & $|E_3,1_v\rangle$ & 0.15 \\
$|E_9,0_v\rangle$ & 0.05 & $|E_9,0_v\rangle$ & 0.12 \\
$|E_1,2_v\rangle$ & 0.02 & $|E_1,2_v\rangle$ & 0.04 \\
$|E_2,1_v\rangle$ & 0.02 & $|E_{10},0_v\rangle$ & 0.03 \\
$|E_1,1_v\rangle$ & 0.01 & $|E_2,1_v\rangle$ & 0.02 \\
 & & $|E_5,1\rangle$ & 0.01 \\
\hline\hline
$|\psi_3\rangle$ & \\
\hline
$|E_{\alpha},N_{v}\rangle$ & $P_{3}(E_{\alpha},N_{v})$ \\
\hline
$|E_4,1_v\rangle$ & 0.81 \\
$|E_8,0_v\rangle$ & 0.09 \\
$|E_1,2_v\rangle$ & 0.05 \\
\hline
\end{tabular}
\end{ruledtabular}
\end{table}

To identify the origin of the vibronic enhancement of charge separation, we consider vibronic eigenstates of the polaron-transformed Hamiltonian $UH_{r}U^{\dagger}$, represented by
\begin{equation}
	\ket{\psi_j}=\sum_{\alpha=1}^{N}\sum_{\vec{n}_v}\psi_{j}(E_\alpha,\vec{n}_v)\ket{E_\alpha,\vec{n}_v},
\end{equation}
where $\ket{E_\alpha}$ denote the eigenstates of electronic Hamiltonian $H_e$, while $\vec{n}_{v}$ describe the distributions of vibrational excitations of the high-frequency modes. The populations of the vibronic eigenstates $\ket{\psi_j}$ in the electronic basis $\ket{k}$, where a hole is fixed at donor and an electron is localised at the donor ($k=0$) or at the $k$-th acceptor ($k\ge 1$), are defined by $P_{j}(k)={\rm Tr}_v[\langle k|\psi_j\rangle\langle\psi_j|k\rangle]$ where ${\rm Tr}_v$ denotes the partial trace over vibrational degrees of freedom. Similarly we consider the populations of the vibronic eigenstates in the electronic eigenbasis $\ket{E_\alpha}$, defined by $P_{j}(E_\alpha)={\rm Tr}_v[\langle E_\alpha|\psi_j\rangle\langle\psi_j|E_\alpha\rangle]$. In Fig.\ref{figSE}(b), the overlap $|\langle \psi_j |\psi_{\rm initial}\rangle|^{2}$ between vibronic eigenstates $\ket{\psi_j}$ and vibrationally cold initial exciton state $\ket{\psi_{\rm initial}}=U |0,0_v \rangle$ in the polaron basis is shown as a function of electron-hole distance of the vibronic eigenstates, defined by $\langle x_j\rangle=\sum_{k=0}^{N-1}k P_{j}(k)$. It is notable that the overlap with the initial state is dominated by three vibronic eigenstates $\ket{\psi_{j=1,2,3}}$ with comparable electron-hole distances of $\langle x_j\rangle\approx 4$, which is close to the maximum electron-hole distance $\langle x(t)\rangle$ shown in Fig.\ref{figSE}(a). The populations $P_j(E_\alpha)$ of $\ket{\psi_{j=1,2,3}}$ in the electronic eigenbasis are shown in Fig.\ref{figSE}(d) as a function of the energy-levels $E_{\alpha}$ of electronic eigenstates. It is notable that the vibronic eigenstates are dominated by the populations of the exciton state $\ket{E_{\rm XT}}=\ket{E_8}$ and lower-energy CT states $\ket{E_{\rm CT}}\in\{\ket{E_4},\ket{E_5}\}$ with a relatively small contribution of a high-lying CT state $\ket{E_{\rm CT}'}=\ket{E_9}$ near-resonant with the exciton state. These results demonstrate that the exciton state is vibronically mixed with delocalised CT states.

The presence of the vibronic mixing can be demonstrated more clearly by analysing both electronic and vibrational states of the vibronic eigenstates $\ket{\psi_j}$. For a linear chain consisting of $N=10$ sites, the electronic states are coupled to ten high-frequency vibrational modes in total. To simplify analysis, for each vibronic eigenstate $\ket{\psi_j}$, we investigate the overlap with electronic eigenstates $\ket{E_{\alpha}}$ in the presence of $N_v$ vibrational excitations, defined by $P_{j}(E_{\alpha},N_{v})=\sum_{{\rm sum}(\vec{n}_v)=N_v}|\psi_{j}(E_{\alpha},\vec{n}_v)|^{2}$ where the summation runs over all possible vibrational states with $N_v$ excitations distributed amongst the ten high-frequency modes. In Table \ref{tableS1}, all $P_{j}(E_{\alpha},N_{v})$ being larger than 0.01 are shown, demonstrating that the vibronic eigenstates $\ket{\psi_{j=1,2,3}}$ governing initial charge separation dynamics are well described by the superpositions of vibrationally cold exciton state $\ket{E_{\rm XT},0_v}$, vibrationally hot $\ket{E_{\rm CT},1_v}$ and cold CT states $\ket{E_{\rm CT}',0_v}$ delocalised in the acceptor domain.

\begin{figure}
\includegraphics[width=0.48\textwidth]{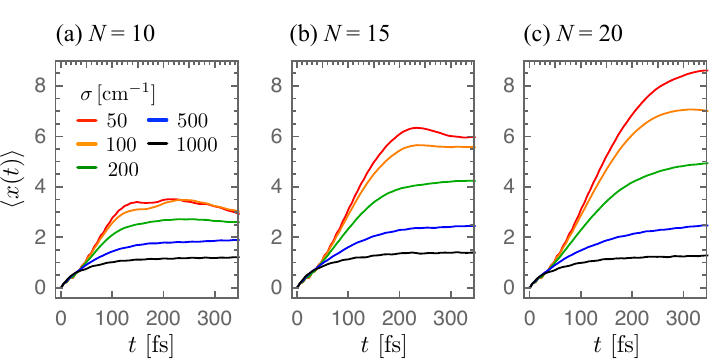}
\caption{Charge separation dynamics in the presence of static disorder. (a-c) Ensemble-averaged mean electron-hole distance $\langle x(t)\rangle$ of one-dimensional chains consisting of (a) $N=10$, (b) $N=20$, (c) $N=30$ sites. The degree of static disorder in the energy-levels $\Omega_k$ of Coulomb potentials is quantified by the standard deviation $\sigma$ of independent Gaussian distributions, which is taken to be $\sigma=50$, $100$, $200$, $500$, $1000\,{\rm cm}^{-1}$, respectively, shown in red, orange, green, blue, black. For each ensemble-averaged transient, 1000 randomly generated sets of energy-levels $\Omega_k$ were considered.}
\label{figSF}
\end{figure}

\section{Static disorder}\label{appendix:static_disorder}

To investigate how static disorder in the energy-levels $\Omega_k$ of Coulomb potentials affects charge separation dynamics, we consider one-dimensional chains consisting of a donor and $(N-1)$ acceptors. The energy-levels $\Omega_k$ of electronic states $\ket{k}$ are modelled by independent Gaussian distributions with mean values taken to be $\langle \Delta\rangle = \langle \Omega_0-\Omega_1 \rangle = 0.3\,$eV and $\langle \Omega_k\rangle = -V/k$ for $k\ge 1$. For simplicity, we employ the reduced vibronic model in Appendix \ref{appendix:reduced_model}.

In Fig.\ref{figSF}(a), where a linear chain consisting of $N=10$ sites is considered, the ensemble-averaged mean electron-hole distance $\langle x(t)\rangle$ is shown for several values of the standard deviation $\sigma$ of the Gaussian distributions. For each value of $\sigma$, 1000 randomly generated sets of the energy-levels $\Omega_k$ were considered. It is notable that the electron-hole distance is suppressed as the standard deviation $\sigma$ of the Gaussian distributions is increased from 50, via 100, 200, 500, to $1000\,{\rm cm}^{-1}$. This implies that charge separation is suppressed as delocalization lengths of the CT states are decreased by static disorder. In Fig.\ref{figSF}(b) and (c), longer linear chains with $N=15$ and $N=20$ are considered, respectively, where vibronic charge separation is also found to be suppressed by static disorder. Note that as the degree of static disorder, quantified by $\sigma$, is increased, the charge separation dynamics becomes almost independent of the size $(N-1)$ of acceptor aggregates (see blue lines in Fig.\ref{figSF}(a)-(c) where $\sigma=J_{k,k+1}=500\,{\rm cm}^{-1}$).

\bibliography{library2}

\begin{thebibliography}{80}%
\makeatletter
\providecommand \@ifxundefined [1]{%
 \@ifx{#1\undefined}
}%
\providecommand \@ifnum [1]{%
 \ifnum #1\expandafter \@firstoftwo
 \else \expandafter \@secondoftwo
 \fi
}%
\providecommand \@ifx [1]{%
 \ifx #1\expandafter \@firstoftwo
 \else \expandafter \@secondoftwo
 \fi
}%
\providecommand \natexlab [1]{#1}%
\providecommand \enquote  [1]{``#1''}%
\providecommand \bibnamefont  [1]{#1}%
\providecommand \bibfnamefont [1]{#1}%
\providecommand \citenamefont [1]{#1}%
\providecommand \href@noop [0]{\@secondoftwo}%
\providecommand \href [0]{\begingroup \@sanitize@url \@href}%
\providecommand \@href[1]{\@@startlink{#1}\@@href}%
\providecommand \@@href[1]{\endgroup#1\@@endlink}%
\providecommand \@sanitize@url [0]{\catcode `\\12\catcode `\$12\catcode
  `\&12\catcode `\#12\catcode `\^12\catcode `\_12\catcode `\%12\relax}%
\providecommand \@@startlink[1]{}%
\providecommand \@@endlink[0]{}%
\providecommand \url  [0]{\begingroup\@sanitize@url \@url }%
\providecommand \@url [1]{\endgroup\@href {#1}{\urlprefix }}%
\providecommand \urlprefix  [0]{URL }%
\providecommand \Eprint [0]{\href }%
\providecommand \doibase [0]{https://doi.org/}%
\providecommand \selectlanguage [0]{\@gobble}%
\providecommand \bibinfo  [0]{\@secondoftwo}%
\providecommand \bibfield  [0]{\@secondoftwo}%
\providecommand \translation [1]{[#1]}%
\providecommand \BibitemOpen [0]{}%
\providecommand \bibitemStop [0]{}%
\providecommand \bibitemNoStop [0]{.\EOS\space}%
\providecommand \EOS [0]{\spacefactor3000\relax}%
\providecommand \BibitemShut  [1]{\csname bibitem#1\endcsname}%
\let\auto@bib@innerbib\@empty
\bibitem [{\citenamefont {Scholes}\ and\ \citenamefont
  {Rumbles}(2006)}]{Scholes2006}%
  \BibitemOpen
  \bibfield  {author} {\bibinfo {author} {\bibfnamefont {G.~D.}\ \bibnamefont
  {Scholes}}\ and\ \bibinfo {author} {\bibfnamefont {G.}~\bibnamefont
  {Rumbles}},\ }\bibfield  {title} {\bibinfo {title} {{Excitons in Nanoscale
  Systems}},\ }\href {https://doi.org/10.1038/nmat1756} {\bibfield  {journal}
  {\bibinfo  {journal} {Nat. Mater.}\ }\textbf {\bibinfo {volume} {5}},\
  \bibinfo {pages} {920} (\bibinfo {year} {2006})}\BibitemShut {NoStop}%
\bibitem [{\citenamefont {Hedley}\ \emph {et~al.}(2017)\citenamefont {Hedley},
  \citenamefont {Ruseckas},\ and\ \citenamefont {Samuel}}]{Hedley2017}%
  \BibitemOpen
  \bibfield  {author} {\bibinfo {author} {\bibfnamefont {G.~J.}\ \bibnamefont
  {Hedley}}, \bibinfo {author} {\bibfnamefont {A.}~\bibnamefont {Ruseckas}},\
  and\ \bibinfo {author} {\bibfnamefont {I.~D.~W.}\ \bibnamefont {Samuel}},\
  }\bibfield  {title} {\bibinfo {title} {{Light Harvesting for Organic
  Photovoltaics}},\ }\href {https://doi.org/10.1021/acs.chemrev.6b00215}
  {\bibfield  {journal} {\bibinfo  {journal} {Chem. Rev.}\ }\textbf {\bibinfo
  {volume} {117}},\ \bibinfo {pages} {796} (\bibinfo {year}
  {2017})}\BibitemShut {NoStop}%
\bibitem [{\citenamefont {Vandewal}(2016)}]{Vandewal2016}%
  \BibitemOpen
  \bibfield  {author} {\bibinfo {author} {\bibfnamefont {K.}~\bibnamefont
  {Vandewal}},\ }\bibfield  {title} {\bibinfo {title} {{Interfacial Charge
  Transfer States in Condensed Phase Systems}},\ }\href
  {https://doi.org/10.1146/annurev-physchem-040215-112144} {\bibfield
  {journal} {\bibinfo  {journal} {Annu. Rev. Phys. Chem.}\ }\textbf {\bibinfo
  {volume} {67}},\ \bibinfo {pages} {113} (\bibinfo {year} {2016})}\BibitemShut
  {NoStop}%
\bibitem [{\citenamefont {Coropceanu}\ \emph {et~al.}(2019)\citenamefont
  {Coropceanu}, \citenamefont {Chen}, \citenamefont {Wang}, \citenamefont
  {Zheng},\ and\ \citenamefont {Br{\'{e}}das}}]{Coropceanu2019}%
  \BibitemOpen
  \bibfield  {author} {\bibinfo {author} {\bibfnamefont {V.}~\bibnamefont
  {Coropceanu}}, \bibinfo {author} {\bibfnamefont {X.-K.}\ \bibnamefont
  {Chen}}, \bibinfo {author} {\bibfnamefont {T.}~\bibnamefont {Wang}}, \bibinfo
  {author} {\bibfnamefont {Z.}~\bibnamefont {Zheng}},\ and\ \bibinfo {author}
  {\bibfnamefont {J.-L.}\ \bibnamefont {Br{\'{e}}das}},\ }\bibfield  {title}
  {\bibinfo {title} {{Charge-transfer Electronic States in Organic Solar
  Cells}},\ }\href {https://doi.org/10.1038/s41578-019-0137-9} {\bibfield
  {journal} {\bibinfo  {journal} {Nat. Rev. Mater.}\ }\textbf {\bibinfo
  {volume} {4}},\ \bibinfo {pages} {689} (\bibinfo {year} {2019})}\BibitemShut
  {NoStop}%
\bibitem [{\citenamefont {van Amerongen}\ \emph {et~al.}(2000)\citenamefont
  {van Amerongen}, \citenamefont {van Grondelle},\ and\ \citenamefont
  {Valkunas}}]{Valkunas_book}%
  \BibitemOpen
  \bibfield  {author} {\bibinfo {author} {\bibfnamefont {H.}~\bibnamefont {van
  Amerongen}}, \bibinfo {author} {\bibfnamefont {R.}~\bibnamefont {van
  Grondelle}},\ and\ \bibinfo {author} {\bibfnamefont {L.}~\bibnamefont
  {Valkunas}},\ }\href {https://doi.org/10.1142/3609} {\emph {\bibinfo {title}
  {{Photosynthetic Excitons}}}}\ (\bibinfo  {publisher} {World Scientific},\
  \bibinfo {year} {2000})\BibitemShut {NoStop}%
\bibitem [{\citenamefont {Huelga}\ and\ \citenamefont
  {Plenio}(2013)}]{Huelga2013}%
  \BibitemOpen
  \bibfield  {author} {\bibinfo {author} {\bibfnamefont {S.~F.}\ \bibnamefont
  {Huelga}}\ and\ \bibinfo {author} {\bibfnamefont {M.~B.}\ \bibnamefont
  {Plenio}},\ }\bibfield  {title} {\bibinfo {title} {{Vibrations, Quanta and
  Biology}},\ }\href {https://doi.org/10.1080/00405000.2013.829687} {\bibfield
  {journal} {\bibinfo  {journal} {Contemp. Phys.}\ }\textbf {\bibinfo {volume}
  {54}},\ \bibinfo {pages} {181} (\bibinfo {year} {2013})}\BibitemShut
  {NoStop}%
\bibitem [{\citenamefont {Romero}\ \emph {et~al.}(2017)\citenamefont {Romero},
  \citenamefont {Novoderezhkin},\ and\ \citenamefont {van
  Grondelle}}]{Romero2017}%
  \BibitemOpen
  \bibfield  {author} {\bibinfo {author} {\bibfnamefont {E.}~\bibnamefont
  {Romero}}, \bibinfo {author} {\bibfnamefont {V.~I.}\ \bibnamefont
  {Novoderezhkin}},\ and\ \bibinfo {author} {\bibfnamefont {R.}~\bibnamefont
  {van Grondelle}},\ }\bibfield  {title} {\bibinfo {title} {{Quantum Design of
  Photosynthesis for Bio-Inspired Solar-Energy Conversion}},\ }\href
  {https://doi.org/10.1038/nature22012} {\bibfield  {journal} {\bibinfo
  {journal} {Nature}\ }\textbf {\bibinfo {volume} {543}},\ \bibinfo {pages}
  {355} (\bibinfo {year} {2017})}\BibitemShut {NoStop}%
\bibitem [{\citenamefont {Br{\'{e}}das}\ \emph {et~al.}(2009)\citenamefont
  {Br{\'{e}}das}, \citenamefont {Norton}, \citenamefont {Cornil},\ and\
  \citenamefont {Coropceanu}}]{Bredas2009}%
  \BibitemOpen
  \bibfield  {author} {\bibinfo {author} {\bibfnamefont {J.-L.}\ \bibnamefont
  {Br{\'{e}}das}}, \bibinfo {author} {\bibfnamefont {J.~E.}\ \bibnamefont
  {Norton}}, \bibinfo {author} {\bibfnamefont {J.}~\bibnamefont {Cornil}},\
  and\ \bibinfo {author} {\bibfnamefont {V.}~\bibnamefont {Coropceanu}},\
  }\bibfield  {title} {\bibinfo {title} {{Molecular Understanding of Organic
  Solar Cells, The Challenges}},\ }\href {https://doi.org/10.1021/ar900099h}
  {\bibfield  {journal} {\bibinfo  {journal} {Acc. Chem. Res.}\ }\textbf
  {\bibinfo {volume} {42}},\ \bibinfo {pages} {1691} (\bibinfo {year}
  {2009})}\BibitemShut {NoStop}%
\bibitem [{\citenamefont {Park}\ \emph {et~al.}(2009)\citenamefont {Park},
  \citenamefont {Roy}, \citenamefont {Beaupr{\'{e}}}, \citenamefont {Cho},
  \citenamefont {Coates}, \citenamefont {Moon}, \citenamefont {Moses},
  \citenamefont {Leclerc}, \citenamefont {Lee},\ and\ \citenamefont
  {Heeger}}]{Park2009}%
  \BibitemOpen
  \bibfield  {author} {\bibinfo {author} {\bibfnamefont {S.~H.}\ \bibnamefont
  {Park}}, \bibinfo {author} {\bibfnamefont {A.}~\bibnamefont {Roy}}, \bibinfo
  {author} {\bibfnamefont {S.}~\bibnamefont {Beaupr{\'{e}}}}, \bibinfo {author}
  {\bibfnamefont {S.}~\bibnamefont {Cho}}, \bibinfo {author} {\bibfnamefont
  {N.}~\bibnamefont {Coates}}, \bibinfo {author} {\bibfnamefont {J.~S.}\
  \bibnamefont {Moon}}, \bibinfo {author} {\bibfnamefont {D.}~\bibnamefont
  {Moses}}, \bibinfo {author} {\bibfnamefont {M.}~\bibnamefont {Leclerc}},
  \bibinfo {author} {\bibfnamefont {K.}~\bibnamefont {Lee}},\ and\ \bibinfo
  {author} {\bibfnamefont {A.~J.}\ \bibnamefont {Heeger}},\ }\bibfield  {title}
  {\bibinfo {title} {{Bulk Heterojunction Solar Cells with Internal Quantum
  Efficiency Approaching 100{\%}}},\ }\href
  {https://doi.org/10.1038/nphoton.2009.69} {\bibfield  {journal} {\bibinfo
  {journal} {Nat. Photonics}\ }\textbf {\bibinfo {volume} {3}},\ \bibinfo
  {pages} {297} (\bibinfo {year} {2009})}\BibitemShut {NoStop}%
\bibitem [{\citenamefont {Clarke}\ and\ \citenamefont
  {Durrant}(2010)}]{Clarke2010}%
  \BibitemOpen
  \bibfield  {author} {\bibinfo {author} {\bibfnamefont {T.~M.}\ \bibnamefont
  {Clarke}}\ and\ \bibinfo {author} {\bibfnamefont {J.~R.}\ \bibnamefont
  {Durrant}},\ }\bibfield  {title} {\bibinfo {title} {{Charge Photogeneration
  in Organic Solar Cells}},\ }\href {https://doi.org/10.1021/cr900271s}
  {\bibfield  {journal} {\bibinfo  {journal} {Chem. Rev.}\ }\textbf {\bibinfo
  {volume} {110}},\ \bibinfo {pages} {6736} (\bibinfo {year}
  {2010})}\BibitemShut {NoStop}%
\bibitem [{\citenamefont {Gao}\ and\ \citenamefont
  {Ingan{\"{a}}s}(2014)}]{Gao2014}%
  \BibitemOpen
  \bibfield  {author} {\bibinfo {author} {\bibfnamefont {F.}~\bibnamefont
  {Gao}}\ and\ \bibinfo {author} {\bibfnamefont {O.}~\bibnamefont
  {Ingan{\"{a}}s}},\ }\bibfield  {title} {\bibinfo {title} {{Charge Generation
  in Polymer–Fullerene Bulk-Heterojunction Solar Cells}},\ }\href
  {https://doi.org/10.1039/C4CP01814A} {\bibfield  {journal} {\bibinfo
  {journal} {Phys. Chem. Chem. Phys.}\ }\textbf {\bibinfo {volume} {16}},\
  \bibinfo {pages} {20291} (\bibinfo {year} {2014})}\BibitemShut {NoStop}%
\bibitem [{\citenamefont {Faist}\ \emph {et~al.}(2011)\citenamefont {Faist},
  \citenamefont {Kirchartz}, \citenamefont {Gong}, \citenamefont {Ashraf},
  \citenamefont {McCulloch}, \citenamefont {de~Mello}, \citenamefont
  {Ekins-Daukes}, \citenamefont {Bradley},\ and\ \citenamefont
  {Nelson}}]{Faist2011}%
  \BibitemOpen
  \bibfield  {author} {\bibinfo {author} {\bibfnamefont {M.~A.}\ \bibnamefont
  {Faist}}, \bibinfo {author} {\bibfnamefont {T.}~\bibnamefont {Kirchartz}},
  \bibinfo {author} {\bibfnamefont {W.}~\bibnamefont {Gong}}, \bibinfo {author}
  {\bibfnamefont {R.~S.}\ \bibnamefont {Ashraf}}, \bibinfo {author}
  {\bibfnamefont {I.}~\bibnamefont {McCulloch}}, \bibinfo {author}
  {\bibfnamefont {J.~C.}\ \bibnamefont {de~Mello}}, \bibinfo {author}
  {\bibfnamefont {N.~J.}\ \bibnamefont {Ekins-Daukes}}, \bibinfo {author}
  {\bibfnamefont {D.~D.~C.}\ \bibnamefont {Bradley}},\ and\ \bibinfo {author}
  {\bibfnamefont {J.}~\bibnamefont {Nelson}},\ }\bibfield  {title} {\bibinfo
  {title} {{Competition Between the Charge Transfer State and the Singlet
  States of Donor or Acceptor Limiting the Efficiency in Polymer:Fullerene
  Solar Cells}},\ }\href {https://doi.org/10.1021/ja210029w} {\bibfield
  {journal} {\bibinfo  {journal} {J. Am. Chem. Soc.}\ }\textbf {\bibinfo
  {volume} {134}},\ \bibinfo {pages} {685} (\bibinfo {year}
  {2011})}\BibitemShut {NoStop}%
\bibitem [{\citenamefont {Nan}\ \emph {et~al.}(2016)\citenamefont {Nan},
  \citenamefont {Zhang},\ and\ \citenamefont {Lu}}]{Nan2016}%
  \BibitemOpen
  \bibfield  {author} {\bibinfo {author} {\bibfnamefont {G.}~\bibnamefont
  {Nan}}, \bibinfo {author} {\bibfnamefont {X.}~\bibnamefont {Zhang}},\ and\
  \bibinfo {author} {\bibfnamefont {G.}~\bibnamefont {Lu}},\ }\bibfield
  {title} {\bibinfo {title} {{The Lowest-Energy Charge-Transfer State and its
  Role in Charge Separation in Organic Photovoltaics}},\ }\href
  {https://doi.org/10.1039/C6CP01622G} {\bibfield  {journal} {\bibinfo
  {journal} {Phys. Chem. Chem. Phys.}\ }\textbf {\bibinfo {volume} {18}},\
  \bibinfo {pages} {17546} (\bibinfo {year} {2016})}\BibitemShut {NoStop}%
\bibitem [{\citenamefont {Azzouzi}\ \emph {et~al.}(2018)\citenamefont
  {Azzouzi}, \citenamefont {Yan}, \citenamefont {Kirchartz}, \citenamefont
  {Liu}, \citenamefont {Wang}, \citenamefont {Wu},\ and\ \citenamefont
  {Nelson}}]{Azzouzi2018}%
  \BibitemOpen
  \bibfield  {author} {\bibinfo {author} {\bibfnamefont {M.}~\bibnamefont
  {Azzouzi}}, \bibinfo {author} {\bibfnamefont {J.}~\bibnamefont {Yan}},
  \bibinfo {author} {\bibfnamefont {T.}~\bibnamefont {Kirchartz}}, \bibinfo
  {author} {\bibfnamefont {K.}~\bibnamefont {Liu}}, \bibinfo {author}
  {\bibfnamefont {J.}~\bibnamefont {Wang}}, \bibinfo {author} {\bibfnamefont
  {H.}~\bibnamefont {Wu}},\ and\ \bibinfo {author} {\bibfnamefont
  {J.}~\bibnamefont {Nelson}},\ }\bibfield  {title} {\bibinfo {title}
  {{Nonradiative Energy Losses in Bulk-Heterojunction Organic Photovoltaics}},\
  }\href {https://doi.org/10.1103/PhysRevX.8.031055} {\bibfield  {journal}
  {\bibinfo  {journal} {Phys. Rev. X}\ }\textbf {\bibinfo {volume} {8}},\
  \bibinfo {pages} {31055} (\bibinfo {year} {2018})}\BibitemShut {NoStop}%
\bibitem [{\citenamefont {Eisner}\ \emph {et~al.}(2019)\citenamefont {Eisner},
  \citenamefont {Azzouzi}, \citenamefont {Fei}, \citenamefont {Hou},
  \citenamefont {Anthopoulos}, \citenamefont {Dennis}, \citenamefont {Heeney},\
  and\ \citenamefont {Nelson}}]{Eisner2019}%
  \BibitemOpen
  \bibfield  {author} {\bibinfo {author} {\bibfnamefont {F.~D.}\ \bibnamefont
  {Eisner}}, \bibinfo {author} {\bibfnamefont {M.}~\bibnamefont {Azzouzi}},
  \bibinfo {author} {\bibfnamefont {Z.}~\bibnamefont {Fei}}, \bibinfo {author}
  {\bibfnamefont {X.}~\bibnamefont {Hou}}, \bibinfo {author} {\bibfnamefont
  {T.~D.}\ \bibnamefont {Anthopoulos}}, \bibinfo {author} {\bibfnamefont
  {T.~J.~S.}\ \bibnamefont {Dennis}}, \bibinfo {author} {\bibfnamefont
  {M.}~\bibnamefont {Heeney}},\ and\ \bibinfo {author} {\bibfnamefont
  {J.}~\bibnamefont {Nelson}},\ }\bibfield  {title} {\bibinfo {title}
  {{Hybridization of Local Exciton and Charge-Transfer States Reduces
  Nonradiative Voltage Losses in Organic Solar Cells}},\ }\href
  {https://doi.org/10.1021/jacs.9b01465} {\bibfield  {journal} {\bibinfo
  {journal} {J. Am. Chem. Soc.}\ }\textbf {\bibinfo {volume} {141}},\ \bibinfo
  {pages} {6362} (\bibinfo {year} {2019})}\BibitemShut {NoStop}%
\bibitem [{\citenamefont {Burke}\ \emph {et~al.}(2015)\citenamefont {Burke},
  \citenamefont {Sweetnam}, \citenamefont {Vandewal},\ and\ \citenamefont
  {McGehee}}]{Burke2015}%
  \BibitemOpen
  \bibfield  {author} {\bibinfo {author} {\bibfnamefont {T.~M.}\ \bibnamefont
  {Burke}}, \bibinfo {author} {\bibfnamefont {S.}~\bibnamefont {Sweetnam}},
  \bibinfo {author} {\bibfnamefont {K.}~\bibnamefont {Vandewal}},\ and\
  \bibinfo {author} {\bibfnamefont {M.~D.}\ \bibnamefont {McGehee}},\
  }\bibfield  {title} {\bibinfo {title} {{Beyond Langevin Recombination: How
  Equilibrium Between Free Carriers and Charge Transfer States Determines the
  Open-Circuit Voltage of Organic Solar Cells}},\ }\href
  {https://doi.org/10.1002/aenm.201500123} {\bibfield  {journal} {\bibinfo
  {journal} {Adv. Energy Mater.}\ }\textbf {\bibinfo {volume} {5}},\ \bibinfo
  {pages} {1500123} (\bibinfo {year} {2015})}\BibitemShut {NoStop}%
\bibitem [{\citenamefont {Menke}\ \emph
  {et~al.}(2018{\natexlab{a}})\citenamefont {Menke}, \citenamefont {Ran},
  \citenamefont {Bazan},\ and\ \citenamefont {Friend}}]{Menke2018Review}%
  \BibitemOpen
  \bibfield  {author} {\bibinfo {author} {\bibfnamefont {S.~M.}\ \bibnamefont
  {Menke}}, \bibinfo {author} {\bibfnamefont {N.~A.}\ \bibnamefont {Ran}},
  \bibinfo {author} {\bibfnamefont {G.~C.}\ \bibnamefont {Bazan}},\ and\
  \bibinfo {author} {\bibfnamefont {R.~H.}\ \bibnamefont {Friend}},\ }\bibfield
   {title} {\bibinfo {title} {{Understanding Energy Loss in Organic Solar
  Cells: Toward a New Efficiency Regime}},\ }\href
  {https://doi.org/10.1016/j.joule.2017.09.020} {\bibfield  {journal} {\bibinfo
   {journal} {Joule}\ }\textbf {\bibinfo {volume} {2}},\ \bibinfo {pages} {25}
  (\bibinfo {year} {2018}{\natexlab{a}})}\BibitemShut {NoStop}%
\bibitem [{\citenamefont {Liu}\ \emph {et~al.}(2019)\citenamefont {Liu},
  \citenamefont {Li}, \citenamefont {Ding},\ and\ \citenamefont
  {Forrest}}]{Liu2019PRApp}%
  \BibitemOpen
  \bibfield  {author} {\bibinfo {author} {\bibfnamefont {X.}~\bibnamefont
  {Liu}}, \bibinfo {author} {\bibfnamefont {Y.}~\bibnamefont {Li}}, \bibinfo
  {author} {\bibfnamefont {K.}~\bibnamefont {Ding}},\ and\ \bibinfo {author}
  {\bibfnamefont {S.}~\bibnamefont {Forrest}},\ }\bibfield  {title} {\bibinfo
  {title} {{Energy Loss in Organic Photovoltaics: Nonfullerene Versus Fullerene
  Acceptors}},\ }\href {https://doi.org/10.1103/PhysRevApplied.11.024060}
  {\bibfield  {journal} {\bibinfo  {journal} {Phys. Rev. Appl.}\ }\textbf
  {\bibinfo {volume} {11}},\ \bibinfo {pages} {024060} (\bibinfo {year}
  {2019})}\BibitemShut {NoStop}%
\bibitem [{\citenamefont {Azzouzi}\ \emph {et~al.}(2019)\citenamefont
  {Azzouzi}, \citenamefont {Kirchartz},\ and\ \citenamefont
  {Nelson}}]{Azzouzi2019}%
  \BibitemOpen
  \bibfield  {author} {\bibinfo {author} {\bibfnamefont {M.}~\bibnamefont
  {Azzouzi}}, \bibinfo {author} {\bibfnamefont {T.}~\bibnamefont {Kirchartz}},\
  and\ \bibinfo {author} {\bibfnamefont {J.}~\bibnamefont {Nelson}},\
  }\bibfield  {title} {\bibinfo {title} {{Factors Controlling Open-Circuit
  Voltage Losses in Organic Solar Cells}},\ }\href
  {https://doi.org/10.1016/j.trechm.2019.01.010} {\bibfield  {journal}
  {\bibinfo  {journal} {Trends Chem.}\ }\textbf {\bibinfo {volume} {1}},\
  \bibinfo {pages} {49} (\bibinfo {year} {2019})}\BibitemShut {NoStop}%
\bibitem [{\citenamefont {Troisi}(2011)}]{Troisi2011Chem}%
  \BibitemOpen
  \bibfield  {author} {\bibinfo {author} {\bibfnamefont {A.}~\bibnamefont
  {Troisi}},\ }\bibfield  {title} {\bibinfo {title} {{Charge Transport in High
  Mobility Molecular Semiconductors: Classical Models and New Theories}},\
  }\href {https://doi.org/10.1039/c0cs00198h} {\bibfield  {journal} {\bibinfo
  {journal} {Chem. Soc. Rev.}\ }\textbf {\bibinfo {volume} {40}},\ \bibinfo
  {pages} {2347} (\bibinfo {year} {2011})}\BibitemShut {NoStop}%
\bibitem [{\citenamefont {Scharber}\ \emph {et~al.}(2006)\citenamefont
  {Scharber}, \citenamefont {M{\"{u}}hlbacher}, \citenamefont {Koppe},
  \citenamefont {Denk}, \citenamefont {Waldauf}, \citenamefont {Heeger},\ and\
  \citenamefont {Brabec}}]{Scharber2006}%
  \BibitemOpen
  \bibfield  {author} {\bibinfo {author} {\bibfnamefont {M.~C.}\ \bibnamefont
  {Scharber}}, \bibinfo {author} {\bibfnamefont {D.}~\bibnamefont
  {M{\"{u}}hlbacher}}, \bibinfo {author} {\bibfnamefont {M.}~\bibnamefont
  {Koppe}}, \bibinfo {author} {\bibfnamefont {P.}~\bibnamefont {Denk}},
  \bibinfo {author} {\bibfnamefont {C.}~\bibnamefont {Waldauf}}, \bibinfo
  {author} {\bibfnamefont {A.~J.}\ \bibnamefont {Heeger}},\ and\ \bibinfo
  {author} {\bibfnamefont {C.~J.}\ \bibnamefont {Brabec}},\ }\bibfield  {title}
  {\bibinfo {title} {{Design Rules for Donors in Bulk-Heterojunction Solar
  Cells - Towards 10 {\%} Energy-Conversion Efficiency}},\ }\href
  {https://doi.org/10.1002/adma.200501717} {\bibfield  {journal} {\bibinfo
  {journal} {Adv. Mater.}\ }\textbf {\bibinfo {volume} {18}},\ \bibinfo {pages}
  {789} (\bibinfo {year} {2006})}\BibitemShut {NoStop}%
\bibitem [{\citenamefont {Veldman}\ \emph {et~al.}(2009)\citenamefont
  {Veldman}, \citenamefont {Meskers},\ and\ \citenamefont
  {Janssen}}]{Veldman2009}%
  \BibitemOpen
  \bibfield  {author} {\bibinfo {author} {\bibfnamefont {D.}~\bibnamefont
  {Veldman}}, \bibinfo {author} {\bibfnamefont {S.~C.~J.}\ \bibnamefont
  {Meskers}},\ and\ \bibinfo {author} {\bibfnamefont {R.~A.}\ \bibnamefont
  {Janssen}},\ }\bibfield  {title} {\bibinfo {title} {{The Energy of
  Charge‐Transfer States in Electron Donor–Acceptor Blends: Insight into
  the Energy Losses in Organic Solar Cells}},\ }\href
  {https://doi.org/10.1002/adfm.200900090} {\bibfield  {journal} {\bibinfo
  {journal} {Adv. Funct. Mater.}\ }\textbf {\bibinfo {volume} {19}},\ \bibinfo
  {pages} {1939} (\bibinfo {year} {2009})}\BibitemShut {NoStop}%
\bibitem [{\citenamefont {Hallermann}\ \emph {et~al.}(2009)\citenamefont
  {Hallermann}, \citenamefont {Kriegel}, \citenamefont {{Da Como}},
  \citenamefont {Berger}, \citenamefont {von Hauff},\ and\ \citenamefont
  {Feldmann}}]{Hallerman2009}%
  \BibitemOpen
  \bibfield  {author} {\bibinfo {author} {\bibfnamefont {M.}~\bibnamefont
  {Hallermann}}, \bibinfo {author} {\bibfnamefont {I.}~\bibnamefont {Kriegel}},
  \bibinfo {author} {\bibfnamefont {E.}~\bibnamefont {{Da Como}}}, \bibinfo
  {author} {\bibfnamefont {J.~M.}\ \bibnamefont {Berger}}, \bibinfo {author}
  {\bibfnamefont {E.}~\bibnamefont {von Hauff}},\ and\ \bibinfo {author}
  {\bibfnamefont {J.}~\bibnamefont {Feldmann}},\ }\bibfield  {title} {\bibinfo
  {title} {{Charge Transfer Excitons in Polymer/Fullerene Blends: The Role of
  Morphology and Polymer Chain Conformation}},\ }\href
  {https://doi.org/10.1002/adfm.200901398} {\bibfield  {journal} {\bibinfo
  {journal} {Adv. Funct. Mater.}\ }\textbf {\bibinfo {volume} {19}},\ \bibinfo
  {pages} {3662} (\bibinfo {year} {2009})}\BibitemShut {NoStop}%
\bibitem [{\citenamefont {Deibel}\ and\ \citenamefont
  {Dyakonov}(2010)}]{Deibel2010}%
  \BibitemOpen
  \bibfield  {author} {\bibinfo {author} {\bibfnamefont {C.}~\bibnamefont
  {Deibel}}\ and\ \bibinfo {author} {\bibfnamefont {V.}~\bibnamefont
  {Dyakonov}},\ }\bibfield  {title} {\bibinfo {title} {{Polymer–fullerene
  Bulk Heterojunction Solar Cells}},\ }\href
  {https://doi.org/10.1088/0034-4885/73/9/096401} {\bibfield  {journal}
  {\bibinfo  {journal} {Rep. Prog. Phys.}\ }\textbf {\bibinfo {volume} {73}},\
  \bibinfo {pages} {096401} (\bibinfo {year} {2010})}\BibitemShut {NoStop}%
\bibitem [{\citenamefont {Coffey}\ \emph {et~al.}(2012)\citenamefont {Coffey},
  \citenamefont {Larson}, \citenamefont {Hains}, \citenamefont {Whitaker},
  \citenamefont {Kopidakis}, \citenamefont {Boltalina}, \citenamefont
  {Strauss},\ and\ \citenamefont {Rumbles}}]{Coffey2012}%
  \BibitemOpen
  \bibfield  {author} {\bibinfo {author} {\bibfnamefont {D.~C.}\ \bibnamefont
  {Coffey}}, \bibinfo {author} {\bibfnamefont {B.~W.}\ \bibnamefont {Larson}},
  \bibinfo {author} {\bibfnamefont {A.~W.}\ \bibnamefont {Hains}}, \bibinfo
  {author} {\bibfnamefont {J.~B.}\ \bibnamefont {Whitaker}}, \bibinfo {author}
  {\bibfnamefont {N.}~\bibnamefont {Kopidakis}}, \bibinfo {author}
  {\bibfnamefont {O.~V.}\ \bibnamefont {Boltalina}}, \bibinfo {author}
  {\bibfnamefont {S.~H.}\ \bibnamefont {Strauss}},\ and\ \bibinfo {author}
  {\bibfnamefont {G.}~\bibnamefont {Rumbles}},\ }\bibfield  {title} {\bibinfo
  {title} {{An Optimal Driving Force for Converting Excitons into Free Carriers
  in Excitonic Solar Cells}},\ }\href {https://doi.org/10.1021/jp302275z}
  {\bibfield  {journal} {\bibinfo  {journal} {J. Phys. Chem. C}\ }\textbf
  {\bibinfo {volume} {116}},\ \bibinfo {pages} {8916} (\bibinfo {year}
  {2012})}\BibitemShut {NoStop}%
\bibitem [{\citenamefont {Wang}\ \emph {et~al.}(2018)\citenamefont {Wang},
  \citenamefont {Qian}, \citenamefont {Cui}, \citenamefont {Zhang},
  \citenamefont {Hou}, \citenamefont {Vandewal}, \citenamefont {Kirchartz},\
  and\ \citenamefont {Gao}}]{Wang2018}%
  \BibitemOpen
  \bibfield  {author} {\bibinfo {author} {\bibfnamefont {Y.}~\bibnamefont
  {Wang}}, \bibinfo {author} {\bibfnamefont {D.}~\bibnamefont {Qian}}, \bibinfo
  {author} {\bibfnamefont {Y.}~\bibnamefont {Cui}}, \bibinfo {author}
  {\bibfnamefont {H.}~\bibnamefont {Zhang}}, \bibinfo {author} {\bibfnamefont
  {J.}~\bibnamefont {Hou}}, \bibinfo {author} {\bibfnamefont {K.}~\bibnamefont
  {Vandewal}}, \bibinfo {author} {\bibfnamefont {T.}~\bibnamefont
  {Kirchartz}},\ and\ \bibinfo {author} {\bibfnamefont {F.}~\bibnamefont
  {Gao}},\ }\bibfield  {title} {\bibinfo {title} {{Optical Gaps of Organic
  Solar Cells as a Reference for Comparing Voltage Losses}},\ }\href
  {https://doi.org/10.1002/aenm.201801352} {\bibfield  {journal} {\bibinfo
  {journal} {Adv. Energy Mater.}\ }\textbf {\bibinfo {volume} {8}},\ \bibinfo
  {pages} {1801352} (\bibinfo {year} {2018})}\BibitemShut {NoStop}%
\bibitem [{\citenamefont {Qian}\ \emph {et~al.}(2018)\citenamefont {Qian},
  \citenamefont {Zheng}, \citenamefont {Yao}, \citenamefont {Tress},
  \citenamefont {Hopper}, \citenamefont {Chen}, \citenamefont {Li},
  \citenamefont {Liu}, \citenamefont {Chen}, \citenamefont {Zhang},
  \citenamefont {Liu}, \citenamefont {Gao}, \citenamefont {Ouyang},
  \citenamefont {Jin}, \citenamefont {Pozina}, \citenamefont {Buyanova},
  \citenamefont {Chen}, \citenamefont {Ingan{\"{a}}s}, \citenamefont
  {Coropceanu}, \citenamefont {Br{\'{e}}das}, \citenamefont {Yan}, \citenamefont
  {Hou}, \citenamefont {Zhang}, \citenamefont {Bakulin},\ and\ \citenamefont
  {Gao}}]{Qian2018}%
  \BibitemOpen
  \bibfield  {author} {\bibinfo {author} {\bibfnamefont {D.}~\bibnamefont
  {Qian}}, \bibinfo {author} {\bibfnamefont {Z.}~\bibnamefont {Zheng}},
  \bibinfo {author} {\bibfnamefont {H.}~\bibnamefont {Yao}}, \bibinfo {author}
  {\bibfnamefont {W.}~\bibnamefont {Tress}}, \bibinfo {author} {\bibfnamefont
  {T.~R.}\ \bibnamefont {Hopper}}, \bibinfo {author} {\bibfnamefont
  {S.}~\bibnamefont {Chen}}, \bibinfo {author} {\bibfnamefont {S.}~\bibnamefont
  {Li}}, \bibinfo {author} {\bibfnamefont {J.}~\bibnamefont {Liu}}, \bibinfo
  {author} {\bibfnamefont {S.}~\bibnamefont {Chen}}, \bibinfo {author}
  {\bibfnamefont {J.}~\bibnamefont {Zhang}}, \bibinfo {author} {\bibfnamefont
  {X.-K.}\ \bibnamefont {Liu}}, \bibinfo {author} {\bibfnamefont
  {B.}~\bibnamefont {Gao}}, \bibinfo {author} {\bibfnamefont {L.}~\bibnamefont
  {Ouyang}}, \bibinfo {author} {\bibfnamefont {Y.}~\bibnamefont {Jin}},
  \bibinfo {author} {\bibfnamefont {G.}~\bibnamefont {Pozina}}, \bibinfo
  {author} {\bibfnamefont {I.~A.}\ \bibnamefont {Buyanova}}, \bibinfo {author}
  {\bibfnamefont {W.~M.}\ \bibnamefont {Chen}}, \bibinfo {author}
  {\bibfnamefont {O.}~\bibnamefont {Ingan{\"{a}}s}}, \bibinfo {author}
  {\bibfnamefont {V.}~\bibnamefont {Coropceanu}}, \bibinfo {author}
  {\bibfnamefont {J.-L.}\ \bibnamefont {Br{\'{e}}das}}, \bibinfo {author}
  {\bibfnamefont {H.}~\bibnamefont {Yan}}, \bibinfo {author} {\bibfnamefont
  {J.}~\bibnamefont {Hou}}, \bibinfo {author} {\bibfnamefont {F.}~\bibnamefont
  {Zhang}}, \bibinfo {author} {\bibfnamefont {A.~A.}\ \bibnamefont {Bakulin}},\
  and\ \bibinfo {author} {\bibfnamefont {F.}~\bibnamefont {Gao}},\ }\bibfield
  {title} {\bibinfo {title} {{Design Rules for Minimizing Voltage Losses in
  High-Efficiency Organic Solar Cells}},\ }\href
  {https://doi.org/10.1038/s41563-018-0128-z} {\bibfield  {journal} {\bibinfo
  {journal} {Nat. Mater.}\ }\textbf {\bibinfo {volume} {17}},\ \bibinfo {pages}
  {703} (\bibinfo {year} {2018})}\BibitemShut {NoStop}%
\bibitem [{\citenamefont {Yang}\ \emph {et~al.}(2018)\citenamefont {Yang},
  \citenamefont {Yao}, \citenamefont {Guo}, \citenamefont {Sun},\ and\
  \citenamefont {Luo}}]{Yang2018}%
  \BibitemOpen
  \bibfield  {author} {\bibinfo {author} {\bibfnamefont {W.}~\bibnamefont
  {Yang}}, \bibinfo {author} {\bibfnamefont {Y.}~\bibnamefont {Yao}}, \bibinfo
  {author} {\bibfnamefont {P.}~\bibnamefont {Guo}}, \bibinfo {author}
  {\bibfnamefont {H.}~\bibnamefont {Sun}},\ and\ \bibinfo {author}
  {\bibfnamefont {Y.}~\bibnamefont {Luo}},\ }\bibfield  {title} {\bibinfo
  {title} {{Optimum Driving Energy for Achieving Balanced Open-Circuit Voltage
  and Short-Circuit Current Density in Organic Bulk Heterojunction Solar
  Cells}},\ }\href {https://doi.org/10.1039/C8CP05145C} {\bibfield  {journal}
  {\bibinfo  {journal} {Phys. Chem. Chem. Phys.}\ }\textbf {\bibinfo {volume}
  {20}},\ \bibinfo {pages} {29866} (\bibinfo {year} {2018})}\BibitemShut
  {NoStop}%
\bibitem [{\citenamefont {Zhang}\ \emph {et~al.}(2015)\citenamefont {Zhang},
  \citenamefont {Kan}, \citenamefont {Liu}, \citenamefont {Long}, \citenamefont
  {Wan}, \citenamefont {Chen}, \citenamefont {Zuo}, \citenamefont {Ni},
  \citenamefont {Zhang}, \citenamefont {Li}, \citenamefont {Hu}, \citenamefont
  {Huang}, \citenamefont {Cao}, \citenamefont {Liang}, \citenamefont {Zhang},
  \citenamefont {Russell},\ and\ \citenamefont {Chen}}]{Zhang2014}%
  \BibitemOpen
  \bibfield  {author} {\bibinfo {author} {\bibfnamefont {Q.}~\bibnamefont
  {Zhang}}, \bibinfo {author} {\bibfnamefont {B.}~\bibnamefont {Kan}}, \bibinfo
  {author} {\bibfnamefont {F.}~\bibnamefont {Liu}}, \bibinfo {author}
  {\bibfnamefont {G.}~\bibnamefont {Long}}, \bibinfo {author} {\bibfnamefont
  {X.}~\bibnamefont {Wan}}, \bibinfo {author} {\bibfnamefont {X.}~\bibnamefont
  {Chen}}, \bibinfo {author} {\bibfnamefont {Y.}~\bibnamefont {Zuo}}, \bibinfo
  {author} {\bibfnamefont {W.}~\bibnamefont {Ni}}, \bibinfo {author}
  {\bibfnamefont {H.}~\bibnamefont {Zhang}}, \bibinfo {author} {\bibfnamefont
  {M.}~\bibnamefont {Li}}, \bibinfo {author} {\bibfnamefont {Z.}~\bibnamefont
  {Hu}}, \bibinfo {author} {\bibfnamefont {F.}~\bibnamefont {Huang}}, \bibinfo
  {author} {\bibfnamefont {Y.}~\bibnamefont {Cao}}, \bibinfo {author}
  {\bibfnamefont {Z.}~\bibnamefont {Liang}}, \bibinfo {author} {\bibfnamefont
  {M.}~\bibnamefont {Zhang}}, \bibinfo {author} {\bibfnamefont {T.~P.}\
  \bibnamefont {Russell}},\ and\ \bibinfo {author} {\bibfnamefont
  {Y.}~\bibnamefont {Chen}},\ }\bibfield  {title} {\bibinfo {title}
  {{Small-Molecule Solar Cells with Efficiency over 9{\%}}},\ }\href
  {https://doi.org/10.1038/nphoton.2014.269} {\bibfield  {journal} {\bibinfo
  {journal} {Nat. Photonics}\ }\textbf {\bibinfo {volume} {9}},\ \bibinfo
  {pages} {35} (\bibinfo {year} {2015})}\BibitemShut {NoStop}%
\bibitem [{\citenamefont {Jing}\ \emph {et~al.}(2016)\citenamefont {Jing},
  \citenamefont {Chen}, \citenamefont {Qian}, \citenamefont {Gautam},
  \citenamefont {Guofang}, \citenamefont {Zhao}, \citenamefont {Bergqvist},
  \citenamefont {Zhang}, \citenamefont {Ma}, \citenamefont {Ade}, \citenamefont
  {Inganas}, \citenamefont {Gundogdu}, \citenamefont {Gao},\ and\ \citenamefont
  {Yan}}]{Liu2016NatEner}%
  \BibitemOpen
  \bibfield  {author} {\bibinfo {author} {\bibfnamefont {L.}~\bibnamefont
  {Jing}}, \bibinfo {author} {\bibfnamefont {S.}~\bibnamefont {Chen}}, \bibinfo
  {author} {\bibfnamefont {D.}~\bibnamefont {Qian}}, \bibinfo {author}
  {\bibfnamefont {B.}~\bibnamefont {Gautam}}, \bibinfo {author} {\bibfnamefont
  {Y.}~\bibnamefont {Guofang}}, \bibinfo {author} {\bibfnamefont
  {J.}~\bibnamefont {Zhao}}, \bibinfo {author} {\bibfnamefont {J.}~\bibnamefont
  {Bergqvist}}, \bibinfo {author} {\bibfnamefont {F.}~\bibnamefont {Zhang}},
  \bibinfo {author} {\bibfnamefont {W.}~\bibnamefont {Ma}}, \bibinfo {author}
  {\bibfnamefont {H.}~\bibnamefont {Ade}}, \bibinfo {author} {\bibfnamefont
  {O.}~\bibnamefont {Inganas}}, \bibinfo {author} {\bibfnamefont
  {K.}~\bibnamefont {Gundogdu}}, \bibinfo {author} {\bibfnamefont
  {F.}~\bibnamefont {Gao}},\ and\ \bibinfo {author} {\bibfnamefont
  {H.}~\bibnamefont {Yan}},\ }\bibfield  {title} {\bibinfo {title} {{Fast
  Charge Separation in a Non-Fullerene Organic Solar Cell with a Small Driving
  Force}},\ }\href {https://doi.org/10.1038/nenergy.2016.89} {\bibfield
  {journal} {\bibinfo  {journal} {Nat. Energy}\ }\textbf {\bibinfo {volume}
  {1}},\ \bibinfo {pages} {16089} (\bibinfo {year} {2016})}\BibitemShut
  {NoStop}%
\bibitem [{\citenamefont {Meng}\ \emph {et~al.}(2018)\citenamefont {Meng},
  \citenamefont {Zhang}, \citenamefont {Wan}, \citenamefont {Li}, \citenamefont
  {Zhang}, \citenamefont {Wang}, \citenamefont {Ke}, \citenamefont {Xiao},
  \citenamefont {Ding}, \citenamefont {Xia}, \citenamefont {Yip}, \citenamefont
  {Cao},\ and\ \citenamefont {Chen}}]{Meng2018}%
  \BibitemOpen
  \bibfield  {author} {\bibinfo {author} {\bibfnamefont {L.}~\bibnamefont
  {Meng}}, \bibinfo {author} {\bibfnamefont {Y.}~\bibnamefont {Zhang}},
  \bibinfo {author} {\bibfnamefont {X.}~\bibnamefont {Wan}}, \bibinfo {author}
  {\bibfnamefont {C.}~\bibnamefont {Li}}, \bibinfo {author} {\bibfnamefont
  {X.}~\bibnamefont {Zhang}}, \bibinfo {author} {\bibfnamefont
  {Y.}~\bibnamefont {Wang}}, \bibinfo {author} {\bibfnamefont {X.}~\bibnamefont
  {Ke}}, \bibinfo {author} {\bibfnamefont {Z.}~\bibnamefont {Xiao}}, \bibinfo
  {author} {\bibfnamefont {L.}~\bibnamefont {Ding}}, \bibinfo {author}
  {\bibfnamefont {R.}~\bibnamefont {Xia}}, \bibinfo {author} {\bibfnamefont
  {H.-L.}\ \bibnamefont {Yip}}, \bibinfo {author} {\bibfnamefont
  {Y.}~\bibnamefont {Cao}},\ and\ \bibinfo {author} {\bibfnamefont
  {Y.}~\bibnamefont {Chen}},\ }\bibfield  {title} {\bibinfo {title} {{Organic
  and Solution-Processed Tandem Solar Cells with 17.3{\%} efficiency}},\ }\href
  {https://doi.org/10.1126/science.aat2612} {\bibfield  {journal} {\bibinfo
  {journal} {Science}\ }\textbf {\bibinfo {volume} {361}},\ \bibinfo {pages}
  {1094} (\bibinfo {year} {2018})}\BibitemShut {NoStop}%
\bibitem [{\citenamefont {Cui}\ \emph {et~al.}(2019)\citenamefont {Cui},
  \citenamefont {Yao}, \citenamefont {Zhang}, \citenamefont {Zhang},
  \citenamefont {Wang}, \citenamefont {Hong}, \citenamefont {Xian},
  \citenamefont {Xu}, \citenamefont {Zhang}, \citenamefont {Peng},
  \citenamefont {Wei}, \citenamefont {Gao},\ and\ \citenamefont
  {Hou}}]{Cui2019}%
  \BibitemOpen
  \bibfield  {author} {\bibinfo {author} {\bibfnamefont {Y.}~\bibnamefont
  {Cui}}, \bibinfo {author} {\bibfnamefont {H.}~\bibnamefont {Yao}}, \bibinfo
  {author} {\bibfnamefont {J.}~\bibnamefont {Zhang}}, \bibinfo {author}
  {\bibfnamefont {T.}~\bibnamefont {Zhang}}, \bibinfo {author} {\bibfnamefont
  {Y.}~\bibnamefont {Wang}}, \bibinfo {author} {\bibfnamefont {L.}~\bibnamefont
  {Hong}}, \bibinfo {author} {\bibfnamefont {K.}~\bibnamefont {Xian}}, \bibinfo
  {author} {\bibfnamefont {B.}~\bibnamefont {Xu}}, \bibinfo {author}
  {\bibfnamefont {S.}~\bibnamefont {Zhang}}, \bibinfo {author} {\bibfnamefont
  {J.}~\bibnamefont {Peng}}, \bibinfo {author} {\bibfnamefont {Z.}~\bibnamefont
  {Wei}}, \bibinfo {author} {\bibfnamefont {F.}~\bibnamefont {Gao}},\ and\
  \bibinfo {author} {\bibfnamefont {J.}~\bibnamefont {Hou}},\ }\bibfield
  {title} {\bibinfo {title} {{Over 16{\%} Efficiency Organic Photovoltaic Cells
  Enabled by a Chlorinated Acceptor with Increased Open-Circuit Voltages}},\
  }\href {https://doi.org/10.1038/s41467-019-10351-5} {\bibfield  {journal}
  {\bibinfo  {journal} {Nat. Commun.}\ }\textbf {\bibinfo {volume} {10}},\
  \bibinfo {pages} {2515} (\bibinfo {year} {2019})}\BibitemShut {NoStop}%
\bibitem [{\citenamefont {Liu}\ \emph {et~al.}(2020)\citenamefont {Liu},
  \citenamefont {Yuan}, \citenamefont {Deng}, \citenamefont {Luo},
  \citenamefont {Xie}, \citenamefont {Liang}, \citenamefont {Zou},
  \citenamefont {He}, \citenamefont {Wu},\ and\ \citenamefont {Cao}}]{Liu2020}%
  \BibitemOpen
  \bibfield  {author} {\bibinfo {author} {\bibfnamefont {S.}~\bibnamefont
  {Liu}}, \bibinfo {author} {\bibfnamefont {J.}~\bibnamefont {Yuan}}, \bibinfo
  {author} {\bibfnamefont {W.}~\bibnamefont {Deng}}, \bibinfo {author}
  {\bibfnamefont {M.}~\bibnamefont {Luo}}, \bibinfo {author} {\bibfnamefont
  {Y.}~\bibnamefont {Xie}}, \bibinfo {author} {\bibfnamefont {Q.}~\bibnamefont
  {Liang}}, \bibinfo {author} {\bibfnamefont {Y.}~\bibnamefont {Zou}}, \bibinfo
  {author} {\bibfnamefont {Z.}~\bibnamefont {He}}, \bibinfo {author}
  {\bibfnamefont {H.}~\bibnamefont {Wu}},\ and\ \bibinfo {author}
  {\bibfnamefont {Y.}~\bibnamefont {Cao}},\ }\bibfield  {title} {\bibinfo
  {title} {{High-Efficiency Organic Solar Cells with Low Non-Radiative
  Recombination Loss and Low Energetic Disorder}},\ }\href
  {https://doi.org/10.1038/s41566-019-0573-5} {\bibfield  {journal} {\bibinfo
  {journal} {Nat. Photonics}\ }\textbf {\bibinfo {volume} {14}},\ \bibinfo
  {pages} {300} (\bibinfo {year} {2020})}\BibitemShut {NoStop}%
\bibitem [{\citenamefont {Zhang}\ \emph {et~al.}(2020)\citenamefont {Zhang},
  \citenamefont {Chen}, \citenamefont {Xiao}, \citenamefont {Chow},
  \citenamefont {Ren}, \citenamefont {Kupgan}, \citenamefont {Jiao},
  \citenamefont {Chan}, \citenamefont {Du}, \citenamefont {Xia}, \citenamefont
  {Chen}, \citenamefont {Yuan}, \citenamefont {Zhang}, \citenamefont {Zhang},
  \citenamefont {Liu}, \citenamefont {Zou}, \citenamefont {Yan}, \citenamefont
  {Wong}, \citenamefont {Coropceanu}, \citenamefont {Li}, \citenamefont
  {Brabec}, \citenamefont {Br{\'{e}}das}, \citenamefont {Yip},\ and\ \citenamefont
  {Cao}}]{Zhang2020}%
  \BibitemOpen
  \bibfield  {author} {\bibinfo {author} {\bibfnamefont {G.}~\bibnamefont
  {Zhang}}, \bibinfo {author} {\bibfnamefont {X.-K.}\ \bibnamefont {Chen}},
  \bibinfo {author} {\bibfnamefont {J.}~\bibnamefont {Xiao}}, \bibinfo {author}
  {\bibfnamefont {P.~C.~Y.}\ \bibnamefont {Chow}}, \bibinfo {author}
  {\bibfnamefont {M.}~\bibnamefont {Ren}}, \bibinfo {author} {\bibfnamefont
  {G.}~\bibnamefont {Kupgan}}, \bibinfo {author} {\bibfnamefont
  {X.}~\bibnamefont {Jiao}}, \bibinfo {author} {\bibfnamefont {C.~C.~S.}\
  \bibnamefont {Chan}}, \bibinfo {author} {\bibfnamefont {X.}~\bibnamefont
  {Du}}, \bibinfo {author} {\bibfnamefont {R.}~\bibnamefont {Xia}}, \bibinfo
  {author} {\bibfnamefont {Z.}~\bibnamefont {Chen}}, \bibinfo {author}
  {\bibfnamefont {J.}~\bibnamefont {Yuan}}, \bibinfo {author} {\bibfnamefont
  {Y.}~\bibnamefont {Zhang}}, \bibinfo {author} {\bibfnamefont
  {S.}~\bibnamefont {Zhang}}, \bibinfo {author} {\bibfnamefont
  {Y.}~\bibnamefont {Liu}}, \bibinfo {author} {\bibfnamefont {Y.}~\bibnamefont
  {Zou}}, \bibinfo {author} {\bibfnamefont {H.}~\bibnamefont {Yan}}, \bibinfo
  {author} {\bibfnamefont {K.~S.}\ \bibnamefont {Wong}}, \bibinfo {author}
  {\bibfnamefont {V.}~\bibnamefont {Coropceanu}}, \bibinfo {author}
  {\bibfnamefont {N.}~\bibnamefont {Li}}, \bibinfo {author} {\bibfnamefont
  {C.~J.}\ \bibnamefont {Brabec}}, \bibinfo {author} {\bibfnamefont {J.-L.}\
  \bibnamefont {Br{\'{e}}das}}, \bibinfo {author} {\bibfnamefont {H.-L.}\
  \bibnamefont {Yip}},\ and\ \bibinfo {author} {\bibfnamefont {Y.}~\bibnamefont
  {Cao}},\ }\bibfield  {title} {\bibinfo {title} {{Delocalization of Exciton
  and Electron Wavefunction in Non-Fullerene Acceptor Molecules Enables
  Efficient Organic Solar Cells}},\ }\href
  {https://doi.org/10.1038/s41467-020-17867-1} {\bibfield  {journal} {\bibinfo
  {journal} {Nat. Commun.}\ }\textbf {\bibinfo {volume} {11}},\ \bibinfo
  {pages} {3943} (\bibinfo {year} {2020})}\BibitemShut {NoStop}%
\bibitem [{\citenamefont {Zhong}\ \emph {et~al.}(2020)\citenamefont {Zhong},
  \citenamefont {Causa'}, \citenamefont {Moore}, \citenamefont {Krauspe},
  \citenamefont {Xiao}, \citenamefont {G{\"{u}}nther}, \citenamefont
  {Kublitski}, \citenamefont {Shivhare}, \citenamefont {Benduhn}, \citenamefont
  {BarOr}, \citenamefont {Mukherjee}, \citenamefont {Yallum}, \citenamefont
  {R{\'{e}}hault}, \citenamefont {Mannsfeld}, \citenamefont {Neher},
  \citenamefont {Richter}, \citenamefont {DeLongchamp}, \citenamefont
  {Ortmann}, \citenamefont {Vandewal}, \citenamefont {Zhou},\ and\
  \citenamefont {Banerji}}]{Zhong2020}%
  \BibitemOpen
  \bibfield  {author} {\bibinfo {author} {\bibfnamefont {Y.}~\bibnamefont
  {Zhong}}, \bibinfo {author} {\bibfnamefont {M.}~\bibnamefont {Causa'}},
  \bibinfo {author} {\bibfnamefont {G.~J.}\ \bibnamefont {Moore}}, \bibinfo
  {author} {\bibfnamefont {P.}~\bibnamefont {Krauspe}}, \bibinfo {author}
  {\bibfnamefont {B.}~\bibnamefont {Xiao}}, \bibinfo {author} {\bibfnamefont
  {F.}~\bibnamefont {G{\"{u}}nther}}, \bibinfo {author} {\bibfnamefont
  {J.}~\bibnamefont {Kublitski}}, \bibinfo {author} {\bibfnamefont
  {R.}~\bibnamefont {Shivhare}}, \bibinfo {author} {\bibfnamefont
  {J.}~\bibnamefont {Benduhn}}, \bibinfo {author} {\bibfnamefont
  {E.}~\bibnamefont {BarOr}}, \bibinfo {author} {\bibfnamefont
  {S.}~\bibnamefont {Mukherjee}}, \bibinfo {author} {\bibfnamefont {K.~M.}\
  \bibnamefont {Yallum}}, \bibinfo {author} {\bibfnamefont {J.}~\bibnamefont
  {R{\'{e}}hault}}, \bibinfo {author} {\bibfnamefont {S.~C.~B.}\ \bibnamefont
  {Mannsfeld}}, \bibinfo {author} {\bibfnamefont {D.}~\bibnamefont {Neher}},
  \bibinfo {author} {\bibfnamefont {L.~J.}\ \bibnamefont {Richter}}, \bibinfo
  {author} {\bibfnamefont {D.~M.}\ \bibnamefont {DeLongchamp}}, \bibinfo
  {author} {\bibfnamefont {F.}~\bibnamefont {Ortmann}}, \bibinfo {author}
  {\bibfnamefont {K.}~\bibnamefont {Vandewal}}, \bibinfo {author}
  {\bibfnamefont {E.}~\bibnamefont {Zhou}},\ and\ \bibinfo {author}
  {\bibfnamefont {N.}~\bibnamefont {Banerji}},\ }\bibfield  {title} {\bibinfo
  {title} {{Sub-Picosecond Charge-Transfer at Near-Zero Driving Force in
  Polymer:Non-Fullerene Acceptor Blends and Bilayers}},\ }\href
  {https://doi.org/10.1038/s41467-020-14549-w} {\bibfield  {journal} {\bibinfo
  {journal} {Nat. Commun.}\ }\textbf {\bibinfo {volume} {11}},\ \bibinfo
  {pages} {833} (\bibinfo {year} {2020})}\BibitemShut {NoStop}%
\bibitem [{\citenamefont {Chen}\ \emph
  {et~al.}(2018{\natexlab{a}})\citenamefont {Chen}, \citenamefont
  {Coropceanu},\ and\ \citenamefont {Br{\'{e}}das}}]{Chen2018NatComms}%
  \BibitemOpen
  \bibfield  {author} {\bibinfo {author} {\bibfnamefont {X.-K.}\ \bibnamefont
  {Chen}}, \bibinfo {author} {\bibfnamefont {V.}~\bibnamefont {Coropceanu}},\
  and\ \bibinfo {author} {\bibfnamefont {J.-L.}\ \bibnamefont {Br{\'{e}}das}},\
  }\bibfield  {title} {\bibinfo {title} {{Assessing the Nature of the
  Charge-Transfer Electronic States in Organic Solar Cells}},\ }\href
  {https://doi.org/10.1038/s41467-018-07707-8} {\bibfield  {journal} {\bibinfo
  {journal} {Nat. Commun.}\ }\textbf {\bibinfo {volume} {9}},\ \bibinfo {pages}
  {5295} (\bibinfo {year} {2018}{\natexlab{a}})}\BibitemShut {NoStop}%
\bibitem [{\citenamefont {Firdaus}\ \emph {et~al.}(2019)\citenamefont
  {Firdaus}, \citenamefont {{Le Corre}}, \citenamefont {Khan}, \citenamefont
  {Kan}, \citenamefont {Laquai}, \citenamefont {Beaujuge},\ and\ \citenamefont
  {Anthopoulos}}]{Firdaus2019}%
  \BibitemOpen
  \bibfield  {author} {\bibinfo {author} {\bibfnamefont {Y.}~\bibnamefont
  {Firdaus}}, \bibinfo {author} {\bibfnamefont {V.~M.}\ \bibnamefont {{Le
  Corre}}}, \bibinfo {author} {\bibfnamefont {J.~I.}\ \bibnamefont {Khan}},
  \bibinfo {author} {\bibfnamefont {Z.}~\bibnamefont {Kan}}, \bibinfo {author}
  {\bibfnamefont {F.}~\bibnamefont {Laquai}}, \bibinfo {author} {\bibfnamefont
  {P.~M.}\ \bibnamefont {Beaujuge}},\ and\ \bibinfo {author} {\bibfnamefont
  {T.~D.}\ \bibnamefont {Anthopoulos}},\ }\bibfield  {title} {\bibinfo {title}
  {{Key Parameters Requirements for Non‐Fullerene‐Based Organic Solar Cells
  with Power Conversion Efficiency {\textgreater}20{\%}}},\ }\href
  {https://doi.org/10.1002/advs.201802028} {\bibfield  {journal} {\bibinfo
  {journal} {Adv. Sci.}\ }\textbf {\bibinfo {volume} {6}},\ \bibinfo {pages}
  {1802028} (\bibinfo {year} {2019})}\BibitemShut {NoStop}%
\bibitem [{\citenamefont {Karki}\ \emph {et~al.}(2020)\citenamefont {Karki},
  \citenamefont {Vollbrecht}, \citenamefont {Gillett}, \citenamefont {Selter},
  \citenamefont {Lee}, \citenamefont {Peng}, \citenamefont {Schopp},
  \citenamefont {Dixon}, \citenamefont {Schrock}, \citenamefont
  {N{\'{a}}da{\v{z}}dy}, \citenamefont {Schauer}, \citenamefont {Ade},
  \citenamefont {Chmelka}, \citenamefont {Bazan}, \citenamefont {Friend},\ and\
  \citenamefont {Nguyen}}]{Karki2020}%
  \BibitemOpen
  \bibfield  {author} {\bibinfo {author} {\bibfnamefont {A.}~\bibnamefont
  {Karki}}, \bibinfo {author} {\bibfnamefont {J.}~\bibnamefont {Vollbrecht}},
  \bibinfo {author} {\bibfnamefont {A.~J.}\ \bibnamefont {Gillett}}, \bibinfo
  {author} {\bibfnamefont {P.}~\bibnamefont {Selter}}, \bibinfo {author}
  {\bibfnamefont {J.}~\bibnamefont {Lee}}, \bibinfo {author} {\bibfnamefont
  {Z.}~\bibnamefont {Peng}}, \bibinfo {author} {\bibfnamefont {N.}~\bibnamefont
  {Schopp}}, \bibinfo {author} {\bibfnamefont {A.~L.}\ \bibnamefont {Dixon}},
  \bibinfo {author} {\bibfnamefont {M.}~\bibnamefont {Schrock}}, \bibinfo
  {author} {\bibfnamefont {V.}~\bibnamefont {N{\'{a}}da{\v{z}}dy}}, \bibinfo
  {author} {\bibfnamefont {F.}~\bibnamefont {Schauer}}, \bibinfo {author}
  {\bibfnamefont {H.}~\bibnamefont {Ade}}, \bibinfo {author} {\bibfnamefont
  {B.~F.}\ \bibnamefont {Chmelka}}, \bibinfo {author} {\bibfnamefont {G.~C.}\
  \bibnamefont {Bazan}}, \bibinfo {author} {\bibfnamefont {R.~H.}\ \bibnamefont
  {Friend}},\ and\ \bibinfo {author} {\bibfnamefont {T.}~\bibnamefont
  {Nguyen}},\ }\bibfield  {title} {\bibinfo {title} {{Unifying Charge
  Generation, Recombination, and Extraction in Low‐Offset Non‐Fullerene
  Acceptor Organic Solar Cells}},\ }\href
  {https://doi.org/10.1002/aenm.202001203} {\bibfield  {journal} {\bibinfo
  {journal} {Adv. Energy Mater.}\ }\textbf {\bibinfo {volume} {10}},\ \bibinfo
  {pages} {2001203} (\bibinfo {year} {2020})}\BibitemShut {NoStop}%
\bibitem [{\citenamefont {Alvertis}\ \emph {et~al.}(2020)\citenamefont
  {Alvertis}, \citenamefont {Pandya}, \citenamefont {Muscarella}, \citenamefont
  {Sawhney}, \citenamefont {Nguyen}, \citenamefont {Ehrler}, \citenamefont
  {Rao}, \citenamefont {Friend}, \citenamefont {Chin},\ and\ \citenamefont
  {Monserrat}}]{Alvertis2020}%
  \BibitemOpen
  \bibfield  {author} {\bibinfo {author} {\bibfnamefont {A.~M.}\ \bibnamefont
  {Alvertis}}, \bibinfo {author} {\bibfnamefont {R.}~\bibnamefont {Pandya}},
  \bibinfo {author} {\bibfnamefont {L.~A.}\ \bibnamefont {Muscarella}},
  \bibinfo {author} {\bibfnamefont {N.}~\bibnamefont {Sawhney}}, \bibinfo
  {author} {\bibfnamefont {M.}~\bibnamefont {Nguyen}}, \bibinfo {author}
  {\bibfnamefont {B.}~\bibnamefont {Ehrler}}, \bibinfo {author} {\bibfnamefont
  {A.}~\bibnamefont {Rao}}, \bibinfo {author} {\bibfnamefont {R.~H.}\
  \bibnamefont {Friend}}, \bibinfo {author} {\bibfnamefont {A.~W.}\
  \bibnamefont {Chin}},\ and\ \bibinfo {author} {\bibfnamefont
  {B.}~\bibnamefont {Monserrat}},\ }\bibfield  {title} {\bibinfo {title}
  {{Impact of Exciton Delocalization on Exciton-Vibration Interactions in
  Organic Semiconductors}},\ }\href
  {https://doi.org/10.1103/PhysRevB.102.081122} {\bibfield  {journal} {\bibinfo
   {journal} {Phys. Rev. B}\ }\textbf {\bibinfo {volume} {102}},\ \bibinfo
  {pages} {081122} (\bibinfo {year} {2020})}\BibitemShut {NoStop}%
\bibitem [{\citenamefont {Pensack}\ and\ \citenamefont
  {Asbury}(2009)}]{Pensack2009}%
  \BibitemOpen
  \bibfield  {author} {\bibinfo {author} {\bibfnamefont {R.~D.}\ \bibnamefont
  {Pensack}}\ and\ \bibinfo {author} {\bibfnamefont {J.~B.}\ \bibnamefont
  {Asbury}},\ }\bibfield  {title} {\bibinfo {title} {{Barrierless Free Carrier
  Formation in an Organic Photovoltaic Material Measured with Ultrafast
  Vibrational Spectroscopy}},\ }\href {https://doi.org/10.1021/ja906293q}
  {\bibfield  {journal} {\bibinfo  {journal} {J. Am. Chem. Soc.}\ }\textbf
  {\bibinfo {volume} {131}},\ \bibinfo {pages} {15986} (\bibinfo {year}
  {2009})}\BibitemShut {NoStop}%
\bibitem [{\citenamefont {Vandewal}\ \emph {et~al.}(2014)\citenamefont
  {Vandewal}, \citenamefont {Albrecht}, \citenamefont {Hoke}, \citenamefont
  {Graham}, \citenamefont {Widmer}, \citenamefont {Douglas}, \citenamefont
  {Schubert}, \citenamefont {Mateker}, \citenamefont {Bloking}, \citenamefont
  {Burkhard}, \citenamefont {Sellinger}, \citenamefont {Fr{\'{e}}chet},
  \citenamefont {Amassian}, \citenamefont {Riede}, \citenamefont {McGehee},
  \citenamefont {Neher},\ and\ \citenamefont {Salleo}}]{Vandewal2014}%
  \BibitemOpen
  \bibfield  {author} {\bibinfo {author} {\bibfnamefont {K.}~\bibnamefont
  {Vandewal}}, \bibinfo {author} {\bibfnamefont {S.}~\bibnamefont {Albrecht}},
  \bibinfo {author} {\bibfnamefont {E.~T.}\ \bibnamefont {Hoke}}, \bibinfo
  {author} {\bibfnamefont {K.~R.}\ \bibnamefont {Graham}}, \bibinfo {author}
  {\bibfnamefont {J.}~\bibnamefont {Widmer}}, \bibinfo {author} {\bibfnamefont
  {J.~D.}\ \bibnamefont {Douglas}}, \bibinfo {author} {\bibfnamefont
  {M.}~\bibnamefont {Schubert}}, \bibinfo {author} {\bibfnamefont {W.~R.}\
  \bibnamefont {Mateker}}, \bibinfo {author} {\bibfnamefont {J.~T.}\
  \bibnamefont {Bloking}}, \bibinfo {author} {\bibfnamefont {G.~F.}\
  \bibnamefont {Burkhard}}, \bibinfo {author} {\bibfnamefont {A.}~\bibnamefont
  {Sellinger}}, \bibinfo {author} {\bibfnamefont {J.~M.~J.}\ \bibnamefont
  {Fr{\'{e}}chet}}, \bibinfo {author} {\bibfnamefont {A.}~\bibnamefont
  {Amassian}}, \bibinfo {author} {\bibfnamefont {M.~K.}\ \bibnamefont {Riede}},
  \bibinfo {author} {\bibfnamefont {M.~D.}\ \bibnamefont {McGehee}}, \bibinfo
  {author} {\bibfnamefont {D.}~\bibnamefont {Neher}},\ and\ \bibinfo {author}
  {\bibfnamefont {A.}~\bibnamefont {Salleo}},\ }\bibfield  {title} {\bibinfo
  {title} {{Efficient Charge Generation by Relaxed Charge-Transfer States at
  Organic Interfaces}},\ }\href {https://doi.org/10.1038/nmat3807} {\bibfield
  {journal} {\bibinfo  {journal} {Nat. Mater.}\ }\textbf {\bibinfo {volume}
  {13}},\ \bibinfo {pages} {63} (\bibinfo {year} {2014})}\BibitemShut {NoStop}%
\bibitem [{\citenamefont {B{\"{a}}ssler}\ and\ \citenamefont
  {K{\"{o}}hler}(2015)}]{Bassler2015}%
  \BibitemOpen
  \bibfield  {author} {\bibinfo {author} {\bibfnamefont {H.}~\bibnamefont
  {B{\"{a}}ssler}}\ and\ \bibinfo {author} {\bibfnamefont {A.}~\bibnamefont
  {K{\"{o}}hler}},\ }\bibfield  {title} {\bibinfo {title} {{“Hot or cold”:
  How do Charge Transfer States at the Donor–Acceptor Interface of an Organic
  Solar Cell Dissociate?}},\ }\href {https://doi.org/10.1039/C5CP04110D}
  {\bibfield  {journal} {\bibinfo  {journal} {Phys. Chem. Chem. Phys.}\
  }\textbf {\bibinfo {volume} {17}},\ \bibinfo {pages} {28451} (\bibinfo {year}
  {2015})}\BibitemShut {NoStop}%
\bibitem [{\citenamefont {Falke}\ \emph {et~al.}(2014)\citenamefont {Falke},
  \citenamefont {Rozzi}, \citenamefont {Brida}, \citenamefont {Maiuri},
  \citenamefont {Amato}, \citenamefont {Sommer}, \citenamefont {Sio},
  \citenamefont {Rubio}, \citenamefont {Cerullo}, \citenamefont {Molinari},\
  and\ \citenamefont {Lienau}}]{Falke2014}%
  \BibitemOpen
  \bibfield  {author} {\bibinfo {author} {\bibfnamefont {S.~M.}\ \bibnamefont
  {Falke}}, \bibinfo {author} {\bibfnamefont {C.~A.}\ \bibnamefont {Rozzi}},
  \bibinfo {author} {\bibfnamefont {D.}~\bibnamefont {Brida}}, \bibinfo
  {author} {\bibfnamefont {M.}~\bibnamefont {Maiuri}}, \bibinfo {author}
  {\bibfnamefont {M.}~\bibnamefont {Amato}}, \bibinfo {author} {\bibfnamefont
  {E.}~\bibnamefont {Sommer}}, \bibinfo {author} {\bibfnamefont {A.~D.}\
  \bibnamefont {Sio}}, \bibinfo {author} {\bibfnamefont {A.}~\bibnamefont
  {Rubio}}, \bibinfo {author} {\bibfnamefont {G.}~\bibnamefont {Cerullo}},
  \bibinfo {author} {\bibfnamefont {E.}~\bibnamefont {Molinari}},\ and\
  \bibinfo {author} {\bibfnamefont {C.}~\bibnamefont {Lienau}},\ }\bibfield
  {title} {\bibinfo {title} {{Coherent ultrafast charge transfer in an organic
  photovoltaic blend}},\ }\href {https://doi.org/10.1126/science.1249771}
  {\bibfield  {journal} {\bibinfo  {journal} {Science}\ }\textbf {\bibinfo
  {volume} {344}},\ \bibinfo {pages} {1001} (\bibinfo {year}
  {2014})}\BibitemShut {NoStop}%
\bibitem [{\citenamefont {Song}\ \emph {et~al.}(2014)\citenamefont {Song},
  \citenamefont {Clafton}, \citenamefont {Pensack}, \citenamefont {Kee},\ and\
  \citenamefont {Scholes}}]{Song2014}%
  \BibitemOpen
  \bibfield  {author} {\bibinfo {author} {\bibfnamefont {Y.}~\bibnamefont
  {Song}}, \bibinfo {author} {\bibfnamefont {S.~N.}\ \bibnamefont {Clafton}},
  \bibinfo {author} {\bibfnamefont {R.~D.}\ \bibnamefont {Pensack}}, \bibinfo
  {author} {\bibfnamefont {T.~W.}\ \bibnamefont {Kee}},\ and\ \bibinfo {author}
  {\bibfnamefont {G.~D.}\ \bibnamefont {Scholes}},\ }\bibfield  {title}
  {\bibinfo {title} {{Vibrational coherence probes the mechanism of ultrafast
  electron transfer in polymer–fullerene blends}},\ }\href
  {https://doi.org/10.1038/ncomms5933} {\bibfield  {journal} {\bibinfo
  {journal} {Nat. Commun.}\ }\textbf {\bibinfo {volume} {5}},\ \bibinfo {pages}
  {4933} (\bibinfo {year} {2014})}\BibitemShut {NoStop}%
\bibitem [{\citenamefont {Rafiq}\ \emph {et~al.}(2015)\citenamefont {Rafiq},
  \citenamefont {Dean},\ and\ \citenamefont {Scholes}}]{Rafiq2015}%
  \BibitemOpen
  \bibfield  {author} {\bibinfo {author} {\bibfnamefont {S.}~\bibnamefont
  {Rafiq}}, \bibinfo {author} {\bibfnamefont {J.~C.}\ \bibnamefont {Dean}},\
  and\ \bibinfo {author} {\bibfnamefont {G.~D.}\ \bibnamefont {Scholes}},\
  }\bibfield  {title} {\bibinfo {title} {{Observing Vibrational Wavepackets
  during an Ultrafast Electron Transfer Reaction}},\ }\href
  {https://doi.org/10.1021/acs.jpca.5b09390} {\bibfield  {journal} {\bibinfo
  {journal} {J. Phys. Chem. A}\ }\textbf {\bibinfo {volume} {119}},\ \bibinfo
  {pages} {11837} (\bibinfo {year} {2015})}\BibitemShut {NoStop}%
\bibitem [{\citenamefont {Sio}\ \emph {et~al.}(2016)\citenamefont {Sio},
  \citenamefont {Troiani}, \citenamefont {Maiuri}, \citenamefont
  {R{\'{e}}hault}, \citenamefont {Sommer}, \citenamefont {Lim}, \citenamefont
  {Huelga}, \citenamefont {Plenio}, \citenamefont {Rozzi}, \citenamefont
  {Cerullo}, \citenamefont {Molinari},\ and\ \citenamefont
  {Lienau}}]{DeSio2016}%
  \BibitemOpen
  \bibfield  {author} {\bibinfo {author} {\bibfnamefont {A.~D.}\ \bibnamefont
  {Sio}}, \bibinfo {author} {\bibfnamefont {F.}~\bibnamefont {Troiani}},
  \bibinfo {author} {\bibfnamefont {M.}~\bibnamefont {Maiuri}}, \bibinfo
  {author} {\bibfnamefont {J.}~\bibnamefont {R{\'{e}}hault}}, \bibinfo {author}
  {\bibfnamefont {E.}~\bibnamefont {Sommer}}, \bibinfo {author} {\bibfnamefont
  {J.}~\bibnamefont {Lim}}, \bibinfo {author} {\bibfnamefont {S.~F.}\
  \bibnamefont {Huelga}}, \bibinfo {author} {\bibfnamefont {M.~B.}\
  \bibnamefont {Plenio}}, \bibinfo {author} {\bibfnamefont {C.~A.}\
  \bibnamefont {Rozzi}}, \bibinfo {author} {\bibfnamefont {G.}~\bibnamefont
  {Cerullo}}, \bibinfo {author} {\bibfnamefont {E.}~\bibnamefont {Molinari}},\
  and\ \bibinfo {author} {\bibfnamefont {C.}~\bibnamefont {Lienau}},\
  }\bibfield  {title} {\bibinfo {title} {{Tracking the Coherent Generation of
  Polaron Pairs in Conjugated Polymers}},\ }\href
  {https://doi.org/10.1038/ncomms13742} {\bibfield  {journal} {\bibinfo
  {journal} {Nat. Commun.}\ }\textbf {\bibinfo {volume} {7}},\ \bibinfo {pages}
  {13742} (\bibinfo {year} {2016})}\BibitemShut {NoStop}%
\bibitem [{\citenamefont {Br{\'{e}}das}\ \emph {et~al.}(2017)\citenamefont
  {Br{\'{e}}das}, \citenamefont {Sargent},\ and\ \citenamefont
  {Scholes}}]{Bredas2017}%
  \BibitemOpen
  \bibfield  {author} {\bibinfo {author} {\bibfnamefont {J.-L.}\ \bibnamefont
  {Br{\'{e}}das}}, \bibinfo {author} {\bibfnamefont {E.~H.}\ \bibnamefont
  {Sargent}},\ and\ \bibinfo {author} {\bibfnamefont {G.~D.}\ \bibnamefont
  {Scholes}},\ }\bibfield  {title} {\bibinfo {title} {{Photovoltaic Concepts
  Inspired by Coherence Effects in Photosynthetic Systems}},\ }\href
  {https://doi.org/10.1038/nmat4767} {\bibfield  {journal} {\bibinfo  {journal}
  {Nat. Mater.}\ }\textbf {\bibinfo {volume} {16}},\ \bibinfo {pages} {35}
  (\bibinfo {year} {2017})}\BibitemShut {NoStop}%
\bibitem [{\citenamefont {Sio}\ \emph {et~al.}(2018)\citenamefont {Sio},
  \citenamefont {Camargo}, \citenamefont {Winte}, \citenamefont {Sommer},
  \citenamefont {Branchi}, \citenamefont {Cerullo},\ and\ \citenamefont
  {Lienau}}]{DeSio2018}%
  \BibitemOpen
  \bibfield  {author} {\bibinfo {author} {\bibfnamefont {A.~D.}\ \bibnamefont
  {Sio}}, \bibinfo {author} {\bibfnamefont {F.~V. d.~A.}\ \bibnamefont
  {Camargo}}, \bibinfo {author} {\bibfnamefont {K.}~\bibnamefont {Winte}},
  \bibinfo {author} {\bibfnamefont {E.}~\bibnamefont {Sommer}}, \bibinfo
  {author} {\bibfnamefont {F.}~\bibnamefont {Branchi}}, \bibinfo {author}
  {\bibfnamefont {G.}~\bibnamefont {Cerullo}},\ and\ \bibinfo {author}
  {\bibfnamefont {C.}~\bibnamefont {Lienau}},\ }\bibfield  {title} {\bibinfo
  {title} {{Ultrafast Relaxation Dynamics in a Polymer: Fullerene Blend for
  Organic Photovoltaics Probed by Two-Dimensional Electronic Spectroscopy}},\
  }\href {https://doi.org/10.1140/epjb/e2018-90216-4} {\bibfield  {journal}
  {\bibinfo  {journal} {Eur. Phys. J. B}\ }\textbf {\bibinfo {volume} {91}},\
  \bibinfo {pages} {236} (\bibinfo {year} {2018})}\BibitemShut {NoStop}%
\bibitem [{\citenamefont {Sio}\ \emph {et~al.}(2021)\citenamefont {Sio},
  \citenamefont {Sommer}, \citenamefont {Nguyen}, \citenamefont {Gro{\ss}},
  \citenamefont {Popovi{\'{c}}}, \citenamefont {Nebgen}, \citenamefont
  {Fernandez-Alberti}, \citenamefont {Pittalis}, \citenamefont {Rozzi},
  \citenamefont {Molinari}, \citenamefont {Mena-Osteritz}, \citenamefont
  {B{\"{a}}uerle}, \citenamefont {Frauenheim}, \citenamefont {Tretiak},\ and\
  \citenamefont {Lienau}}]{DeSio2021}%
  \BibitemOpen
  \bibfield  {author} {\bibinfo {author} {\bibfnamefont {A.~D.}\ \bibnamefont
  {Sio}}, \bibinfo {author} {\bibfnamefont {E.}~\bibnamefont {Sommer}},
  \bibinfo {author} {\bibfnamefont {X.~T.}\ \bibnamefont {Nguyen}}, \bibinfo
  {author} {\bibfnamefont {L.}~\bibnamefont {Gro{\ss}}}, \bibinfo {author}
  {\bibfnamefont {D.}~\bibnamefont {Popovi{\'{c}}}}, \bibinfo {author}
  {\bibfnamefont {B.~T.}\ \bibnamefont {Nebgen}}, \bibinfo {author}
  {\bibfnamefont {S.}~\bibnamefont {Fernandez-Alberti}}, \bibinfo {author}
  {\bibfnamefont {S.}~\bibnamefont {Pittalis}}, \bibinfo {author}
  {\bibfnamefont {C.~A.}\ \bibnamefont {Rozzi}}, \bibinfo {author}
  {\bibfnamefont {E.}~\bibnamefont {Molinari}}, \bibinfo {author}
  {\bibfnamefont {E.}~\bibnamefont {Mena-Osteritz}}, \bibinfo {author}
  {\bibfnamefont {P.}~\bibnamefont {B{\"{a}}uerle}}, \bibinfo {author}
  {\bibfnamefont {T.}~\bibnamefont {Frauenheim}}, \bibinfo {author}
  {\bibfnamefont {S.}~\bibnamefont {Tretiak}},\ and\ \bibinfo {author}
  {\bibfnamefont {C.}~\bibnamefont {Lienau}},\ }\bibfield  {title} {\bibinfo
  {title} {{Intermolecular Conical Intersections in Molecular Aggregates}},\
  }\href {https://doi.org/10.1038/s41565-020-00791-2} {\bibfield  {journal}
  {\bibinfo  {journal} {Nat. Nanotechnol.}\ }\textbf {\bibinfo {volume} {16}},\
  \bibinfo {pages} {63} (\bibinfo {year} {2021})}\BibitemShut {NoStop}%
\bibitem [{\citenamefont {Smith}\ and\ \citenamefont {Chin}(2015)}]{Smith2015}%
  \BibitemOpen
  \bibfield  {author} {\bibinfo {author} {\bibfnamefont {S.~L.}\ \bibnamefont
  {Smith}}\ and\ \bibinfo {author} {\bibfnamefont {A.~W.}\ \bibnamefont
  {Chin}},\ }\bibfield  {title} {\bibinfo {title} {{Phonon-Assisted Ultrafast
  Charge Separation in the PCBM Band Structure}},\ }\href
  {https://doi.org/10.1103/PhysRevB.91.201302} {\bibfield  {journal} {\bibinfo
  {journal} {Phys. Rev. B}\ }\textbf {\bibinfo {volume} {91}},\ \bibinfo
  {pages} {201302} (\bibinfo {year} {2015})}\BibitemShut {NoStop}%
\bibitem [{\citenamefont {Lee}\ \emph {et~al.}(2015)\citenamefont {Lee},
  \citenamefont {Moix},\ and\ \citenamefont {Cao}}]{Lee2015}%
  \BibitemOpen
  \bibfield  {author} {\bibinfo {author} {\bibfnamefont {C.~K.}\ \bibnamefont
  {Lee}}, \bibinfo {author} {\bibfnamefont {J.}~\bibnamefont {Moix}},\ and\
  \bibinfo {author} {\bibfnamefont {J.}~\bibnamefont {Cao}},\ }\bibfield
  {title} {\bibinfo {title} {{Coherent Quantum Transport in Disordered Systems:
  A Unified Polaron Treatment of Hopping and Band-like Transport}},\ }\href
  {https://doi.org/10.1063/1.4918736} {\bibfield  {journal} {\bibinfo
  {journal} {J. Chem. Phys.}\ }\textbf {\bibinfo {volume} {142}},\ \bibinfo
  {pages} {164103} (\bibinfo {year} {2015})}\BibitemShut {NoStop}%
\bibitem [{\citenamefont {Kato}\ and\ \citenamefont
  {Ishizaki}(2018)}]{Kato2018}%
  \BibitemOpen
  \bibfield  {author} {\bibinfo {author} {\bibfnamefont {A.}~\bibnamefont
  {Kato}}\ and\ \bibinfo {author} {\bibfnamefont {A.}~\bibnamefont
  {Ishizaki}},\ }\bibfield  {title} {\bibinfo {title} {{Non-Markovian
  Quantum-Classical Ratchet for Ultrafast Long-Range Electron-Hole Separation
  in Condensed Phases}},\ }\href
  {https://doi.org/10.1103/PhysRevLett.121.026001} {\bibfield  {journal}
  {\bibinfo  {journal} {Phys. Rev. Lett.}\ }\textbf {\bibinfo {volume} {121}},\
  \bibinfo {pages} {026001} (\bibinfo {year} {2018})}\BibitemShut {NoStop}%
\bibitem [{\citenamefont {Tamura}\ and\ \citenamefont
  {Burghardt}(2013{\natexlab{a}})}]{Tamura2013JACS}%
  \BibitemOpen
  \bibfield  {author} {\bibinfo {author} {\bibfnamefont {H.}~\bibnamefont
  {Tamura}}\ and\ \bibinfo {author} {\bibfnamefont {I.}~\bibnamefont
  {Burghardt}},\ }\bibfield  {title} {\bibinfo {title} {{Ultrafast Charge
  Separation in Organic Photovoltaics Enhanced by Charge Delocalization and
  Vibronically Hot Exciton Dissociation}},\ }\href
  {https://doi.org/10.1021/ja4093874} {\bibfield  {journal} {\bibinfo
  {journal} {J. Am. Chem. Soc.}\ }\textbf {\bibinfo {volume} {135}},\ \bibinfo
  {pages} {16364} (\bibinfo {year} {2013}{\natexlab{a}})}\BibitemShut {NoStop}%
\bibitem [{\citenamefont {Tamura}\ and\ \citenamefont
  {Burghardt}(2013{\natexlab{b}})}]{Tamura2013JPCC}%
  \BibitemOpen
  \bibfield  {author} {\bibinfo {author} {\bibfnamefont {H.}~\bibnamefont
  {Tamura}}\ and\ \bibinfo {author} {\bibfnamefont {I.}~\bibnamefont
  {Burghardt}},\ }\bibfield  {title} {\bibinfo {title} {{Potential Barrier and
  Excess Energy for Electron–Hole Separation from the Charge-Transfer Exciton
  at Donor–Acceptor Heterojunctions of Organic Solar Cells}},\ }\href
  {https://doi.org/10.1021/jp406224a} {\bibfield  {journal} {\bibinfo
  {journal} {J. Phys. Chem. C}\ }\textbf {\bibinfo {volume} {117}},\ \bibinfo
  {pages} {15020} (\bibinfo {year} {2013}{\natexlab{b}})}\BibitemShut {NoStop}%
\bibitem [{\citenamefont {Hughes}\ \emph {et~al.}(2014)\citenamefont {Hughes},
  \citenamefont {Cahier}, \citenamefont {Martinazzo}, \citenamefont {Tamura},\
  and\ \citenamefont {Burghardt}}]{Hughes2014}%
  \BibitemOpen
  \bibfield  {author} {\bibinfo {author} {\bibfnamefont {K.~H.}\ \bibnamefont
  {Hughes}}, \bibinfo {author} {\bibfnamefont {B.}~\bibnamefont {Cahier}},
  \bibinfo {author} {\bibfnamefont {R.}~\bibnamefont {Martinazzo}}, \bibinfo
  {author} {\bibfnamefont {H.}~\bibnamefont {Tamura}},\ and\ \bibinfo {author}
  {\bibfnamefont {I.}~\bibnamefont {Burghardt}},\ }\bibfield  {title} {\bibinfo
  {title} {{Non-Markovian Reduced Dynamics of Ultrafast Charge Transfer at an
  Oligothiophene-Fullerene Heterojunction}},\ }\href
  {https://doi.org/10.1016/j.chemphys.2014.06.015} {\bibfield  {journal}
  {\bibinfo  {journal} {Chem. Phys.}\ }\textbf {\bibinfo {volume} {442}},\
  \bibinfo {pages} {111} (\bibinfo {year} {2014})}\BibitemShut {NoStop}%
\bibitem [{\citenamefont {Chenel}\ \emph {et~al.}(2014)\citenamefont {Chenel},
  \citenamefont {Mangaud}, \citenamefont {Burghardt}, \citenamefont {Meier},\
  and\ \citenamefont {Desouter-Lecomte}}]{Chenel2014}%
  \BibitemOpen
  \bibfield  {author} {\bibinfo {author} {\bibfnamefont {A.}~\bibnamefont
  {Chenel}}, \bibinfo {author} {\bibfnamefont {E.}~\bibnamefont {Mangaud}},
  \bibinfo {author} {\bibfnamefont {I.}~\bibnamefont {Burghardt}}, \bibinfo
  {author} {\bibfnamefont {C.}~\bibnamefont {Meier}},\ and\ \bibinfo {author}
  {\bibfnamefont {M.}~\bibnamefont {Desouter-Lecomte}},\ }\bibfield  {title}
  {\bibinfo {title} {{Exciton Dissociation at Donor-Acceptor Heterojunctions:
  Dynamics Using the Collective Effective Mode Representation of the Spin-Boson
  Model}},\ }\href {https://doi.org/10.1063/1.4861853} {\bibfield  {journal}
  {\bibinfo  {journal} {J. Chem. Phys.}\ }\textbf {\bibinfo {volume} {140}},\
  \bibinfo {pages} {044104} (\bibinfo {year} {2014})}\BibitemShut {NoStop}%
\bibitem [{\citenamefont {Huix-Rotllant}\ \emph {et~al.}(2015)\citenamefont
  {Huix-Rotllant}, \citenamefont {Tamura},\ and\ \citenamefont
  {Burghardt}}]{Rotllant2015}%
  \BibitemOpen
  \bibfield  {author} {\bibinfo {author} {\bibfnamefont {M.}~\bibnamefont
  {Huix-Rotllant}}, \bibinfo {author} {\bibfnamefont {H.}~\bibnamefont
  {Tamura}},\ and\ \bibinfo {author} {\bibfnamefont {I.}~\bibnamefont
  {Burghardt}},\ }\bibfield  {title} {\bibinfo {title} {{Concurrent Effects of
  Delocalization and Internal Conversion Tune Charge Separation at Regioregular
  Polythiophene–Fullerene Heterojunctions}},\ }\href
  {https://doi.org/10.1021/acs.jpclett.5b00336} {\bibfield  {journal} {\bibinfo
   {journal} {J. Phys. Chem. Lett.}\ }\textbf {\bibinfo {volume} {6}},\
  \bibinfo {pages} {1702} (\bibinfo {year} {2015})}\BibitemShut {NoStop}%
\bibitem [{\citenamefont {Chaudhuri}\ \emph {et~al.}(2017)\citenamefont
  {Chaudhuri}, \citenamefont {Hedstr{\"{o}}m}, \citenamefont
  {M{\'{e}}ndez-Hern{\'{a}}ndez}, \citenamefont {Hendrickson}, \citenamefont
  {Jung}, \citenamefont {Ho},\ and\ \citenamefont {Batista}}]{Chaudhuri2017}%
  \BibitemOpen
  \bibfield  {author} {\bibinfo {author} {\bibfnamefont {S.}~\bibnamefont
  {Chaudhuri}}, \bibinfo {author} {\bibfnamefont {S.}~\bibnamefont
  {Hedstr{\"{o}}m}}, \bibinfo {author} {\bibfnamefont {D.~D.}\ \bibnamefont
  {M{\'{e}}ndez-Hern{\'{a}}ndez}}, \bibinfo {author} {\bibfnamefont {H.~P.}\
  \bibnamefont {Hendrickson}}, \bibinfo {author} {\bibfnamefont {K.~A.}\
  \bibnamefont {Jung}}, \bibinfo {author} {\bibfnamefont {J.}~\bibnamefont
  {Ho}},\ and\ \bibinfo {author} {\bibfnamefont {V.~S.}\ \bibnamefont
  {Batista}},\ }\bibfield  {title} {\bibinfo {title} {{Electron Transfer
  Assisted by Vibronic Coupling from Multiple Modes}},\ }\href
  {https://doi.org/10.1021/acs.jctc.7b00513} {\bibfield  {journal} {\bibinfo
  {journal} {J. Chem. Theory Comput.}\ }\textbf {\bibinfo {volume} {13}},\
  \bibinfo {pages} {6000} (\bibinfo {year} {2017})}\BibitemShut {NoStop}%
\bibitem [{\citenamefont {Polkehn}\ \emph {et~al.}(2018)\citenamefont
  {Polkehn}, \citenamefont {Tamura},\ and\ \citenamefont
  {Burghardt}}]{Polkehn2018}%
  \BibitemOpen
  \bibfield  {author} {\bibinfo {author} {\bibfnamefont {M.}~\bibnamefont
  {Polkehn}}, \bibinfo {author} {\bibfnamefont {H.}~\bibnamefont {Tamura}},\
  and\ \bibinfo {author} {\bibfnamefont {I.}~\bibnamefont {Burghardt}},\
  }\bibfield  {title} {\bibinfo {title} {{Impact of Charge-Transfer Excitons in
  Regioregular Polythiophene on the Charge Separation at
  Polythiophene-Fullerene Heterojunctions}},\ }\href
  {https://doi.org/10.1088/1361-6455/aa93d0} {\bibfield  {journal} {\bibinfo
  {journal} {J. Phys.B}\ }\textbf {\bibinfo {volume} {51}},\ \bibinfo {pages}
  {014003} (\bibinfo {year} {2018})}\BibitemShut {NoStop}%
\bibitem [{\citenamefont {Bian}\ \emph {et~al.}(2020)\citenamefont {Bian},
  \citenamefont {Ma}, \citenamefont {Chen}, \citenamefont {Wei}, \citenamefont
  {Su}, \citenamefont {Buyanova}, \citenamefont {Chen}, \citenamefont
  {Ponseca}, \citenamefont {Linares}, \citenamefont {Karki}, \citenamefont
  {Yartsev},\ and\ \citenamefont {Ingan{\"{a}}s}}]{Bian2020}%
  \BibitemOpen
  \bibfield  {author} {\bibinfo {author} {\bibfnamefont {Q.}~\bibnamefont
  {Bian}}, \bibinfo {author} {\bibfnamefont {F.}~\bibnamefont {Ma}}, \bibinfo
  {author} {\bibfnamefont {S.}~\bibnamefont {Chen}}, \bibinfo {author}
  {\bibfnamefont {Q.}~\bibnamefont {Wei}}, \bibinfo {author} {\bibfnamefont
  {X.}~\bibnamefont {Su}}, \bibinfo {author} {\bibfnamefont {I.~A.}\
  \bibnamefont {Buyanova}}, \bibinfo {author} {\bibfnamefont {W.~M.}\
  \bibnamefont {Chen}}, \bibinfo {author} {\bibfnamefont {C.~S.}\ \bibnamefont
  {Ponseca}}, \bibinfo {author} {\bibfnamefont {M.}~\bibnamefont {Linares}},
  \bibinfo {author} {\bibfnamefont {K.~J.}\ \bibnamefont {Karki}}, \bibinfo
  {author} {\bibfnamefont {A.}~\bibnamefont {Yartsev}},\ and\ \bibinfo {author}
  {\bibfnamefont {O.}~\bibnamefont {Ingan{\"{a}}s}},\ }\bibfield  {title}
  {\bibinfo {title} {{Vibronic Coherence Contributes to Photocurrent Generation
  in Organic Semiconductor Heterojunction Diodes}},\ }\href
  {https://doi.org/10.1038/s41467-020-14476-w} {\bibfield  {journal} {\bibinfo
  {journal} {Nat. Commun.}\ }\textbf {\bibinfo {volume} {11}},\ \bibinfo
  {pages} {617} (\bibinfo {year} {2020})}\BibitemShut {NoStop}%
\bibitem [{\citenamefont {Yao}\ \emph {et~al.}(2016)\citenamefont {Yao},
  \citenamefont {Xie},\ and\ \citenamefont {Ma}}]{Yao2016a}%
  \BibitemOpen
  \bibfield  {author} {\bibinfo {author} {\bibfnamefont {Y.}~\bibnamefont
  {Yao}}, \bibinfo {author} {\bibfnamefont {X.}~\bibnamefont {Xie}},\ and\
  \bibinfo {author} {\bibfnamefont {H.}~\bibnamefont {Ma}},\ }\bibfield
  {title} {\bibinfo {title} {{Ultrafast Long-Range Charge Separation in Organic
  Photovoltaics: Promotion by Off-Diagonal Vibronic Couplings and Entropy
  Increase}},\ }\href {https://doi.org/10.1021/acs.jpclett.6b02400} {\bibfield
  {journal} {\bibinfo  {journal} {J. Phys. Chem. Lett.}\ }\textbf {\bibinfo
  {volume} {7}},\ \bibinfo {pages} {4830} (\bibinfo {year} {2016})}\BibitemShut
  {NoStop}%
\bibitem [{\citenamefont {Duan}\ \emph {et~al.}(2020)\citenamefont {Duan},
  \citenamefont {Jha}, \citenamefont {Tiwari}, \citenamefont {Miller},\ and\
  \citenamefont {Thorwart}}]{Duan2020}%
  \BibitemOpen
  \bibfield  {author} {\bibinfo {author} {\bibfnamefont {H.~G.}\ \bibnamefont
  {Duan}}, \bibinfo {author} {\bibfnamefont {A.}~\bibnamefont {Jha}}, \bibinfo
  {author} {\bibfnamefont {V.}~\bibnamefont {Tiwari}}, \bibinfo {author}
  {\bibfnamefont {R.~J.}\ \bibnamefont {Miller}},\ and\ \bibinfo {author}
  {\bibfnamefont {M.}~\bibnamefont {Thorwart}},\ }\bibfield  {title} {\bibinfo
  {title} {{Dissociation and Localization Dynamics of Charge Transfer Excitons
  at a Donor-Acceptor Interface}},\ }\href
  {https://doi.org/10.1016/j.chemphys.2019.110525} {\bibfield  {journal}
  {\bibinfo  {journal} {Chem. Phys.}\ }\textbf {\bibinfo {volume} {528}},\
  \bibinfo {pages} {110525} (\bibinfo {year} {2020})}\BibitemShut {NoStop}%
\bibitem [{\citenamefont {Tamascelli}\ \emph {et~al.}(2018)\citenamefont
  {Tamascelli}, \citenamefont {Smirne}, \citenamefont {Huelga},\ and\
  \citenamefont {Plenio}}]{Tamascelli2018}%
  \BibitemOpen
  \bibfield  {author} {\bibinfo {author} {\bibfnamefont {D.}~\bibnamefont
  {Tamascelli}}, \bibinfo {author} {\bibfnamefont {A.}~\bibnamefont {Smirne}},
  \bibinfo {author} {\bibfnamefont {S.~F.}\ \bibnamefont {Huelga}},\ and\
  \bibinfo {author} {\bibfnamefont {M.~B.}\ \bibnamefont {Plenio}},\ }\bibfield
   {title} {\bibinfo {title} {{Nonperturbative Treatment of Non-Markovian
  Dynamics of Open Quantum Systems}},\ }\href
  {https://doi.org/10.1103/PhysRevLett.120.030402} {\bibfield  {journal}
  {\bibinfo  {journal} {Phys. Rev. Lett.}\ }\textbf {\bibinfo {volume} {120}},\
  \bibinfo {pages} {30402} (\bibinfo {year} {2018})}\BibitemShut {NoStop}%
\bibitem [{\citenamefont {Somoza}\ \emph {et~al.}(2019)\citenamefont {Somoza},
  \citenamefont {Marty}, \citenamefont {Lim}, \citenamefont {Huelga},\ and\
  \citenamefont {Plenio}}]{Somoza2019}%
  \BibitemOpen
  \bibfield  {author} {\bibinfo {author} {\bibfnamefont {A.~D.}\ \bibnamefont
  {Somoza}}, \bibinfo {author} {\bibfnamefont {O.}~\bibnamefont {Marty}},
  \bibinfo {author} {\bibfnamefont {J.}~\bibnamefont {Lim}}, \bibinfo {author}
  {\bibfnamefont {S.~F.}\ \bibnamefont {Huelga}},\ and\ \bibinfo {author}
  {\bibfnamefont {M.~B.}\ \bibnamefont {Plenio}},\ }\bibfield  {title}
  {\bibinfo {title} {{Dissipation-Assisted Matrix Product Factorization}},\
  }\href {https://doi.org/10.1103/PhysRevLett.123.100502} {\bibfield  {journal}
  {\bibinfo  {journal} {Phys. Rev. Lett.}\ }\textbf {\bibinfo {volume} {123}},\
  \bibinfo {pages} {100502} (\bibinfo {year} {2019})}\BibitemShut {NoStop}%
\bibitem [{\citenamefont {Tamascelli}\ \emph {et~al.}(2019)\citenamefont
  {Tamascelli}, \citenamefont {Smirne}, \citenamefont {Lim}, \citenamefont
  {Huelga},\ and\ \citenamefont {Plenio}}]{Tamascelli2019}%
  \BibitemOpen
  \bibfield  {author} {\bibinfo {author} {\bibfnamefont {D.}~\bibnamefont
  {Tamascelli}}, \bibinfo {author} {\bibfnamefont {A.}~\bibnamefont {Smirne}},
  \bibinfo {author} {\bibfnamefont {J.}~\bibnamefont {Lim}}, \bibinfo {author}
  {\bibfnamefont {S.~F.}\ \bibnamefont {Huelga}},\ and\ \bibinfo {author}
  {\bibfnamefont {M.~B.}\ \bibnamefont {Plenio}},\ }\bibfield  {title}
  {\bibinfo {title} {{Efficient Simulation of Finite-Temperature Open Quantum
  Systems}},\ }\href {https://doi.org/10.1103/PhysRevLett.123.090402}
  {\bibfield  {journal} {\bibinfo  {journal} {Phys. Rev. Lett.}\ }\textbf
  {\bibinfo {volume} {123}},\ \bibinfo {pages} {090402} (\bibinfo {year}
  {2019})}\BibitemShut {NoStop}%
\bibitem [{\citenamefont {Mascherpa}\ \emph {et~al.}(2020)\citenamefont
  {Mascherpa}, \citenamefont {Smirne}, \citenamefont {Somoza}, \citenamefont
  {Fern{\'{a}}ndez-Acebal}, \citenamefont {Donadi}, \citenamefont {Tamascelli},
  \citenamefont {Huelga},\ and\ \citenamefont {Plenio}}]{Mascherpa2020}%
  \BibitemOpen
  \bibfield  {author} {\bibinfo {author} {\bibfnamefont {F.}~\bibnamefont
  {Mascherpa}}, \bibinfo {author} {\bibfnamefont {A.}~\bibnamefont {Smirne}},
  \bibinfo {author} {\bibfnamefont {A.~D.}\ \bibnamefont {Somoza}}, \bibinfo
  {author} {\bibfnamefont {P.}~\bibnamefont {Fern{\'{a}}ndez-Acebal}}, \bibinfo
  {author} {\bibfnamefont {S.}~\bibnamefont {Donadi}}, \bibinfo {author}
  {\bibfnamefont {D.}~\bibnamefont {Tamascelli}}, \bibinfo {author}
  {\bibfnamefont {S.~F.}\ \bibnamefont {Huelga}},\ and\ \bibinfo {author}
  {\bibfnamefont {M.~B.}\ \bibnamefont {Plenio}},\ }\bibfield  {title}
  {\bibinfo {title} {{Optimized Auxiliary Oscillators for the Simulation of
  General Open Quantum Systems}},\ }\href
  {https://doi.org/10.1103/PhysRevA.101.052108} {\bibfield  {journal} {\bibinfo
   {journal} {Phys. Rev. A}\ }\textbf {\bibinfo {volume} {101}},\ \bibinfo
  {pages} {052108} (\bibinfo {year} {2020})}\BibitemShut {NoStop}%
\bibitem [{\citenamefont {Gelinas}\ \emph {et~al.}(2014)\citenamefont
  {Gelinas}, \citenamefont {Rao}, \citenamefont {Kumar}, \citenamefont {Smith},
  \citenamefont {Chin}, \citenamefont {Clark}, \citenamefont {van~der Poll},
  \citenamefont {Bazan},\ and\ \citenamefont {Friend}}]{Gelinas2014Science}%
  \BibitemOpen
  \bibfield  {author} {\bibinfo {author} {\bibfnamefont {S.}~\bibnamefont
  {Gelinas}}, \bibinfo {author} {\bibfnamefont {A.}~\bibnamefont {Rao}},
  \bibinfo {author} {\bibfnamefont {A.}~\bibnamefont {Kumar}}, \bibinfo
  {author} {\bibfnamefont {S.~L.}\ \bibnamefont {Smith}}, \bibinfo {author}
  {\bibfnamefont {A.~W.}\ \bibnamefont {Chin}}, \bibinfo {author}
  {\bibfnamefont {J.}~\bibnamefont {Clark}}, \bibinfo {author} {\bibfnamefont
  {T.~S.}\ \bibnamefont {van~der Poll}}, \bibinfo {author} {\bibfnamefont
  {G.~C.}\ \bibnamefont {Bazan}},\ and\ \bibinfo {author} {\bibfnamefont
  {R.~H.}\ \bibnamefont {Friend}},\ }\bibfield  {title} {\bibinfo {title}
  {{Ultrafast Long-Range Charge Separation in Organic Semiconductor
  Photovoltaic Diodes}},\ }\href {https://doi.org/10.1126/science.1246249}
  {\bibfield  {journal} {\bibinfo  {journal} {Science}\ }\textbf {\bibinfo
  {volume} {343}},\ \bibinfo {pages} {512} (\bibinfo {year}
  {2014})}\BibitemShut {NoStop}%
\bibitem [{\citenamefont {Nan}\ \emph {et~al.}(2015)\citenamefont {Nan},
  \citenamefont {Zhang},\ and\ \citenamefont {Lu}}]{Nan2015}%
  \BibitemOpen
  \bibfield  {author} {\bibinfo {author} {\bibfnamefont {G.}~\bibnamefont
  {Nan}}, \bibinfo {author} {\bibfnamefont {X.}~\bibnamefont {Zhang}},\ and\
  \bibinfo {author} {\bibfnamefont {G.}~\bibnamefont {Lu}},\ }\bibfield
  {title} {\bibinfo {title} {{Do “Hot” Charge-Transfer Excitons Promote
  Free Carrier Generation in Organic Photovoltaics?}},\ }\href
  {https://doi.org/10.1021/acs.jpcc.5b04652} {\bibfield  {journal} {\bibinfo
  {journal} {J. Phys. Chem. C}\ }\textbf {\bibinfo {volume} {119}},\ \bibinfo
  {pages} {15028} (\bibinfo {year} {2015})}\BibitemShut {NoStop}%
\bibitem [{\citenamefont {Dennler}\ \emph {et~al.}(2009)\citenamefont
  {Dennler}, \citenamefont {Scharber},\ and\ \citenamefont
  {Brabec}}]{Dennler2009}%
  \BibitemOpen
  \bibfield  {author} {\bibinfo {author} {\bibfnamefont {G.}~\bibnamefont
  {Dennler}}, \bibinfo {author} {\bibfnamefont {M.~C.}\ \bibnamefont
  {Scharber}},\ and\ \bibinfo {author} {\bibfnamefont {C.~J.}\ \bibnamefont
  {Brabec}},\ }\bibfield  {title} {\bibinfo {title} {{Polymer-Fullerene
  Bulk-Heterojunction Solar Cells}},\ }\href
  {https://doi.org/10.1002/adma.200801283} {\bibfield  {journal} {\bibinfo
  {journal} {Adv. Mater.}\ }\textbf {\bibinfo {volume} {21}},\ \bibinfo {pages}
  {1323} (\bibinfo {year} {2009})}\BibitemShut {NoStop}%
\bibitem [{\citenamefont {Cheung}\ and\ \citenamefont
  {Troisi}(2010)}]{Cheung2010}%
  \BibitemOpen
  \bibfield  {author} {\bibinfo {author} {\bibfnamefont {D.~L.}\ \bibnamefont
  {Cheung}}\ and\ \bibinfo {author} {\bibfnamefont {A.}~\bibnamefont
  {Troisi}},\ }\bibfield  {title} {\bibinfo {title} {{Theoretical Study of the
  Organic Photovoltaic Electron Acceptor PCBM: Morphology, Electronic
  Structure, and Charge Localization}},\ }\href
  {https://doi.org/10.1021/jp1049167} {\bibfield  {journal} {\bibinfo
  {journal} {J. Phys. Chem. C}\ }\textbf {\bibinfo {volume} {114}},\ \bibinfo
  {pages} {20479} (\bibinfo {year} {2010})}\BibitemShut {NoStop}%
\bibitem [{\citenamefont {Id{\'{e}}}\ \emph {et~al.}(2014)\citenamefont
  {Id{\'{e}}}, \citenamefont {Fazzi}, \citenamefont {Casalegno}, \citenamefont
  {Meille},\ and\ \citenamefont {Raos}}]{Ide2014}%
  \BibitemOpen
  \bibfield  {author} {\bibinfo {author} {\bibfnamefont {J.}~\bibnamefont
  {Id{\'{e}}}}, \bibinfo {author} {\bibfnamefont {D.}~\bibnamefont {Fazzi}},
  \bibinfo {author} {\bibfnamefont {M.}~\bibnamefont {Casalegno}}, \bibinfo
  {author} {\bibfnamefont {S.~V.}\ \bibnamefont {Meille}},\ and\ \bibinfo
  {author} {\bibfnamefont {G.}~\bibnamefont {Raos}},\ }\bibfield  {title}
  {\bibinfo {title} {{Electron transport in crystalline PCBM-like fullerene
  derivatives: A comparative computational study}},\ }\href
  {https://doi.org/10.1039/c4tc00502c} {\bibfield  {journal} {\bibinfo
  {journal} {J. Mater. Chem. C}\ }\textbf {\bibinfo {volume} {2}},\ \bibinfo
  {pages} {7313} (\bibinfo {year} {2014})}\BibitemShut {NoStop}%
\bibitem [{\citenamefont {Chen}\ \emph
  {et~al.}(2018{\natexlab{b}})\citenamefont {Chen}, \citenamefont
  {Coropceanu},\ and\ \citenamefont {Br{\'{e}}das}}]{Chen2018}%
  \BibitemOpen
  \bibfield  {author} {\bibinfo {author} {\bibfnamefont {X.-K.}\ \bibnamefont
  {Chen}}, \bibinfo {author} {\bibfnamefont {V.}~\bibnamefont {Coropceanu}},\
  and\ \bibinfo {author} {\bibfnamefont {J.-L.}\ \bibnamefont {Br{\'{e}}das}},\
  }\bibfield  {title} {\bibinfo {title} {{Assessing the Nature of the
  Charge-Transfer Electronic States in Organic Solar Cells}},\ }\href
  {https://doi.org/10.1038/s41467-018-07707-8} {\bibfield  {journal} {\bibinfo
  {journal} {Nat. Commun.}\ }\textbf {\bibinfo {volume} {9}},\ \bibinfo {pages}
  {5295} (\bibinfo {year} {2018}{\natexlab{b}})}\BibitemShut {NoStop}%
\bibitem [{\citenamefont {Chen}\ and\ \citenamefont
  {Br{\'{e}}das}(2018)}]{Chen2018AdvMater}%
  \BibitemOpen
  \bibfield  {author} {\bibinfo {author} {\bibfnamefont {X.-K.}\ \bibnamefont
  {Chen}}\ and\ \bibinfo {author} {\bibfnamefont {J.}~\bibnamefont
  {Br{\'{e}}das}},\ }\bibfield  {title} {\bibinfo {title} {{Voltage Losses in
  Organic Solar Cells: Understanding the Contributions of Intramolecular
  Vibrations to Nonradiative Recombinations}},\ }\href
  {https://doi.org/10.1002/aenm.201702227} {\bibfield  {journal} {\bibinfo
  {journal} {Adv. Energy Mater.}\ }\textbf {\bibinfo {volume} {8}},\ \bibinfo
  {pages} {1702227} (\bibinfo {year} {2018})}\BibitemShut {NoStop}%
\bibitem [{\citenamefont {Menke}\ \emph
  {et~al.}(2018{\natexlab{b}})\citenamefont {Menke}, \citenamefont {Cheminal},
  \citenamefont {Conaghan}, \citenamefont {Ran}, \citenamefont {Greehnam},
  \citenamefont {Bazan}, \citenamefont {Nguyen}, \citenamefont {Rao},\ and\
  \citenamefont {Friend}}]{Menke2018}%
  \BibitemOpen
  \bibfield  {author} {\bibinfo {author} {\bibfnamefont {S.~M.}\ \bibnamefont
  {Menke}}, \bibinfo {author} {\bibfnamefont {A.}~\bibnamefont {Cheminal}},
  \bibinfo {author} {\bibfnamefont {P.}~\bibnamefont {Conaghan}}, \bibinfo
  {author} {\bibfnamefont {N.~A.}\ \bibnamefont {Ran}}, \bibinfo {author}
  {\bibfnamefont {N.~C.}\ \bibnamefont {Greehnam}}, \bibinfo {author}
  {\bibfnamefont {G.~C.}\ \bibnamefont {Bazan}}, \bibinfo {author}
  {\bibfnamefont {T.-Q.}\ \bibnamefont {Nguyen}}, \bibinfo {author}
  {\bibfnamefont {A.}~\bibnamefont {Rao}},\ and\ \bibinfo {author}
  {\bibfnamefont {R.~H.}\ \bibnamefont {Friend}},\ }\bibfield  {title}
  {\bibinfo {title} {{Order Enables Efficient Electron-Hole Separation at an
  Organic Heterojunction with a Small Energy Loss}},\ }\href
  {https://doi.org/10.1038/s41467-017-02457-5} {\bibfield  {journal} {\bibinfo
  {journal} {Nat. Commun.}\ }\textbf {\bibinfo {volume} {9}},\ \bibinfo {pages}
  {277} (\bibinfo {year} {2018}{\natexlab{b}})}\BibitemShut {NoStop}%
\bibitem [{\citenamefont {Hinrichsen}\ \emph {et~al.}(2020)\citenamefont
  {Hinrichsen}, \citenamefont {Chan}, \citenamefont {Ma}, \citenamefont
  {Pale{\v{c}}ek}, \citenamefont {Gillett}, \citenamefont {Chen}, \citenamefont
  {Zou}, \citenamefont {Zhang}, \citenamefont {Yip}, \citenamefont {Wong},
  \citenamefont {Friend}, \citenamefont {Yan}, \citenamefont {Rao},\ and\
  \citenamefont {Chow}}]{Hinrichsen2020}%
  \BibitemOpen
  \bibfield  {author} {\bibinfo {author} {\bibfnamefont {T.~F.}\ \bibnamefont
  {Hinrichsen}}, \bibinfo {author} {\bibfnamefont {C.~C.~S.}\ \bibnamefont
  {Chan}}, \bibinfo {author} {\bibfnamefont {C.}~\bibnamefont {Ma}}, \bibinfo
  {author} {\bibfnamefont {D.}~\bibnamefont {Pale{\v{c}}ek}}, \bibinfo {author}
  {\bibfnamefont {A.}~\bibnamefont {Gillett}}, \bibinfo {author} {\bibfnamefont
  {S.}~\bibnamefont {Chen}}, \bibinfo {author} {\bibfnamefont {X.}~\bibnamefont
  {Zou}}, \bibinfo {author} {\bibfnamefont {G.}~\bibnamefont {Zhang}}, \bibinfo
  {author} {\bibfnamefont {H.-L.}\ \bibnamefont {Yip}}, \bibinfo {author}
  {\bibfnamefont {K.~S.}\ \bibnamefont {Wong}}, \bibinfo {author}
  {\bibfnamefont {R.~H.}\ \bibnamefont {Friend}}, \bibinfo {author}
  {\bibfnamefont {H.}~\bibnamefont {Yan}}, \bibinfo {author} {\bibfnamefont
  {A.}~\bibnamefont {Rao}},\ and\ \bibinfo {author} {\bibfnamefont {P.~C.~Y.}\
  \bibnamefont {Chow}},\ }\bibfield  {title} {\bibinfo {title} {{Long-lived and
  Disorder-free Charge Transfer States Enable Endothermic Charge Separation in
  Efficient Non-Fullerene Organic Solar Cells}},\ }\href
  {https://doi.org/10.1038/s41467-020-19332-5} {\bibfield  {journal} {\bibinfo
  {journal} {Nat. Commun.}\ }\textbf {\bibinfo {volume} {11}},\ \bibinfo
  {pages} {5617} (\bibinfo {year} {2020})}\BibitemShut {NoStop}%
\bibitem [{\citenamefont {Gregg}(2011)}]{Gregg2011}%
  \BibitemOpen
  \bibfield  {author} {\bibinfo {author} {\bibfnamefont {B.~A.}\ \bibnamefont
  {Gregg}},\ }\bibfield  {title} {\bibinfo {title} {{Entropy of Charge
  Separation in Organic Photovoltaic Cells: The Benefit of Higher
  Dimensionality}},\ }\href {https://doi.org/10.1021/jz2012403} {\bibfield
  {journal} {\bibinfo  {journal} {J. Phys. Chem. Lett.}\ }\textbf {\bibinfo
  {volume} {2}},\ \bibinfo {pages} {3013} (\bibinfo {year} {2011})}\BibitemShut
  {NoStop}%
\bibitem [{\citenamefont {Ono}\ and\ \citenamefont {Ohno}(2016)}]{Ono2016}%
  \BibitemOpen
  \bibfield  {author} {\bibinfo {author} {\bibfnamefont {S.}~\bibnamefont
  {Ono}}\ and\ \bibinfo {author} {\bibfnamefont {K.}~\bibnamefont {Ohno}},\
  }\bibfield  {title} {\bibinfo {title} {{Combined Impact of Entropy and
  Carrier Delocalization on Charge Transfer Exciton Dissociation at the
  Donor-Acceptor Iolution in multidimensional ultrnterface}},\ }\href
  {https://doi.org/10.1103/PhysRevB.94.075305} {\bibfield  {journal} {\bibinfo
  {journal} {Phys. Rev. B}\ }\textbf {\bibinfo {volume} {94}},\ \bibinfo
  {pages} {075305} (\bibinfo {year} {2016})}\BibitemShut {NoStop}%
\bibitem [{\citenamefont {Tanimura}\ and\ \citenamefont
  {Kubo}(1989)}]{Tanimura1989}%
  \BibitemOpen
  \bibfield  {author} {\bibinfo {author} {\bibfnamefont {Y.}~\bibnamefont
  {Tanimura}}\ and\ \bibinfo {author} {\bibfnamefont {R.}~\bibnamefont
  {Kubo}},\ }\bibfield  {title} {\bibinfo {title} {{Time Evolution of a Quantum
  System in Contact with a Nearly Gaussian-Markoffian Noise Bath}},\ }\href
  {https://doi.org/10.1143/JPSJ.58.101} {\bibfield  {journal} {\bibinfo
  {journal} {J. Phys. Soc. Jpn.}\ }\textbf {\bibinfo {volume} {58}},\ \bibinfo
  {pages} {101} (\bibinfo {year} {1989})}\BibitemShut {NoStop}%
\bibitem [{\citenamefont {Tanimura}(2006)}]{Tanimura2006}%
  \BibitemOpen
  \bibfield  {author} {\bibinfo {author} {\bibfnamefont {Y.}~\bibnamefont
  {Tanimura}},\ }\bibfield  {title} {\bibinfo {title} {{Stochastic Liouville,
  Langevin, Fokker–Planck, and Master Equation Approaches to Quantum
  Dissipative Systems}},\ }\href {https://doi.org/10.1143/JPSJ.75.082001}
  {\bibfield  {journal} {\bibinfo  {journal} {J. Phys. Soc. Jpn.}\ }\textbf
  {\bibinfo {volume} {75}},\ \bibinfo {pages} {082001} (\bibinfo {year}
  {2006})}\BibitemShut {NoStop}%
\bibitem [{\citenamefont {Rivas}\ and\ \citenamefont
  {Huelga}(2012)}]{RivasSusana2012Book}%
  \BibitemOpen
  \bibfield  {author} {\bibinfo {author} {\bibfnamefont {{\'{A}}.}~\bibnamefont
  {Rivas}}\ and\ \bibinfo {author} {\bibfnamefont {S.~F.}\ \bibnamefont
  {Huelga}},\ }\href {http://link.springer.com/10.1007/978-3-642-23354-8}
  {\emph {\bibinfo {title} {{Open Quantum Systems}}}},\ SpringerBriefs in
  Physics\ (\bibinfo  {publisher} {Springer Berlin Heidelberg},\ \bibinfo
  {year} {2012})\BibitemShut {NoStop}%
\end{thebibliography}%


%

\end{document}